\documentclass[aps,prd,superscriptaddress,showkeys,nofootinbib]{revtex4-2}
\usepackage[english]{babel}
\usepackage{amsmath,bm}
\usepackage{amssymb}
\usepackage{enumitem}
\usepackage{textcomp}
\usepackage{graphicx}
\usepackage{dsfont}
\usepackage{color}

\usepackage[breaklinks=true]{hyperref}
\hypersetup{
	colorlinks=true,
	linkcolor=blue,
	filecolor=magenta,      
	urlcolor=blue,
	citecolor=red,
}
\usepackage{float}
\usepackage{cleveref}

\usepackage{mathrsfs}
\usepackage[normalem]{ulem}

\usepackage{placeins}

\graphicspath{{Figures/}}

\begin{document}
	
	\title{Generation of an electromagnetic field nonminimally coupled to gravity during Higgs inflation}
	
	\author{O.O.~Sobol}
	\email{oleksandr.sobol@epfl.ch}
	\affiliation{Institute of Physics, Laboratory for Particle Physics and Cosmology, \'{E}cole Polytechnique F\'{e}d\'{e}rale de Lausanne, CH-1015 Lausanne, Switzerland}
	\affiliation{Physics Faculty, Taras Shevchenko National University of Kyiv, 64/13, Volodymyrska Str., 01601 Kyiv, Ukraine}
	
	\author{E.V.~Gorbar}
	\affiliation{Physics Faculty, Taras Shevchenko National University of Kyiv, 64/13, Volodymyrska Str., 01601 Kyiv, Ukraine}
	\affiliation{Bogolyubov Institute for Theoretical Physics, 14-b, Metrologichna Str., 03143 Kyiv, Ukraine}

	\author{O.M.~Teslyk}
	\affiliation{Physics Faculty, Taras Shevchenko National University of Kyiv, 64/13, Volodymyrska Str., 01601 Kyiv, Ukraine}
	
	\author{S.I.~Vilchinskii}
	\affiliation{Institute of Physics, Laboratory for Particle Physics and Cosmology, \'{E}cole Polytechnique F\'{e}d\'{e}rale de Lausanne, CH-1015 Lausanne, Switzerland}
	\affiliation{Physics Faculty, Taras Shevchenko National University of Kyiv, 64/13, Volodymyrska Str., 01601 Kyiv, Ukraine}
	
	\date{\today}
	
	\begin{abstract}
		In the framework of Higgs inflation, we consider the electromagnetic field nonminimally coupled to gravity via the parity-preserving $\propto RF^{2}$ and parity-violating $\propto RF\tilde{F}$ terms. Using the perturbation theory to the leading order in these couplings, we study the generation of the electromagnetic field during the inflation stage. We derive the effective kinetic and axial coupling functions arising in the Einstein frame in the usual metric as well as Palatini formulations of gravity. For both formulations, we determine the power spectrum, energy density, and helicality of the generated electromagnetic fields for different values of the nonminimal coupling constants. Finally, we estimate the maximal present-day magnitude of the magnetic field as $10^{-14}-10^{-15}\,$G with the correlation length of order 10 pc.
	\end{abstract}

\maketitle

\section{Introduction}
\label{sec-intro}

The presence of magnetic fields with a magnitude of at least $10^{-17}\,$G in cosmic voids \cite{Tavecchio:2010,Ando:2010,Neronov:2010,Dolag:2010,Dermer:2011,Taylor:2011,Caprini:2015}, where there are a few astrophysical objects, is often considered as an indication of their primordial origin. This would also explain the ubiquity of the magnetic field throughout the Universe: From planets and stars to galaxies and their clusters (for a review, see Refs.~\cite{Kronberg:1994,Grasso:2001,Widrow:2002,Giovannini:2004,Kandus:2011,Durrer:2013,Subramanian:2016}), all these fields could have been sourced by the same primordial magnetic field and then amplified by adiabatic contraction and different types of dynamos inside astrophysical objects \cite{Zeldovich:1980book,Lesch:1995,Kulsrud:1997,Colgate:2001}. However, the origin of such hypothetical primordial fields is still unknown.

The inflation stage is one of the most natural places for the generation of cosmological seed magnetic fields. Indeed, primordial perturbations of the energy density and curvature which resulted in the anisotropy of the cosmic microwave background (CMB) and the large-scale structure of the Universe \cite{Harrison:1970,Zeldovich:1972,Chibisov:1982} were generated during inflation (see Refs.~\cite{Mukhanov:1992,Durrer:book} for a review). Magnetogenesis due to the coupling of the electromagnetic field to gravity and/or the inflaton field was first discussed in Refs.~\cite{Turner:1988,Ratra:1992,Garretson:1992,Dolgov:1993} and revisited many times in the literature \cite{Giovannini:2001,Bamba:2004,Martin:2008,Kanno:2009,Demozzi:2009,Ferreira:2013,Ferreira:2014,Vilchinskii:2017,Sharma:2017b,Savchenko:2018,Sobol:2018,Shtanov:2020,Talebian:2020,Durrer:2011,Anber:2006,Anber:2010,Barnaby:2012,Caprini:2014,Anber:2015,Ng:2015,Fujita:2015,Adshead:2015,Adshead:2016,Notari:2016,Domcke:2018eki,Cuissa:2018,Shtanov:2019,Shtanov:2019b,Sobol:2019,Kamarpour:2019,Domcke:2020zez,Bamba:2008,Bamba:2020,Maity:2021}. In many cases, the coupling functions are constructed phenomenologically in order to obtain the desired magnitude and spectrum of the electromagnetic field. The absence of a physical reason for the choice of a specific coupling function is one of the disadvantages of this approach. Moreover, this arbitrariness is multiplied by a potentially infinite number of inflationary models (see, e.g., a list of the most popular ones in Ref.~\cite{Martin:2013}). Even though observations of the CMB anisotropy by the Planck Collaboration \cite{Planck:2018-infl} impose some constraints on the inflaton potential, the number of models which are still in accordance with the observations is very big. Most of them belong to the class of plateau models whose potential becomes very flat for large values of the inflaton. These are, e.g., the Starobinsky $R^{2}$ model \cite{Starobinsky:1980}, $\alpha$ attractors \cite{Ferrara:2013,Kallosh:2013}, and Higgs inflation \cite{Bezrukov:2007,Bauer:2008}. The quartic hilltop model \cite{Linde:1982} with a flat potential for small values of the inflaton is still viable too \cite{Kallosh:2019}.

From the point of view of model building, models with the minimal number of new fields are the most attractive. An ideal situation is when no new fields are introduced. For instance, in the Starobinsky model \cite{Starobinsky:1980}, the usual Einstein-Hilbert action of gravity is extended by the quadratic term $\propto R^{2}$ which naturally appears due to quantum corrections to the gravitational action. Such a quadratic action is equivalent to the usual Einstein gravity with an additional scalar degree of freedom playing the role of the inflaton. 

Another example is the Higgs inflationary model where the role of the inflaton is played by the Standard Model Higgs boson $h$ nonminimally coupled to gravity via the $\propto h^{2}R$ term \cite{Bezrukov:2007,Bauer:2008}.\footnote{We consider any coupling of matter fields to curvature as the nonminimal one because such couplings are absent in the usual Einstein-Hilbert theory. Nonminimal couplings naturally appear due to quantum corrections to the gravitational action and present an interesting possibility to couple scalar and/or vector fields to gravity.} The action can be reduced to the Einstein form by means of the Weyl transformation. It transforms the usual quartic Higgs effective potential into a plateaulike potential which can be in accordance with the CMB observations for a certain range of the nonminimal coupling constant. (Note, that the quartic Higgs potential is not consistent with the Planck observations \cite{Planck:2018-infl} in the absence of nonminimal coupling.) Remarkably, quite different results follow from the same action in the Jordan frame if one chooses the usual (metric) formulation of gravity or the Palatini one. In the former case, the only dynamical gravitational degrees of freedom are the components of the metric tensor $g_{\mu\nu}$, while the connection and curvature are expressed through the metric and its derivatives. On the other hand, in the Palatini case, the metric and connection $\Gamma^{\lambda}_{\mu\nu}$ are treated as independent dynamical variables \cite{Palatini:1919,Einstein:1925}. For the Einstein-Gilbert action, both formulations are totally equivalent. However, in the presence of nonminimal coupling they lead to different theories in the Einstein frame. Both formulations of Higgs inflation have been extensively studied in the literature until now (for a review, see Refs.~\cite{Rubio:2018,Tenkanen:2020,Shaposhnikov:2020}).

A nonminimal coupling of the electromagnetic field to gravity has been already considered for inflationary magnetogenesis in the seminal paper~\cite{Turner:1988}. It has a big advantage compared to the kinetic and axial coupling models: The coupling functions are not guessed or postulated, but deduced from the simple form of the nonminimal coupling with a few relevant parameters. Although this idea is very natural, very little attention was paid to it in the literature (see Refs.~\cite{Savchenko:2018,Kamarpour:2019,Bamba:2008,Bamba:2020} for different scenarios of nonminimal coupling during inflation and Ref.~\cite{Frion:2020} for the case of bouncing cosmology). In the framework of the Starobinsky model, the most general gauge-invariant action quadratic in the electromagnetic field was considered in Ref.~\cite{Savchenko:2018}, where both parity-preserving and parity-violating terms were taken into account. The case of radiatively corrected Higgs inflation was discussed in Ref.~\cite{Kamarpour:2019}. In the present paper, we study the simplest nonminimal coupling of the electromagnetic field to gravity in the Higgs inflationary model.

This paper is organized as follows. In Sec.~\ref{sec-Higgs-infl}, we consider the action for the nonminimally coupled Higgs and electromagnetic fields in the Jordan frame. Performing the Weyl transformation, we deduce the coupling functions of the electromagnetic field to the inflaton in the Einstein frame both in the metric and Palatini formulations. In Sec.~\ref{sec-MG}, we review some basic equations describing inflationary magnetogenesis. In Sec.~\ref{sec-numerical}, we present numerical results for the power spectrum, energy density, and helicity of electromagnetic fields generated during Higgs inflation and estimate the present values of the magnetic field and its correlation length. Section~\ref{sec-concl} is devoted to conclusions. In Appendix~\ref{app-A}, we explain in detail how we choose the initial conditions and parameters of the potential using the results of the CMB observations. In Appendix~\ref{app-B}, we relate the conformal times at the end of reheating and at recombination. Throughout the work we use the natural units and set $\hbar=c=1$.

\section{Metric vs Palatini formulations}
\label{sec-Higgs-infl}

In the unitary gauge, the Standard Model Higgs field is represented by a real scalar field $h$. As we mentioned in the Introduction, the main idea of the Higgs inflation model is to couple this field to the curvature scalar~\cite{Bezrukov:2007,Bauer:2008}. The corresponding action in the Jordan frame reads as
\begin{equation}
	\label{action-Higgs}
	S_{h}=\int d^{4}x \sqrt{-g}\Bigg\{-\frac{M_{p}^{2}}{2}\left(1+\frac{\xi h^{2}}{M_{p}^{2}}\right)R+\frac{1}{2}(\partial_{\mu}h)(\partial^{\mu}h)-U(h)\Bigg\},
\end{equation}
where $g_{\mu\nu}$ is the spacetime metric with the signature $(+,\,-,\,-,\,-)$, $g={\rm det\,}(g_{\mu\nu})$, $R$ is the Ricci curvature scalar, $M_{p}=(8\pi G)^{-1/2}\approx 2.43\times 10^{18}\,{\rm GeV}$ is the reduced Planck mass, and $\xi$ is the nonminimal coupling constant. The Higgs effective potential has the form
\begin{equation}
	U(h)=\frac{\lambda h^{4}}{4},
\end{equation}
where $\lambda$ is dimensionless quartic coupling constant. We neglected the vacuum expectation value $v$ of the Higgs field because it is much smaller than the characteristic value of the field $h$ during inflation.

In addition, there is a massless vector field (which we will interchangeably call the electromagnetic field) that is also coupled to the spacetime curvature. The lowest-order gauge-invariant action has the form
\begin{equation}
	\label{action-gauge}
	S_{v}=\int d^{4}x \sqrt{-g}\Big\{-\frac{1}{4}F_{\lambda\rho}F^{\lambda\rho}+\frac{1}{2M_{p}^{2}}\left(\chi_{1}F_{\lambda\rho}F^{\lambda\rho}+\chi_{2}F_{\lambda\rho}\tilde{F}^{\lambda\rho}\right)R\Big\},
\end{equation}
where $F_{\mu\nu}$ is the electromagnetic field tensor, $\tilde{F}^{\mu\nu}=\frac{1}{2\sqrt{-g}}\epsilon^{\mu\nu\alpha\beta}F_{\alpha\beta}$ is its dual tensor, and $\epsilon^{\mu\nu\alpha\beta}$ is the
totally antisymmetric Levi-Civita symbol with $\epsilon^{0123}=+1$. The dimensionless coupling constant $\chi_1$ ($\chi_{2}$) characterizes the nonminimal coupling of the vector field to curvature which preserves (violates) the parity conservation. Since parity symmetry plays an important role in models of magnetogenesis, let us discuss the symmetry properties of interaction terms in more details. First, $F_{\lambda\rho}F^{\lambda\rho}$ is the standard Maxwell term which is a scalar quantity; therefore, it preserves parity. On the other hand, $F_{\lambda\rho}\tilde{F}^{\lambda\rho}$ is odd under parity transformation; i.e., it is a pseudoscalar quantity.

We would like to mention that one can, in principle, couple the vector field to the Ricci and/or Riemann curvature tensors as well~\cite{Turner:1988}. A comprehensive study of these terms in the Starobinsky inflationary model was performed in Ref.~\cite{Savchenko:2018}, where it was shown that in the Friedmannian universe there is only one additional nontrivial term to (\ref{action-gauge}), $\propto R_{\mu\nu}F^{\mu\alpha}F^{\nu}_{\alpha}$. During slow-roll inflation, however, this term is equivalent (up to terms of higher order in slow-roll parameters) to the $\chi_{1}$ term in Eq.~(\ref{action-gauge}). That is why we restrict ourselves by considering only couplings to the Ricci scalar. (One could also consider other types of quantum corrections on top of Higgs inflation, e.g., an $R^{2}$ term \cite{Calmet:2016,Antoniadis:2018} or derivative coupling of the inflaton to gravity \cite{Gialamas:2020}. However, for the sake of simplicity, we omit such terms in our analysis.)

In this work, we treat the electromagnetic field perturbatively and neglect its backreaction on the evolution of the inflaton field. First, we consider only the gravitational and Higgs action (\ref{action-Higgs}) and rewrite it in the Einstein frame. Second, we translate the vector field action (\ref{action-gauge}) into the same Einstein frame and analyze the generation of electromagnetic fields during inflation.

In order to rewrite the gravitational part of the action (\ref{action-Higgs}) in the canonical form, we perform the Weyl transformation
\begin{equation}
	\label{conformal-transformation}
	g_{\mu\nu}=\Psi^{-1}\bar{g}_{\mu\nu}
\end{equation}
with
\begin{equation}
	\label{Psi-h}
	\Psi=1+\frac{\xi h^{2}}{M_{p}^{2}}.
\end{equation}
As we mentioned in the Introduction, there are two formulations of gravity differing in the way the connection $\Gamma^{\lambda}_{\mu\nu}$ is introduced. In the metric formulation, it is expressed in terms of the metric and its derivatives. Therefore, the only dynamical degrees of freedom are components of the metric $g_{\mu\nu}$. On the other hand, in the Palatini formulation, both $g_{\mu\nu}$ and $\Gamma^{\lambda}_{\mu\nu}$ are treated as independent degrees of freedom. If the action for gravity has the Einstein-Hilbert form, then the equations of motion in both formulations coincide and the physical implications of both approaches are identical. However, they are inequivalent in the presence of nonminimal coupling. In practice, the differences occur when one performs the Weyl transformation to the Einstein frame. Below we consider the metric and Palatini formulations separately. 

We assume that in the Einstein frame the Universe is described by the spatially flat Friedmann-Lema\^{i}tre-Robertson-Walker metric
\begin{equation}
	\bar{g}_{\mu\nu}={\rm diag}(1,\,-a^{2},\,-a^{2},\,-a^{2}).
\end{equation}

\subsection{Metric Higgs inflation}
\label{subsec-metric-Higgs-inflation}

The metric formulation of gravity operates with the Levi-Civita connection, which is expressed in terms of derivatives of the metric tensor
\begin{equation}
\Gamma^{\lambda}_{\mu\nu}=\frac{g^{\lambda\rho}}{2}\big(\partial_{\mu}g_{\rho\nu}+\partial_{\nu}g_{\mu\rho}-\partial_{\rho}g_{\mu\nu}\big).
\end{equation}
This implies that the connection as well as the Ricci scalar nontrivially change under the Weyl transformation (\ref{conformal-transformation}). In particular, the curvature scalar transforms as follows:
\begin{equation}
	\label{Ricci-transformation}
	R=\Psi \bar{R}+3\bar{g}^{\mu\nu}\bar{\nabla}_{\mu}\bar{\nabla}_{\nu}\Psi-\frac{9}{2\Psi}\bar{g}^{\mu\nu}(\partial_{\mu}\Psi)(\partial_{\nu}\Psi).
\end{equation}
Then, after the transformation, we obtain the action in the following form:
\begin{multline}
	\label{action-2}
	S=\int d^{4}x \sqrt{-\bar{g}}\Bigg\{-\frac{M_{p}^{2}}{2}\bar{R}+\frac{3M_{p}^{2}}{4\Psi^{2}}\bar{g}^{\mu\nu}(\partial_{\mu}\Psi)(\partial_{\nu}\Psi)+\frac{1}{2\Psi}\bar{g}^{\mu\nu}(\partial_{\mu}h)(\partial_{\nu}h)-\frac{U(h)}{\Psi^{2}}-\\
	-\frac{1}{4}F_{\lambda\rho}F^{\lambda\rho} +\frac{1}{2M_{p}^{2}}\left(\chi_{1}F_{\lambda\rho}F^{\lambda\rho}+\chi_{2}F_{\lambda\rho}\tilde{F}^{\lambda\rho}\right)\left[\Psi \bar{R}+3\bar{g}^{\mu\nu}\bar{\nabla}_{\mu}\bar{\nabla}_{\nu}\Psi-\frac{9}{2\Psi}\bar{g}^{\mu\nu}(\partial_{\mu}\Psi)(\partial_{\nu}\Psi)\right]\Bigg\}.
\end{multline}
Taking into account Eq.~(\ref{Psi-h}), we get $\partial_{\mu}\Psi=2\xi h(\partial_{\mu} h)/M_{p}^{2}$. Thus, the second and third terms in (\ref{action-2}) can be combined as
\begin{equation}
	\frac{1}{2}\left[\frac{1}{\Psi^{2}(h)}\frac{6\xi^{2}h^{2}}{M_{p}^{2}}+\frac{1}{\Psi(h)}\right](\partial_{\mu}h)(\partial^{\mu}h).
\end{equation}
This is the kinetic part of the Lagrangian of the scalar field which is not canonically normalized. Performing the change of the field variable
\begin{equation}
	\sqrt{\frac{1}{\Psi^{2}(h)}\frac{6\xi^{2}h^{2}}{M_{p}^{2}}+\frac{1}{\Psi(h)}}dh=d\phi,
\end{equation}
we rewrite it in the canonical form. The relation between the old and new fields is the following:
\begin{eqnarray}
	\phi(h)&=&\int_{0}^{h} \frac{dh'}{1+\frac{\xi h^{\prime 2}}{M_{p}^{2}}}\sqrt{1+(1+6\xi)\frac{\xi h^{\prime 2}}{M_{p}^{2}}}=\nonumber\\
	&=&M_{p}\sqrt{6+\frac{1}{\xi}}{\rm arcsinh}\frac{h\sqrt{\xi(1+6\xi)}}{M_{p}}-M_{p}\sqrt{6}{\rm arcsinh}\frac{\xi h \sqrt{6}}{M_{p}\sqrt{1+\frac{\xi h^{2}}{M_{p}^{2}}}}.
\end{eqnarray}
This dependence is rather complicated and cannot be inverted. However, for large coupling constant $\xi\gg 1$ and for the field $h\gg M_{p}/\xi$, this relation can be significantly simplified
\begin{equation}
	\phi(h)\approx \int_{0}^{h} \frac{dh'}{1+\frac{\xi h^{\prime 2}}{M_{p}^{2}}}\sqrt{6}\frac{\xi h^{\prime}}{M_{p}}
	=M_{p}\sqrt{\frac{3}{2}}\ln\left(1+\frac{\xi h^{2}}{M_{p}^{2}}\right).
\end{equation}
Inverting it, we easily get
\begin{equation}
	\label{h-Higgs}
	h(\phi)=\frac{M_{p}}{\sqrt{\xi}}\left[\exp\left(\sqrt{\frac{2}{3}}\frac{\phi}{M_{p}}\right)-1\right]^{1/2}, \qquad \Psi(\phi)=\exp\left(\sqrt{\frac{2}{3}}\frac{\phi}{M_{p}}\right).
\end{equation}
The effective potential then reads as
\begin{equation}
	\label{pot-Higgs}
	V(\phi)=\left.\frac{U(h)}{\Psi^{2}(h)}\right|_{h=h(\phi)}=\frac{\lambda M_{p}^{4}}{4\xi^{2}}\left[1-\exp\left(-\sqrt{\frac{2}{3}}\frac{\phi}{M_{p}}\right)\right]^{2}.
\end{equation}
It has the same form as the potential in the Starobinsky model \cite{Starobinsky:1980}. However, this is not the exact form of the potential. It is only valid for $\phi\gg M_{p}/\xi$ and $\xi\gg 1$. For potential (\ref{pot-Higgs}), inflation ends when the slow-roll parameter $\epsilon$ becomes as large as unity:
\begin{equation}
	\epsilon=\frac{M_{p}^{2}}{2}\left(\frac{V'}{V}\right)^{2}=\frac{4/3}{\left[\exp\left(\sqrt{\frac{2}{3}}\frac{\phi}{M_{p}}\right)-1\right]^{2}}=1, \quad \phi_{e}=M_{p}\sqrt{\frac{3}{2}}\ln\left(1+\frac{2}{\sqrt{3}}\right)\approx 0.94 M_{p}.
\end{equation}
This value of the inflaton is much greater than $M_{p}/\xi$ and, thus, potential (\ref{pot-Higgs}) is valid during the whole inflation stage.

Thus, in terms of the transformed metric $\bar{g}_{\mu\nu}$ and new field $\phi$, the gravitational action takes the Einstein-Hilbert form and the inflaton is canonically normalized. Let us now consider the electromagnetic part of the action
\begin{equation}
	\label{action-EM}
	S_{v}=\int d^{4}x \sqrt{-\bar{g}}\Big\{-\frac{I_{1}}{4}F_{\lambda\rho}F^{\lambda\rho}-\frac{I_{2}}{4}F_{\lambda\rho}\tilde{F}^{\lambda\rho} \Big\},
\end{equation}
where
\begin{equation}
	\label{I1-Higgs}
	I_{j}=\delta_{j1}-\frac{2\chi_{j}}{M_{p}^{2}}\left[\Psi(\phi)\bar{R}+3\bar{\nabla}_{\mu}\bar{\nabla}^{\mu}\Psi(\phi)-\frac{9}{2\Psi}(\partial_{\mu}\Psi)(\partial^{\mu}\Psi)\right].
\end{equation}
Here, $\bar{R}$ must be expressed in terms of the inflaton $\phi$ and its derivatives. In order to do this, we write down the background equations
\begin{eqnarray}
	H^{2}&=&\frac{1}{3M_{p}^{2}}\rho=\frac{1}{3M_{p}^{2}}\left[\frac{1}{2}\dot{\phi}^{2}+V(\phi)\right],\label{Friedmann-1}\\
	\dot{H}&=&-\frac{1}{2M_{p}^{2}}(\rho+P)=-\frac{1}{2M_{p}^{2}}\dot{\phi}^{2},\label{Friedmann-2}\\
	\ddot{\phi}&+&3H\dot{\phi}+V'(\phi)=0.\label{KGF-eq}
\end{eqnarray}
Using these equations together with Eqs.~(\ref{h-Higgs}) and (\ref{pot-Higgs}), we get
\begin{eqnarray}
	\bar{R}&=&-6(\dot{H}+2H^{2})=\frac{1}{M_{p}^{2}}\left[\dot{\phi}^{2}-4V(\phi)\right],\label{Ricci-scalar}\\
	\bar{\nabla}_{\mu}\bar{\nabla}^{\mu}\Psi(\phi)&=&\ddot\Psi+3H\dot{\Psi}=\exp\left(\sqrt{\frac{2}{3}}\frac{\phi}{M_{p}}\right)\left[-\sqrt{\frac{2}{3}}\frac{1}{M_{p}}V'(\phi)+\frac{2}{3}\frac{\dot{\phi}^{2}}{M_{p}^{2}}\right],\\
	(\partial_{\mu}\Psi)(\partial^{\mu}\Psi)&=&\exp\left(2\sqrt{\frac{2}{3}}\frac{\phi}{M_{p}}\right)\frac{2}{3}\frac{\dot{\phi}^{2}}{M_{p}^{2}}.
\end{eqnarray}
Substituting these expressions into Eq.~(\ref{I1-Higgs}), we obtain
\begin{equation}
	\label{I1-Higgs-2}
	I_{j}=\delta_{j1}+\frac{2\lambda \chi_{j}}{\xi^{2}}\left[\exp\left(\sqrt{\frac{2}{3}}\frac{\phi}{M_{p}}\right)-1\right].
\end{equation}
Remarkably, Eqs.~(\ref{pot-Higgs}) and (\ref{I1-Higgs-2}) exactly coincide with the corresponding expressions for the Starobinsky model \cite{Savchenko:2018}, if one replaces $\lambda M_{p}^{4}/(4\xi^{2})$ with the corresponding amplitude of the Starobinsky potential.

\subsection{Palatini Higgs inflation}
\label{subsec-Higgs-Palatini}

Let us now consider the Palatini formulation of gravity. In this case, the components of connection $\Gamma^{\alpha}_{\beta\gamma}$ are treated as independent dynamical variables in addition to the metric 
components $g_{\mu\nu}$. Then, the Ricci scalar reads
\begin{equation}
	R=g^{\mu\nu} R_{\mu\nu}[\Gamma^{\alpha}_{\beta\gamma}].
\end{equation}
Again, we perform the Weyl transformation (\ref{conformal-transformation}) with the same $\Psi$ function (\ref{Psi-h}). The curvature scalar changes as
\begin{equation}
	R=\Psi\bar{R}
\end{equation}
and the action in the Einstein frame reads as
\begin{equation}
	\label{action-Palatini}
	S=\int d^{4}x \sqrt{-\bar{g}}\Bigg\{-\frac{M_{p}^{2}}{2}\bar{R}+\frac{1}{2\Psi}\bar{g}^{\mu\nu}(\partial_{\mu}h)(\partial_{\nu}h)-\frac{U(h)}{\Psi^{2}}
	-\frac{1}{4}F_{\lambda\rho}F^{\lambda\rho} +\frac{\Psi \bar{R}}{2M_{p}^{2}}\left(\chi_{1}F_{\lambda\rho}F^{\lambda\rho}+\chi_{2}F_{\lambda\rho}\tilde{F}^{\lambda\rho}\right)\Bigg\}.
\end{equation}
Since the Higgs field $h$ is not canonically normalized, we introduce a new field $\phi$ defined by the following relation:
\begin{equation}
	\frac{dh}{\sqrt{1+\frac{\xi h^{2}}{M_{p}^{2}}}}=d\phi.
\end{equation}
This equation can be easily integrated and inverted. We obtain
\begin{equation}
	h(\phi)=\frac{M_{p}}{\sqrt{\xi}}{\rm sinh}\frac{\sqrt{\xi}\phi}{M_{p}}, \qquad \Psi(\phi)={\rm cosh}^{2}\frac{\sqrt{\xi}\phi}{M_{p}}.
\end{equation}
The effective potential of the inflaton then reads as
\begin{equation}
	V(\phi)=\left.\frac{U(h)}{\Psi^{2}(h)}\right|_{h=h(\phi)}=\frac{\lambda M_{p}^{4}}{4\xi^{2}}{\rm tanh}^{4}\frac{\sqrt{\xi}\phi}{M_{p}}.
\end{equation}
Again, the effective potential belongs to the class of plateau models; more precisely, it has the same form as in the symmetric $\alpha$-attractor inflationary model \cite{Martin:2013,Ferrara:2013,Kallosh:2013}. 

Let us now return to the electromagnetic part of the action. It gets the form of Eq.~(\ref{action-EM}) with the coupling functions
\begin{equation}
	I_{j}=\delta_{j1}-\frac{2\chi_{j}}{M_{p}^{2}}\Psi \bar{R}.
\end{equation}
The Ricci scalar can be found in a standard way from Eq.~(\ref{Ricci-scalar}), where $\dot{\phi}$ during inflation can be expressed in terms of the potential in the slow-roll approximation
\begin{equation}
	\dot{\phi}\approx -\frac{V'(\phi)}{3H}, \qquad \bar{R}=\frac{1}{M_{p}^{2}}\left[\dot{\phi}^{2}-4V(\phi)\right]\approx -\frac{4V(\phi)}{M_{p}^{2}}\left(1-\frac{\epsilon}{6}\right).
\end{equation}
Finally, the coupling functions take the form
\begin{equation}
	\label{I1-Palatini}
	I_{j}=\delta_{j1}+\frac{2\lambda\chi_{j}}{\xi^{2}}\left[\frac{{\rm sinh}^{4}\frac{\sqrt{\xi}\phi}{M_{p}}}{{\rm cosh}^{2}\frac{\sqrt{\xi}\phi}{M_{p}}}-\frac{4\xi}{3}\frac{{\rm sinh}^{2}\frac{\sqrt{\xi}\phi}{M_{p}}}{{\rm cosh}^{4}\frac{\sqrt{\xi}\phi}{M_{p}}}\right].
\end{equation}

\section{Inflationary magnetogenesis} 
\label{sec-MG}

In the previous section, we showed that the nonminimal coupling of the electromagnetic field to curvature in the Jordan frame boils down to the usual kinetic and axial couplings to the inflaton field in the Einstein frame; cf. Eq.~(\ref{action-EM}). However, the coupling functions are not arbitrary. Their form is fixed by the inflationary model. In this section, we remind of some basic equations determining the generation of the vector field in such a mixed kinetic and axial coupling model.

All computations in the previous section were performed under the assumption that the generated electromagnetic field does not backreact on the inflaton dynamics. In order to be self-consistent, in what follows we should check this condition which can be formulated in terms of energy densities:
\begin{equation}
	\rho_{v}\ll \rho_{\rm inf},
\end{equation}
where $\rho_{\rm inf}=\dot{\phi}^{2}/2+V(\phi)$ is the inflaton contribution, while the vector field counterpart $\rho_{v}$ is determined by the 00 component of the corresponding part of stress-energy tensor
\begin{equation}
	\label{T-mu-nu}
	T_{\mu\nu}^{v}=\frac{2}{\sqrt{-\bar{g}}}\frac{\delta S_{v}}{\delta \bar{g}^{\mu\nu}}=
	I_{1}F_{\mu\lambda}F_{\rho\nu}\bar{g}^{\lambda\rho}+\bar{g}_{\mu\nu}\frac{I_{1}}{4}F_{\lambda\rho}F^{\lambda\rho}.
\end{equation}
Note that the term proportional to $\chi_{2}$ does not depend on the metric, that is why it does not contribute to the stress-energy tensor. The energy density is thus equal to
\begin{equation}
	\rho_{v}=\langle T_{00}\rangle = I_{1}(\phi) \frac{\langle \mathbf{E}^{2}\rangle + \langle \mathbf{B}^{2}\rangle}{2},
\end{equation}
where the angle brackets $\langle \ldots \rangle$ denote the vacuum expectation value, and the electric $\mathbf{E}$ and magnetic $\mathbf{B}$ field three-vectors are defined as follows:
\begin{equation}
	\label{EM-tensors}
	F^{0i}=\frac{1}{a} E^{i},\quad F^{ij}=\frac{1}{a^{2}}\varepsilon^{ijk}B^{k},
	\quad \tilde{F}^{0i}=\frac{1}{a}B^{i}, \quad \tilde{F}^{ij}=-\frac{1}{a^{2}}\varepsilon^{ijk}E^{k},
\end{equation}
where $\varepsilon^{ijk}$ is the three-dimensional Levi-Civita symbol. We would like to note that the electric and magnetic fields defined in Eqs.~(\ref{EM-tensors}) are physical fields measured by a comoving observer.

Therefore, if the electromagnetic field satisfies the condition
\begin{equation}
	\label{weak-field}
	I_{1}(\phi)\langle \mathbf{E}^{2}\rangle,\ I_{1}(\phi)\langle \mathbf{B}^{2}\rangle\ll \rho_{\rm inf}\simeq V(\phi)
\end{equation}
during the whole inflation stage, we are allowed to use the linear approximation and consider the generation of the electromagnetic field in the inflaton background determined by the unperturbed Friedmann and Klein-Gordon equations (\ref{Friedmann-1})--(\ref{KGF-eq}). Were condition (\ref{weak-field}) violated, then the generated electromagnetic field would impact the Universe expansion and the inflaton evolution. In turn, this modifies the coupling functions $I_{1,2}$ and, as a result, influences the evolution of the electromagnetic field. In such a case, the linear approximation is no more valid and a self-consistent treatment of all relevant electromagnetic modes is needed. Moreover, possible higher order corrections to the electromagnetic part of the action would become important. However, this lies beyond the scope of the present study.

Varying action (\ref{action-EM}) with respect to the vector field, we get the following linear system of equations:
\begin{eqnarray}
	&&\dot{\mathbf{E}}+2H\mathbf{E}-\frac{1}{a}{\rm rot\,}\mathbf{B}+\frac{\dot{I}_{1}}{I_{1}}\mathbf{E}+\frac{\dot{I}_{2}}{I_{1}}\mathbf{B}=0,\label{Maxwell-2}\\
	&&\dot{\mathbf{B}}+2H\mathbf{B}+\frac{1}{a}{\rm rot\,}\mathbf{E}=0,\label{Maxwell-second-equation}\\
	&&{\rm div\,}\mathbf{E}=0,\qquad {\rm div\,}\mathbf{B}=0.
\end{eqnarray}
Here we do not take into account the currents of charged matter fields coupled to our vector field because during inflation the contribution of matter fields is small (we assume that charged matter fields 
are in the vacuum state) because the Schwinger pair production is negligibly small in weak electric fields satisfying condition (\ref{weak-field}); see, e.g., Refs.~\cite{Sobol:2018,Sobol:2019,Momot:2019}. 
Compared to the free electromagnetic field case, we got two additional terms on the right-hand side of Eq.~(\ref{Maxwell-2}). Although they arise from the nonminimal coupling to gravity, their form is analogous to well-known matter currents.	The first is similar to the usual Ohm current where $\dot{I}_{1}/I_{1}$ plays the role of electric conductivity $\sigma$. If this coefficient is positive, it leads to the dissipation of the electric field. However, if this coefficient is negative [which is the case for the coupling functions (\ref{I1-Higgs-2}) and (\ref{I1-Palatini})], then it may enhance the electric field and trigger magnetogenesis. The last term in Eq.~(\ref{Maxwell-2}) is analogous to the chiral magnetic effect current	$\mathbf{j}\propto \mu_{5}\mathbf{B}$, with $\dot{I}_{2}/I_{1}$ playing the role of the fermionic chiral chemical potential $\mu_{5}$. It is well known that such a current generates helical electromagnetic fields; see, e.g., Refs.~\cite{Joyce:1997uy,Boyarsky:2011uy}. As we will see below, in our case a similar phenomenon takes place where the parity-violating interaction with curvature (or inflaton) plays the role of the source for the electromagnetic field.

In such a situation, it is convenient to use the Coulomb gauge for the electromagnetic field where ${\rm div\,}\mathbf{A}=0$ and $A_{\mu}=(0,\,\mathbf{A})$. Then, the electric and magnetic fields can be expressed in terms of the vector potential as
\begin{equation}
	\label{fields-E-and-B}
	\mathbf{E}=-\frac{1}{a}\dot{\mathbf{A}}, \qquad \mathbf{B}=\frac{1}{a^{2}} {\rm rot\,}\mathbf{A},
\end{equation}
Then Eq.(\ref{Maxwell-second-equation}) is identically satisfied, and Eq.~(\ref{Maxwell-2}) reads as
\begin{equation}
	\label{Maxwell-A}
	\ddot{\mathbf{A}}+\left(H+\frac{\dot{I}_{1}}{I_{1}}\right)\dot{\mathbf{A}}-\frac{1}{a^{2}}\Delta\mathbf{A}-\frac{1}{a}\frac{\dot{I}_{2}}{I_{1}}{\rm rot\,}\mathbf{A}=0,
\end{equation}
where $\Delta=\partial_{i}^{2}$ is the spatial Laplacian operator.

The electromagnetic field is generated from quantum fluctuations. In order to describe this process, we decompose the field operator over the full set of creation (annihilation) operators $\hat{b}^{\dagger}_{\mathbf{k},\lambda}$ ($\hat{b}_{\mathbf{k},\lambda}$) of the modes with momentum $\mathbf{k}$ and transverse polarization $\lambda$,
\begin{equation}
	\label{quant-operator}
	\hat{\mathbf{A}}(t,\mathbf{x})=\int\frac{d^{3}\mathbf{k}}{(2\pi)^{3/2}}\!\!\sum_{\lambda=\pm}\left\{ \boldsymbol{\varepsilon}_{\lambda}(\mathbf{k})\hat{b}_{\mathbf{k},\lambda}A_{\lambda}(t,\mathbf{k})e^{i\mathbf{k}\cdot\mathbf{x}}+\boldsymbol{\varepsilon}^{*}_{\lambda}(\mathbf{k})\hat{b}^{\dagger}_{\mathbf{k},\lambda}A^{*}_{\lambda}(t,\mathbf{k})e^{-i\mathbf{k}\cdot\mathbf{x}}\right\},
\end{equation}
where $\boldsymbol{\varepsilon}_{\lambda}(\mathbf{k})$ are two circular polarization three-vectors which satisfy the following 
conditions:
\begin{equation}
	\mathbf{k}\cdot\boldsymbol{\varepsilon}_{\lambda}(\mathbf{k})=0,\quad \boldsymbol{\varepsilon}^{*}_{\lambda}(\mathbf{k})=\boldsymbol{\varepsilon}_{-\lambda}(\mathbf{k}), \quad [i\mathbf{k}\times\boldsymbol{\varepsilon}_{\lambda}(\mathbf{k})]=\lambda k \boldsymbol{\varepsilon}_{\lambda}(\mathbf{k}), \quad \boldsymbol{\varepsilon}^{*}_{\lambda}(\mathbf{k})\cdot\boldsymbol{\varepsilon}_{\lambda'}(\mathbf{k})=\delta_{\lambda\lambda'}. 
\end{equation}
The creation and annihilation operators satisfy the canonical commutation relations
\begin{equation}
	[\hat{b}_{\lambda,\mathbf{k}},\,\hat{b}^{\dagger}_{\lambda',\mathbf{k}'}]=\delta_{\lambda\lambda'}\delta^{(3)}(\mathbf{k}-\mathbf{k}').
\end{equation}
Decomposition (\ref{quant-operator}) is convenient because the operator $\hat{\mathbf{A}}$ satisfies the linear equation of motion. Thus, each mode can be described independently from the others.

Substituting decomposition (\ref{quant-operator}) into Eq.~(\ref{Maxwell-A}), we find the following equation governing the evolution of the mode function:
\begin{equation}
	\label{eq-mode-1}
	\ddot{A}_{\lambda}(t,\mathbf{k})+\left(H+\frac{\dot{I}_{1}}{I_{1}}\right)\dot{A}_{\lambda}(t,\mathbf{k})+\left(\frac{k^{2}}{a^{2}}-\lambda\frac{k}{a}\frac{\dot{I}_{2}}{I_{1}}\right)A_{\lambda}(t,\mathbf{k})=0.
\end{equation}
Introducing the new function $\mathcal{B}_{\lambda}(t,\mathbf{k})=\sqrt{I_{1}}A_{\lambda}(t,\mathbf{k})$, we can rewrite it as follows:
\begin{equation}
	\label{eq-mode-2}
	\ddot{\mathcal{B}}_{\lambda}(t,\mathbf{k})+H\dot{\mathcal{B}}_{\lambda}(t,\mathbf{k})+\left(\frac{k^{2}}{a^{2}}-H\frac{\dot{I}_{1}}{2I_{1}}-\frac{1}{\sqrt{I_{1}}}\frac{d^{2}\sqrt{I_{1}}}{dt^{2}}-\lambda\frac{k}{a}\frac{\dot{I}_{2}}{I_{1}}\right)\mathcal{B}_{\lambda}(t,\mathbf{k})=0,
\end{equation}
or in the conformal time $\tau=\int^{t} dt'/a(t')$,
\begin{equation}
	\label{eq-mode-conf}
	\mathcal{B}''_{\lambda}(\tau,\mathbf{k})+\left(k^{2}-\frac{1}{\sqrt{I_{1}}}\frac{d^{2}\sqrt{I_{1}}}{d\tau^{2}}-\lambda k\frac{I'_{2}}{I_{1}}\right)\mathcal{B}_{\lambda}(\tau,\mathbf{k})=0.
\end{equation}
For modes far inside the Hubble horizon, the first term in brackets in Eq.~(\ref{eq-mode-conf}) dominates over the last two. This allows us to choose the Bunch-Davies vacuum initial conditions for these modes \cite{Bunch:1978} 
\begin{equation}
	\label{Bunch-Davies-vacuum}
	\mathcal{B}_{\lambda}(\tau,k)=\frac{1}{\sqrt{2k}}e^{-ik\tau}, \quad k\tau\to-\infty.
\end{equation}

For practical purposes, in particular, for the numerical solution of the mode equation, it is more convenient to represent Eq.~(\ref{eq-mode-2}) as a system of two first-order differential equations. In fact, introducing the auxiliary function $\mathcal{E}_{\lambda}=\sqrt{I_{1}}(a/k)\dot{A}_{\lambda}$, we rewrite Eq.~(\ref{eq-mode-2}) as follows:
\begin{eqnarray}
	\dot{\mathcal{B}}_{\lambda}(t,\mathbf{k})&=&\frac{\dot{I}_{1}}{2I_{1}}\mathcal{B}_{\lambda}(t,k)+\frac{k}{a}\mathcal{E}_{\lambda}(t,k), \label{eq-B}\\
	\dot{\mathcal{E}}_{\lambda}(t,\mathbf{k})&=&-\frac{\dot{I}_{1}}{2I_{1}}\mathcal{E}_{\lambda}(t,k)-\left(\frac{k}{a}-\lambda\frac{\dot{I}_{2}}{I_{1}}\right)\mathcal{B}_{\lambda}(t,k).\label{eq-E}
\end{eqnarray}
The Bunch-Davies boundary condition for the function $\mathcal{E}_{\lambda}$ has the form
\begin{equation}
	\label{Bunch-Davies-vacuum-E}
	\mathcal{E}_{\lambda}(\tau,k)=\frac{1}{\sqrt{2k}}e^{-ik\tau}\left[-i-\frac{1}{2k}\frac{I^{\prime}_{1}}{I_{1}}\right], \quad k\tau\to-\infty.
\end{equation}

Finally, we express the spectral energy densities of the magnetic and electric fields as follows:
\begin{eqnarray}
	\frac{d\rho_{B}}{d\ln \,k}&=&\sum_{\lambda=\pm} \frac{k^{5}}{4\pi^{2}a^{4}}I_{1}|A_{\lambda}(t,k)|^{2}=\sum_{\lambda=\pm} \frac{k^{5}}{4\pi^{2}a^{4}}|\mathcal{B}_{\lambda}(t,k)|^{2}, \label{en-dens-sp-B}\\
	\frac{d\rho_{E}}{d\ln \,k}&=&\sum_{\lambda=\pm} \frac{k^{3}}{4\pi^{2}a^{2}}I_{1}|\dot{A}_{\lambda}(t,k)|^{2}=\sum_{\lambda=\pm} \frac{k^{5}}{4\pi^{2}a^{4}}|\mathcal{E}_{\lambda}(t,k)|^{2}.\label{en-dens-sp-E}
\end{eqnarray}
In order to find the energy densities of the generated electromagnetic fields, we integrate these spectral densities over wave numbers (momenta) of all physically relevant modes. To define these modes, let us consider Eq.~(\ref{eq-mode-2}). If the first term in parentheses dominates over the others, the mode behaves as the Bunch-Davies state (\ref{Bunch-Davies-vacuum}); i.e., it oscillates with constant amplitude. Such a mode describes a vacuum fluctuation and contains no information about the interaction with the inflaton. In other words, it does not ``feel'' this interaction and, therefore, it is not enhanced. Naturally, we must not take such modes into account. In the opposite situation where the first term is negligible compared to other terms the evolution of the corresponding mode function strongly differs from the vacuum solution. Thus, such modes should definitely be taken into account. The formal boundary dividing modes in two classes can be defined, e.g., as follows:
\begin{equation}
	\label{k-h}
	k_{h}(t)=\max\{k_{1}(t), \, k_{2}(t)\},
\end{equation}
where
\begin{equation}
	\label{k-h1}
	k_{1}(t)=
	\underset{t'\leq t}{\max} \Big\{a(t')\Big|\frac{H(t')}{2I_{1}(t')}\frac{dI_{1}(t')}{dt'}+\frac{1}{\sqrt{I_{1}(t')}}\frac{d^{2}\sqrt{I_{1}(t')}}{dt^{'2}}\Big|^{1/2}\Big\},
\end{equation}
\begin{equation}
	\label{k-h2}
	k_{2}(t)=
	\underset{t'\leq t}{\max}\Big\{a(t')\Big|\frac{1}{I_{1}(t')}\frac{dI_{2}(t')}{dt'}\Big|\Big\}.
\end{equation}
If $k=k_{h}(t)$, we say that the mode with momentum $k$ crosses the horizon [not to be confused with crossing the Hubble horizon, $k_{H}=a(t)H(t)$]. Thus, the physically relevant modes at the moment of time $t$ are those which crossed the horizon from the beginning of inflation until the moment $t$. Therefore, the energy densities of the generated fields are defined as follows:
\begin{equation}
	\rho_{E,B}(t)=\int_{0}^{k_{h}(t)}\frac{dk}{k}\frac{d\rho_{E,B}(t)}{d\ln\, k}.\label{en-dens}
\end{equation}

Having determined the magnetic spectral density, we can find the correlation length of the magnetic field
\begin{equation}
	\label{lambda-B}
	\lambda_{B}=\left<\frac{2\pi a}{k}\right>_{B}=\frac{1}{\rho_{B}}\int_{0}^{k_{h}(t)}\frac{dk}{k}\frac{2\pi a}{k}\frac{d\rho_{B}(t)}{d\ln\, k}=\frac{1}{\rho_{B}}\int_{0}^{k_{h}(t)}\!\!\!\!dk\frac{k^{3}}{2\pi a^{3}}\big\{|\mathcal{B}_{+}(t,k)|^{2}+|\mathcal{B}_{-}(t,k)|^{2} \big\}.
\end{equation}

For nonzero $\chi_{2}$, modes with different circular polarizations evolve differently. As a result, the generated fields are helical; i.e., they have nonzero helicity
\begin{equation}
	\label{helicity}
	\mathcal{H}=\frac{I_{1}}{a}\langle\mathbf{A}\cdot\mathbf{B}\rangle=\int_{0}^{k_{h}(t)}\!\!\!\!dk\frac{k^{3}}{2\pi^{2} a^{3}}\big\{|\mathcal{B}_{+}(t,k)|^{2}-|\mathcal{B}_{-}(t,k)|^{2} \big\}.
\end{equation}
Comparing Eqs.~(\ref{lambda-B}) and (\ref{helicity}), one can deduce the following inequality:
\begin{equation}
|\mathcal{H}|\leq \frac{\lambda_{B}\rho_{B}}{\pi},
\end{equation}
which is often called the realizability condition. In the limiting case when one polarization strongly dominates over the other, the equality is reached. Such electromagnetic fields are maximally helical. Thus, the degree of ``helicality'' can be characterized by the quantity
\begin{equation}
	\eta_{h}=\frac{\pi |\mathcal{H}|}{\lambda_{B}\rho_{B}}, \qquad 0\leq \eta_{h}\leq 1.
\end{equation}

\section{Numerical results and discussion}
\label{sec-numerical}

In this section, we present the numerical results for the power spectrum, energy density, and helicality of electromagnetic fields generated during Higgs inflation for different values of coupling parameters $\chi_{1,2}$. Then, we discuss their postinflationary evolution and estimate the upper bound for the magnetic fields at the present time in this model.

In the numerical analysis, we proceed as follows. First, we fix the inflationary model; i.e., we choose the nonminimal coupling $\xi$ and determine the amplitude of the potential which leads to the correct amplitude of the primordial scalar power spectrum measured by the Planck Collaboration \cite{Planck:2018-infl}. The starting point of the simulation is chosen as the moment of time when the CMB pivot scale crosses the horizon, because exactly at this moment of time we normalize the potential. This procedure is discussed in detail in Appendix~\ref{app-A}, and all numerical values of the parameters that we use in our computations are listed in Table~\ref{tab-parameters}. Second, we solve numerically the background equations (\ref{Friedmann-1}) and (\ref{KGF-eq}) which do not take into account the electromagnetic field. Thus, we neglect the backreaction of the generated fields on the background evolution which sets the bound of applicability of our approach given by Eq.~(\ref{weak-field}).

\begin{table}[!h]
	\centering
	\begin{tabular}{ c | c |c }
		\hline\hline
		Parameter & Metric HI & Palatini HI \\ 
		\hline
		$\xi$ & $\sim 10^{3}$ \cite{Bezrukov:2007} & $10^{7}$ \cite{Shaposhnikov:2020}\\ 
		$N_{\star}$ & 55.4 & 50.9 \\  
		$\phi_{\star}/M_{p}$ & 5.36 & $3.72\times 10^{-3}$\\  
		$V_{0}/M_{p}^{4}$ & $1.1\times 10^{-10}$ & $2.3\times 10^{-18}$ \\
		$n_{s}$ & 0.965 & 0.961 \\
		$r$ & $3.5\times 10^{-3}$ & $7.7\times 10^{-11}$ \\
		$T_{\rm reh}$, GeV & $2.4\times 10^{15}$ & $3.9\times 10^{13}$ \\ \hline\hline
	\end{tabular}
	\caption{Numerical values of the parameters that we use in the numerical computations.
		\label{tab-parameters}}
\end{table}

Having determined the time dependences of the scale factor and the inflaton field, we find the solutions of the mode equations (\ref{eq-B}) and (\ref{eq-E}) which satisfy the Bunch-Davies boundary conditions (\ref{Bunch-Davies-vacuum}) and (\ref{Bunch-Davies-vacuum-E}) for given values of the nonminimal coupling constants $\chi_{1,2}$. We do this for all modes which are deep inside the horizon at the beginning of our simulation and cross the horizon during inflation. Further, we calculate the spectral densities of the electric and magnetic energy densities (\ref{en-dens-sp-B}) and (\ref{en-dens-sp-E}), as well as the spectral helicity density which is the integrand in Eq.~(\ref{helicity}). Finally, integrating these spectral densities over the modes which have already crossed the horizon until the moment of time $t$, we get the values of the energy densities $\rho_{E,B}(t)$, magnetic correlation length $\lambda_{B}(t)$, and helicity $\mathcal{H}(t)$ using Eqs.~(\ref{en-dens}), (\ref{lambda-B}), and (\ref{helicity}), respectively.

\subsection{Spectra of generated electromagnetic fields}

Let us begin with the power spectra of generated fields. Here we would like to mention only some general features of the spectra which are common for both formulations of Higgs inflation. That is why we show only the case of metric Higgs inflation in the plots.

\begin{figure}[ht!]
	\centering
	\includegraphics[height=3.6cm]{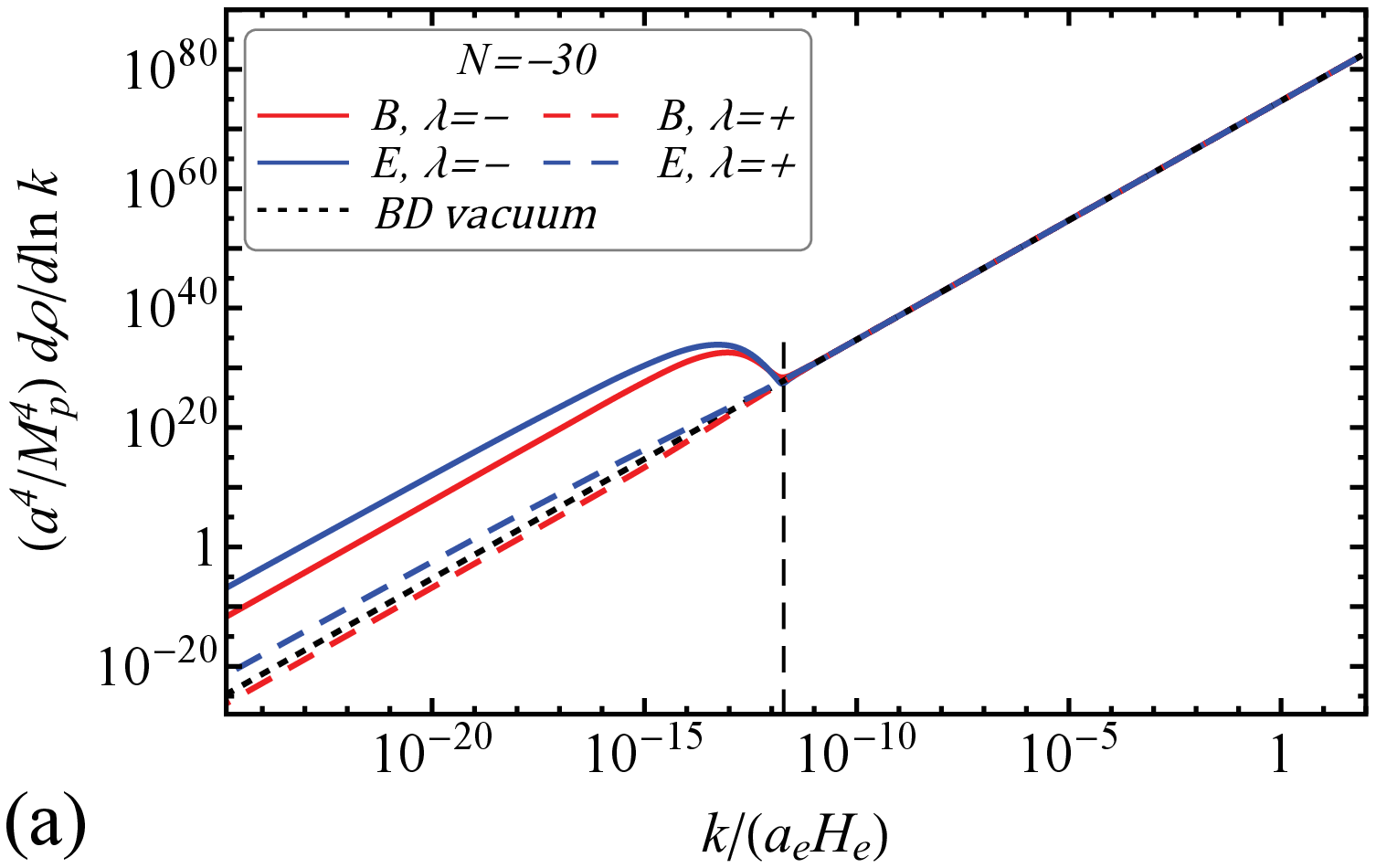}\hspace*{0.1cm}
	\includegraphics[height=3.6cm]{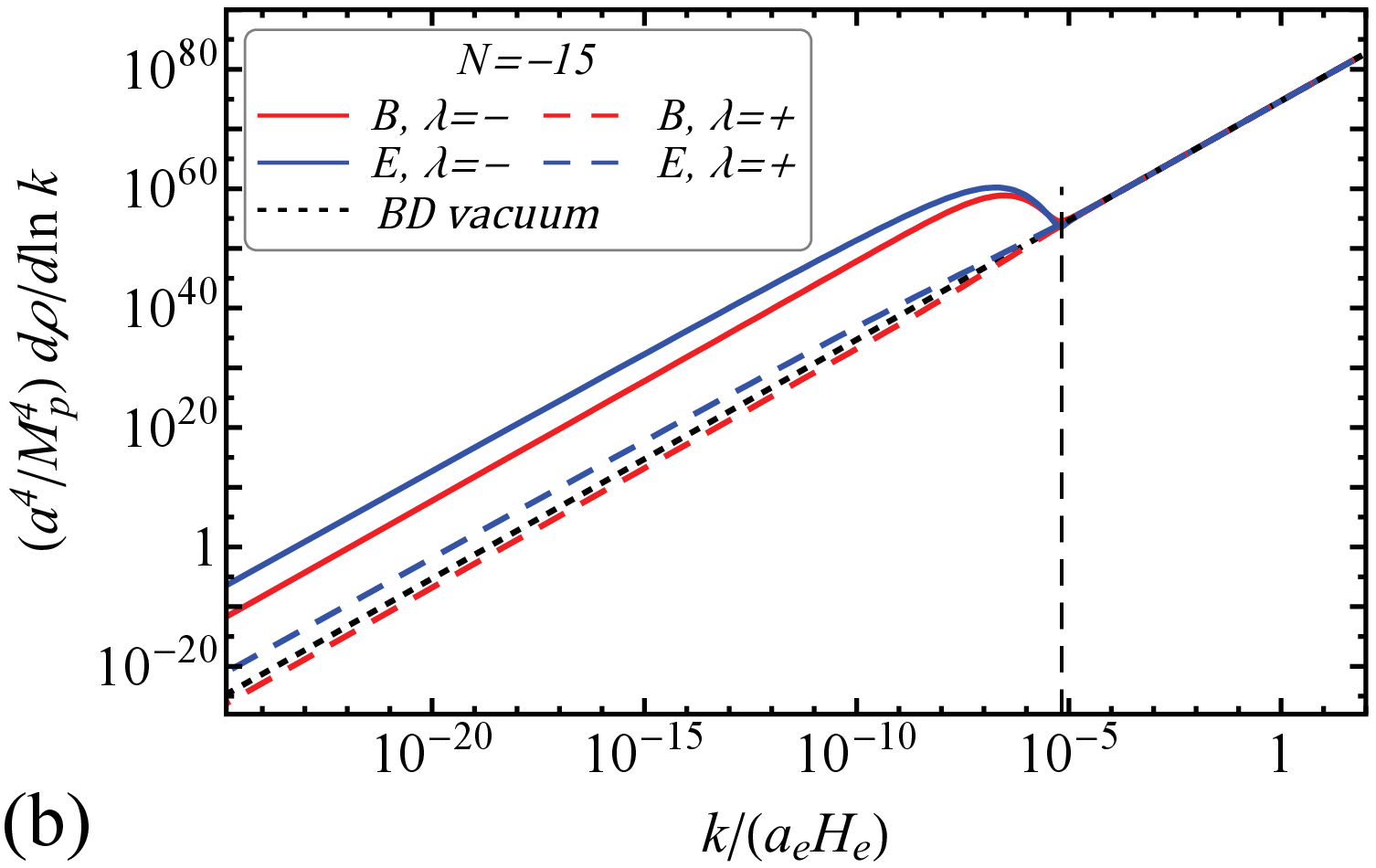}\hspace*{0.1cm}
	\includegraphics[height=3.6cm]{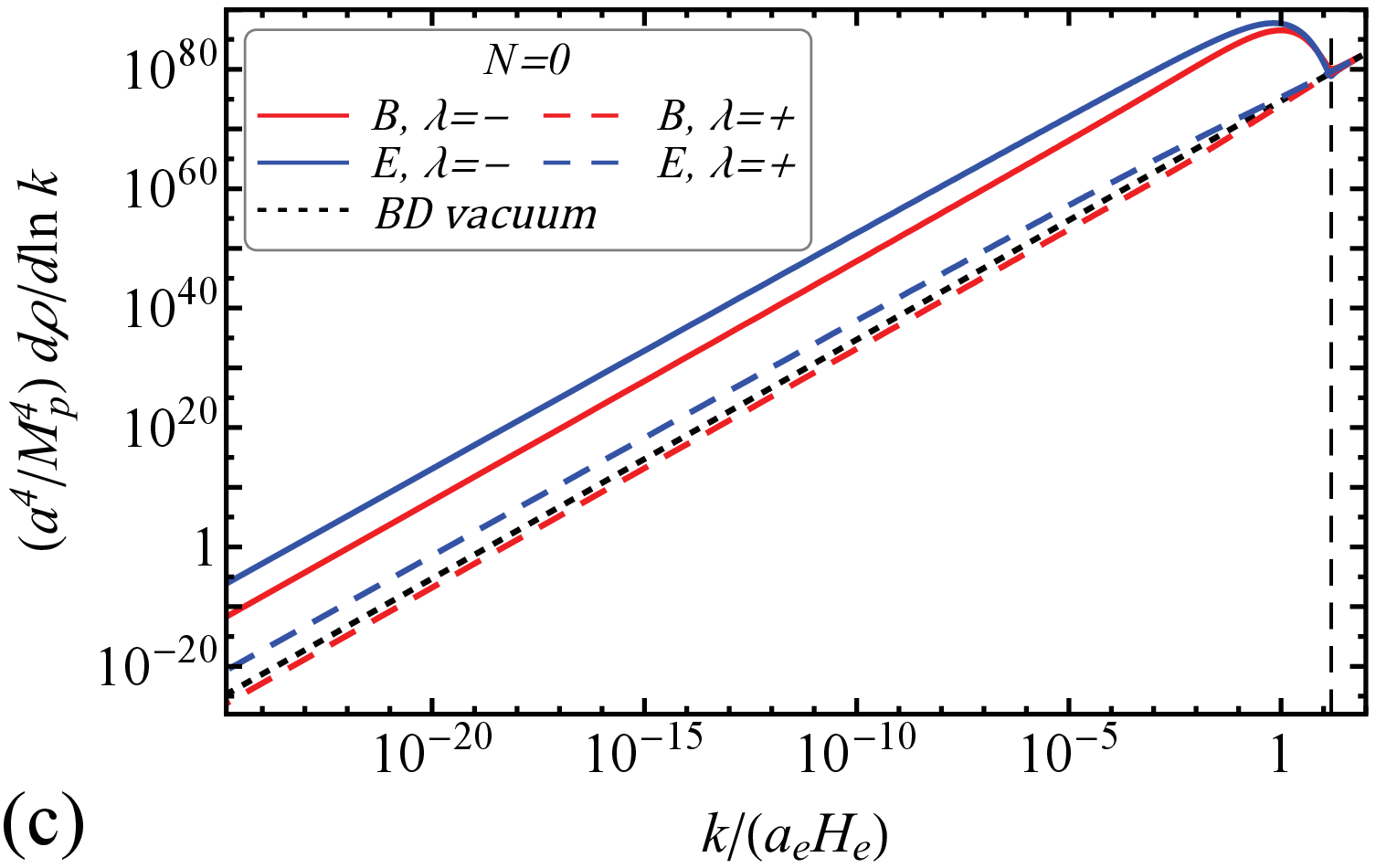}
	\caption{Components of the spectral densities of electric (blue lines) and magnetic (red lines) energy densities with the circular polarization $\lambda=-$ (solid lines) and $\lambda=+$ (dashed lines) for $\chi_{1}=0$ and $\chi_{2}=9\times 10^{9}$ at three moments of time during metric Higgs inflation: (a) $N=-30$, (b) $N=-15$, and (c) $N=0$ $e$-foldings from the end of inflation. The black dotted lines correspond to the unperturbed spectrum of vacuum fluctuations (the Bunch-Davies vacuum solution). Vertical dashed line on each panel shows the momentum of the mode crossing the horizon at the corresponding moment of time.}
	\label{fig-sp-a}
\end{figure}

The three panels in Fig.~\ref{fig-sp-a} show the power spectra of the generated electromagnetic field at three moments of time during inflation, namely, at $N=-30$, $N=-15$, and $N=0$ $e$-foldings from the end of inflation. They are found for the case of purely axial coupling, $\chi_{1}=0$, $\chi_{2}=9\times 10^{9}$ (the latter value is the largest for which the backreaction still can be neglected; see the next subsection). The electric and magnetic components of the energy density are shown in blue and red, respectively, while the contributions from different circular polarizations are shown by solid and dashed lines. Because of the axial coupling, only one polarization is enhanced, $\lambda=-$ in our case, the other remains practically unchanged (compare it with the Bunch-Davies vacuum solution shown by the black dotted line). As it is clear from Fig.~\ref{fig-sp-a}, the range of modes which undergo amplification during inflation is bounded from above by the last mode crossing the horizon. Such modes (at the corresponding moments of time) are shown by the vertical dashed lines in each panel. Clearly, for higher momenta the spectrum almost coincides with the Bunch-Davies vacuum spectrum. Thus, we can conclude that Eqs.~(\ref{k-h})--(\ref{k-h2}) for the momentum of the horizon-crossing mode indeed work satisfactorily.

\begin{figure}[ht!]
	\centering
	\includegraphics[height=3.6cm]{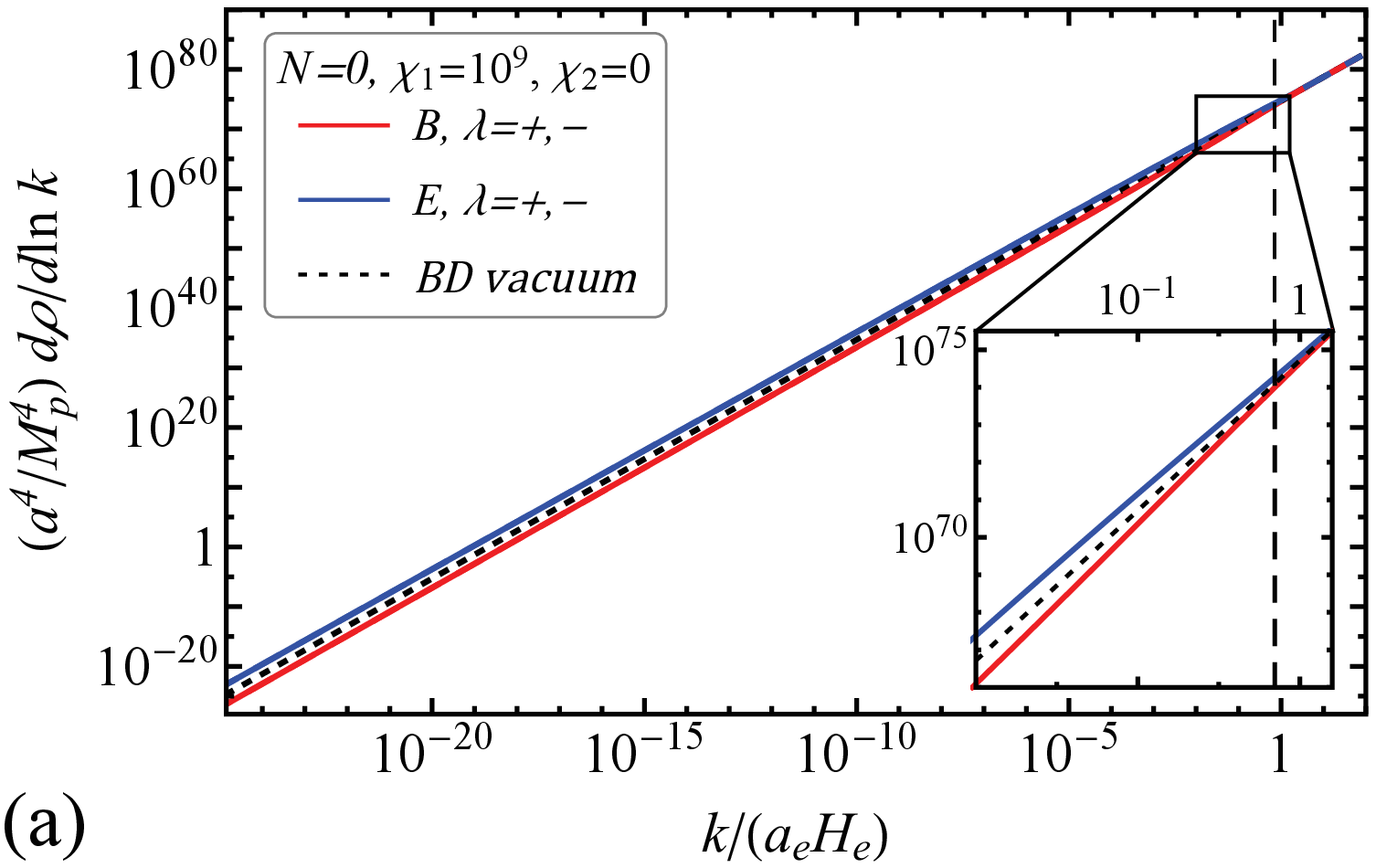}\hspace*{0.1cm}
	\includegraphics[height=3.6cm]{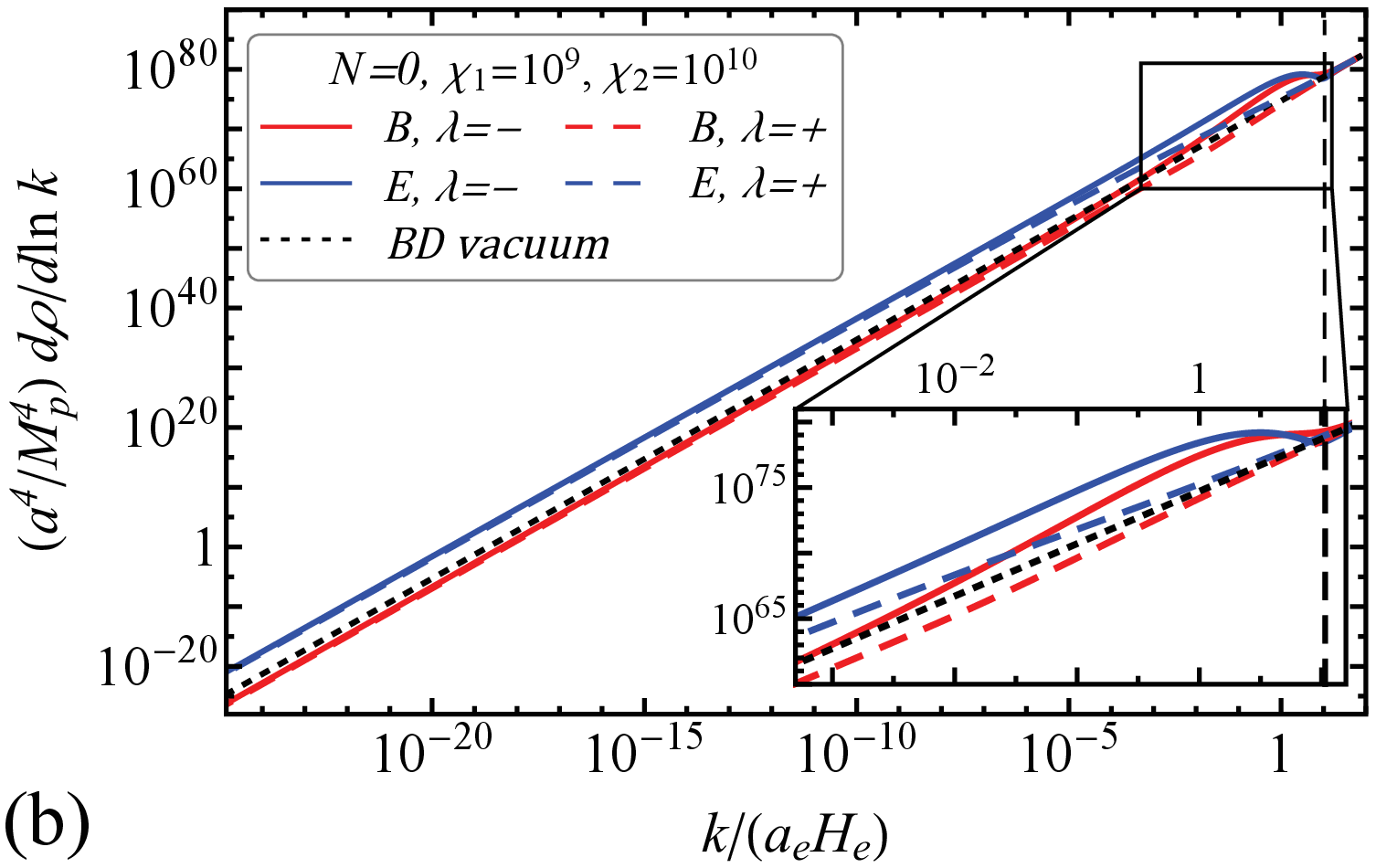}\hspace*{0.1cm}
	\includegraphics[height=3.6cm]{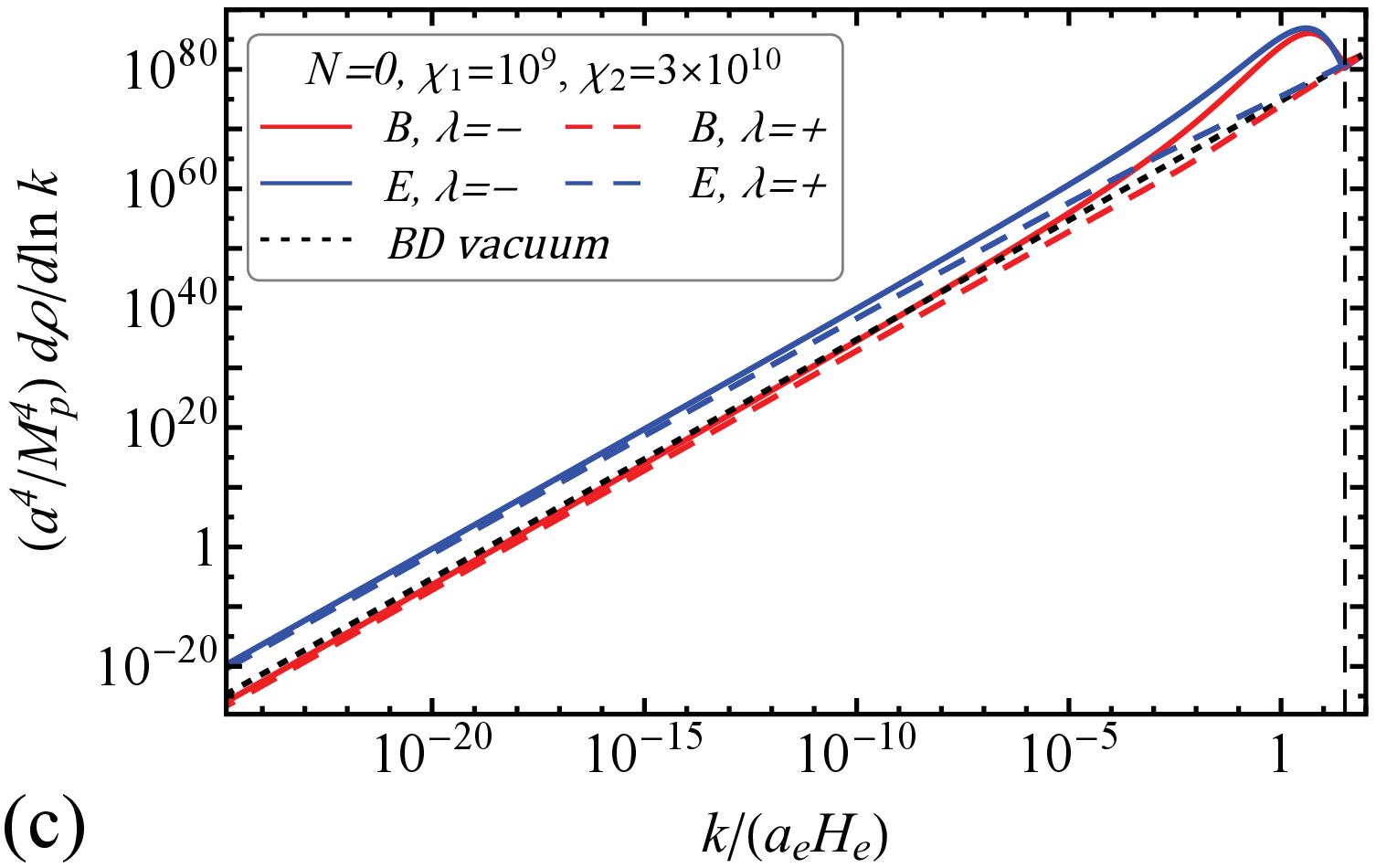}
	\caption{Components of the spectral densities of electric (blue lines) and magnetic (red lines) energy densities with the circular polarization $\lambda=-$ (solid lines) and $\lambda=+$ (dashed lines) at the end of metric Higgs inflation calculated for a fixed value of the parity-preserving coupling $\chi_{1}=10^{9}$ and three values of the parity-violating coupling constant $\chi_{2}$: (a) $\chi_{2}=0$ (contributions of both polarizations coincide), (b) $\chi_{2}=10^{10}$, and (c) $\chi_{2}=3\times 10^{10}$. The black dotted lines correspond to the unperturbed spectrum of vacuum fluctuations (the Bunch-Davies vacuum solution). The vertical dashed line in each panel shows the momentum of the mode crossing the horizon at the corresponding moment of time.}
	\label{fig-sp-mixed}
\end{figure}

Let us now discuss the impact of the kinetic coupling $\chi_{1}$ on the magnetogenesis. Figure~\ref{fig-sp-mixed} shows the power spectra of the generated fields at the end of inflation for nonzero kinetic coupling $\chi_{1}=10^{9}$ and three values of the axial coupling, namely, $\chi_{2}=0$, $\chi_{2}=10^{10}$, and $\chi_{2}=3\times 10^{10}$. First of all, we should mention that in the case of purely kinetic coupling the generation of the electromagnetic field is very weak. The electric component is slightly greater and the magnetic one is a bit less than the Bunch-Davies vacuum spectrum; however, both are of the same order of magnitude. Note that the generated field is the strongest at the end of inflation, and for earlier times the deviation from the vacuum solution is even less. Second, for nonzero axial coupling $\chi_{2}$, the generation is more significant; however, it is still much weaker than in the purely axial case [compare Fig.~\ref{fig-sp-a}(c) for $\chi_{2}=9\times 10^{9}$ and Fig.~\ref{fig-sp-mixed}(c) for $\chi_{2}=3\times 10^{10}$]. Third, the difference in the spectra of different polarizations (for nonzero axial coupling) appears only for the shortest modes which cross the horizon close to the end of inflation. All these features can be understood from the analysis of the coupling functions. We will perform such an analysis in the following subsection.

\subsection{Dynamics of the magnetogenesis during inflation}

Let us discuss the time evolution of the energy density and helicity of the generated field for different values of the coupling constants.

\begin{figure}[ht!]
	\centering
	\includegraphics[height=4.5cm]{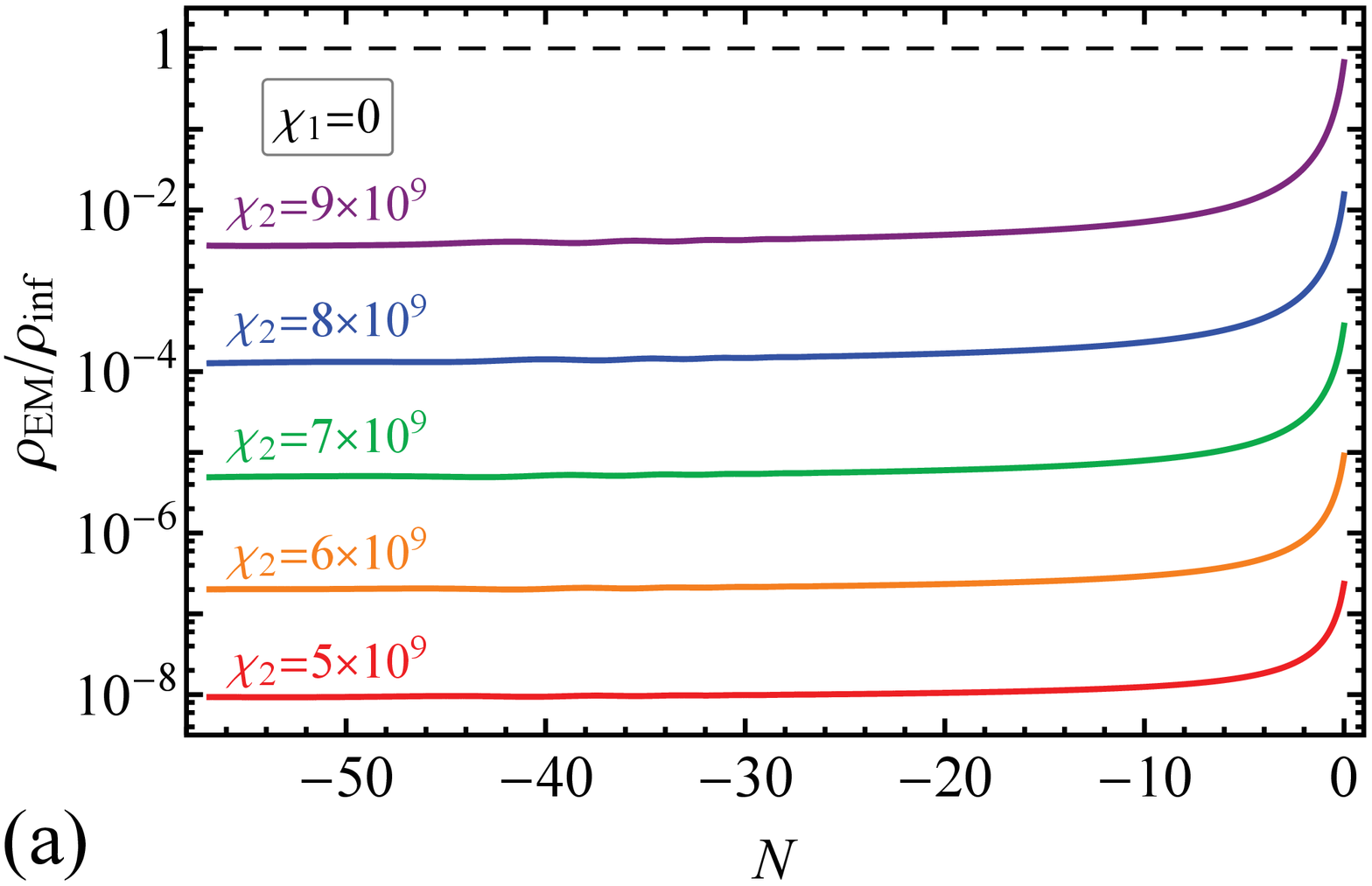}\hspace*{0.5cm}
	\includegraphics[height=4.5cm]{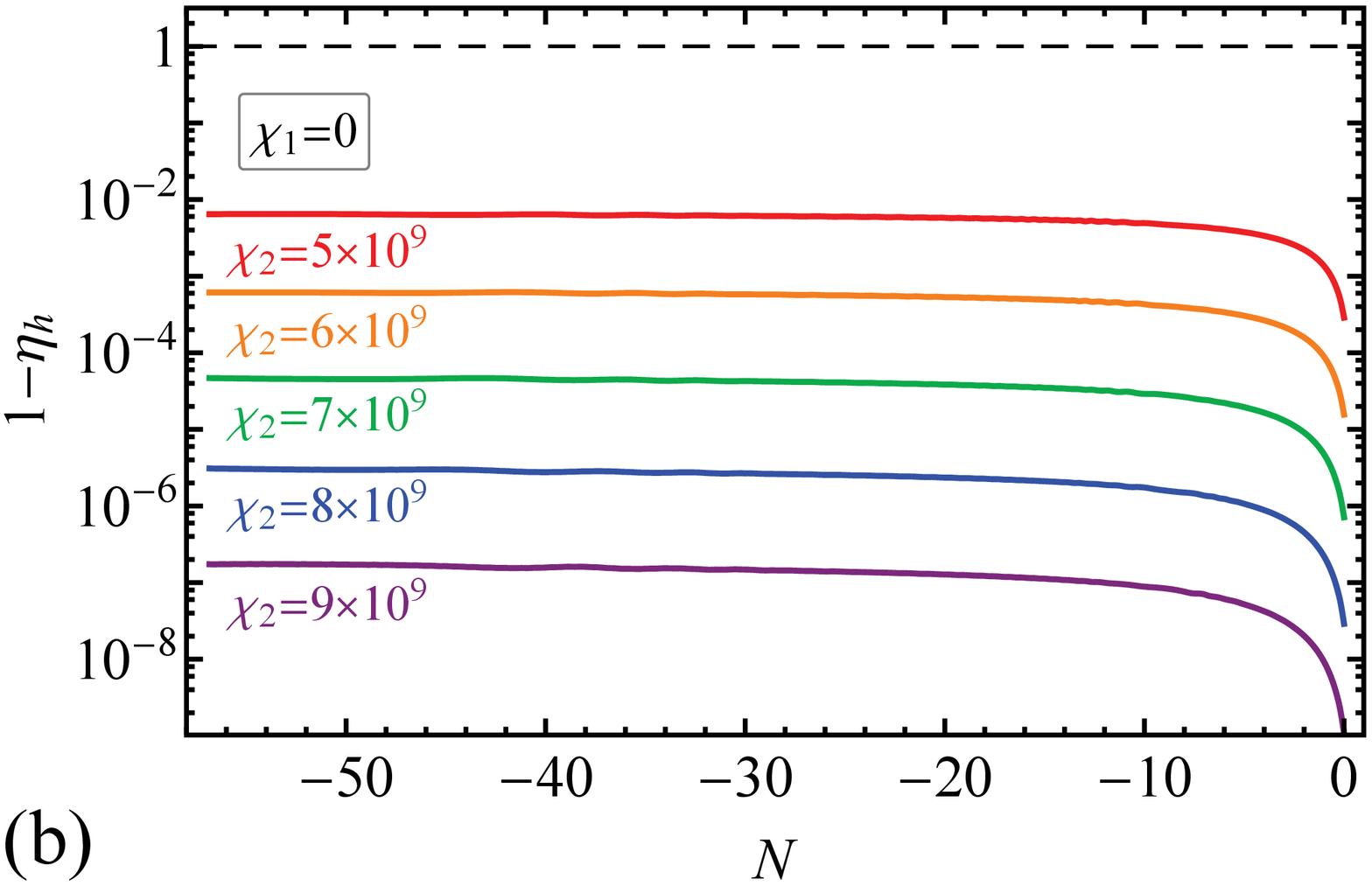}
	\caption{(a) The ratio of the electromagnetic energy density to that of the inflaton and (b) the deviation of the helicality of the electromagnetic field from unity $1-\eta_{h}$ as functions of the $e$-folding number counted from the end of Higgs inflation in metric formulation. The nonminimal coupling parameter $\chi_{1}=0$ and five different values of $\chi_{2}$ are used. The electromagnetic energy density increases in time; therefore, only its final value is important in the analysis of the backreaction. The generated field becomes stronger and more helical when the coupling parameter $\chi_{2}$ increases.}
	\label{fig-m-k0}
\end{figure}

\begin{figure}[ht!]
	\centering
	\includegraphics[height=4.5cm]{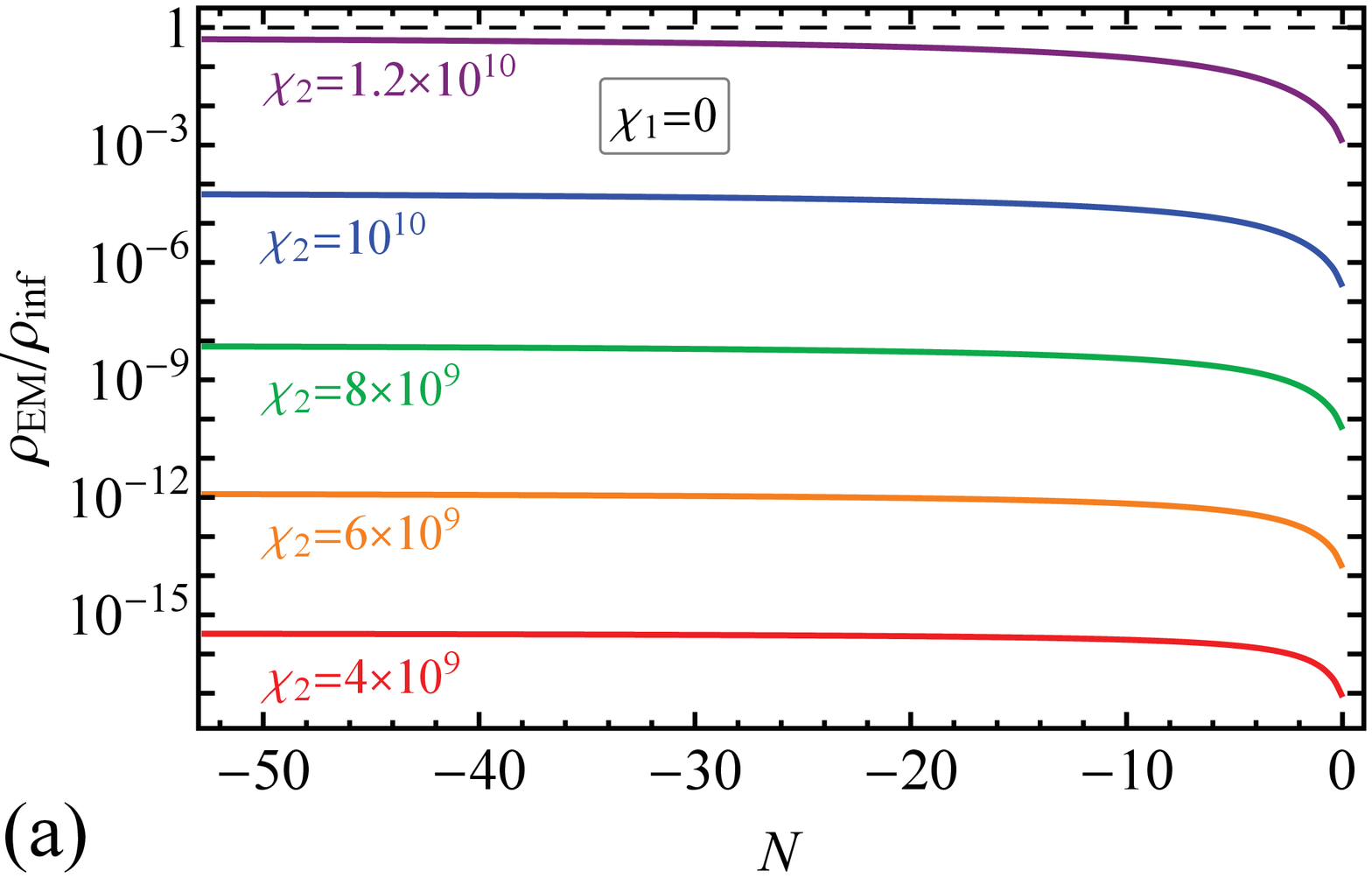}\hspace*{0.5cm}
	\includegraphics[height=4.5cm]{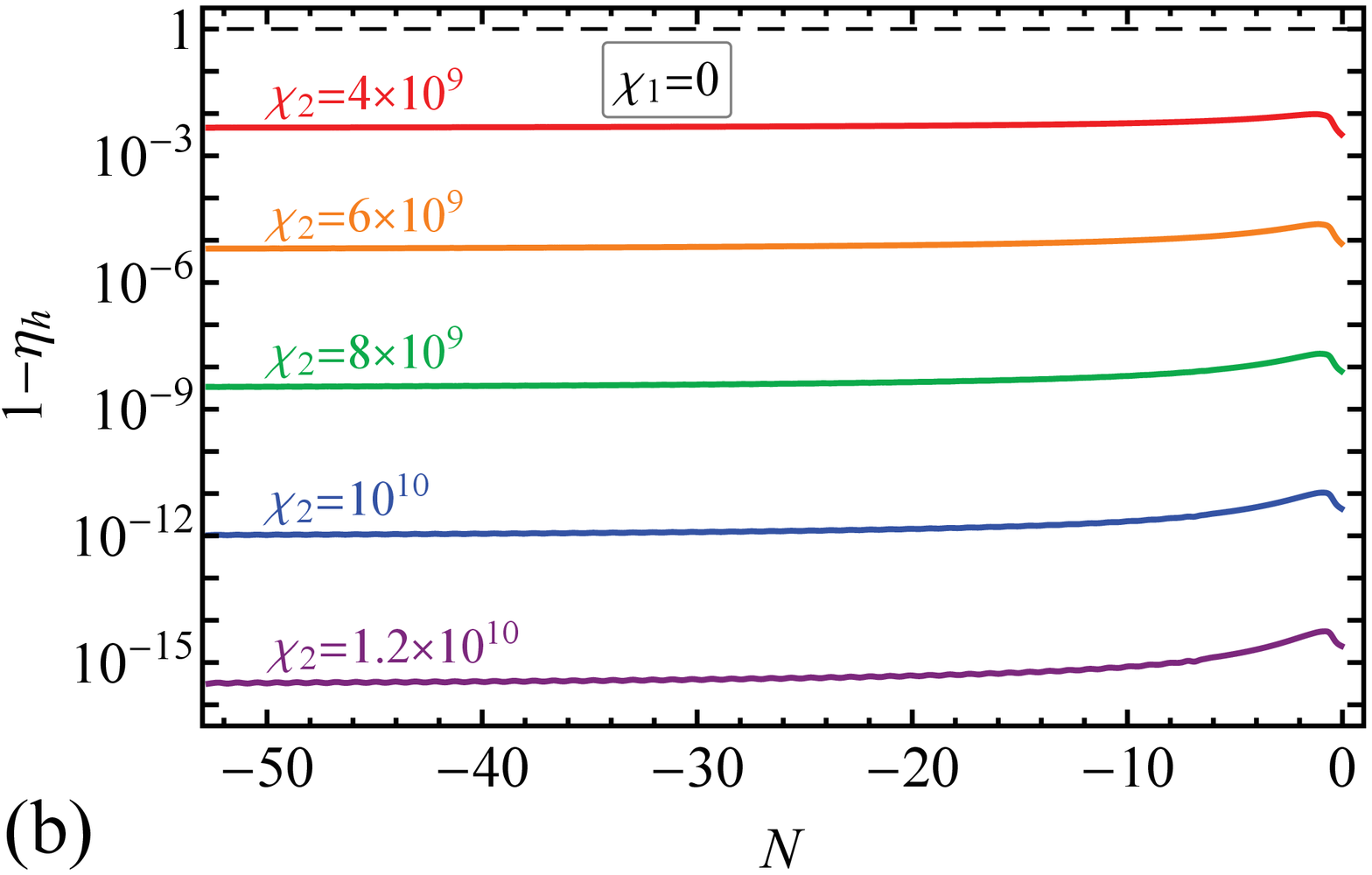}
	\caption{The same quantities as shown in Fig.~\ref{fig-m-k0}(a) and \ref{fig-m-k0}(b), respectively, in the case of Palatini Higgs inflation. In contrast to the metric case, the electromagnetic energy density decreases in time; consequently, its value far from the end of inflation must be taken in the analysis of the backreaction. The generated field becomes larger and more helical when the axial coupling parameter $\chi_{2}$ increases.}
	\label{fig-p-k0}
\end{figure}

We start from a purely helical case ($\chi_{1}=0$) and consider different values of $\chi_{2}$. The results for the electromagnetic energy density  are shown in Fig.~\ref{fig-m-k0} for metric Higgs inflation and Fig.~\ref{fig-p-k0} for the Palatini case.

The common features of both metric and Palatini cases are (i) the energy density of the generated field monotonically grows, and (ii) the helicality becomes closer to unity when the coupling constant $\chi_{2}$ increases. This can be easily understood from the mode equation~(\ref{eq-mode-2}) for $I_{1}\equiv 1$. Indeed, the last term in brackets is responsible for the enhancement of modes of only one circular polarization, because the tachyonic instability occurs only for $\lambda={\rm sign\,}(\dot{I}_{2})$ and $k<k_{h}(t)$. Since this term is proportional to $\chi_{2}$, the amplification is stronger for larger $\chi_{2}$ and, therefore, the relative contribution of the amplified mode also increases.

There are, however, some differences between the metric and Palatini cases. First of all, the energy density has qualitatively different time behavior: It increases in the metric case and decreases in the Palatini case. The reason can be deduced from the explicit expressions for the coupling functions (\ref{I1-Higgs-2}) and (\ref{I1-Palatini}). The rate of generation of the electromagnetic field is determined by the parameter \cite{Durrer:2011,Anber:2006,Anber:2010}
\begin{equation}
	\zeta=\left|\frac{I_{2}'(\phi)\dot{\phi}}{2H}\right|
\end{equation}
(usually, it is denoted by $\xi$ in the literature, but we would like to avoid the confusion with the nonminimal coupling $\xi$). Using the slow-roll approximation $3H\dot{\phi}=-V'(\phi)$, we obtain the following expressions for this parameter in two formulations of Higgs inflation:
\begin{equation}
	\zeta_{\rm MHI}=\frac{16}{3}\frac{V_{0}\chi_{2}}{M_{p}^{4}}\Big(1-e^{-\sqrt{\frac{2}{3}}\frac{\phi}{M_{p}}}\Big)^{-1},\qquad \zeta_{\rm PHI}\approx 32\xi \frac{V_{0}\chi_{2}}{M_{p}^{4}} {\rm tanh}^{4}\Big(\frac{\sqrt{\xi}\phi}{M_{p}}\Big),
\end{equation}
where $V_{0}=\lambda M_{p}^{4}/(4\xi^{2})$ is the amplitude of the inflaton potential. Obviously, the first function increases in time during inflation while the second one decreases (note that the inflaton $\phi$ decreases in time in both cases). The generated field evolves in time correspondingly. Such a decreasing behavior of the energy density in the Palatini case is unfavorable for magnetogenesis because of the following reason. To generate at the end of inflation as large electromagnetic field as possible, we would have in this case an even stronger field far from the end of inflation which would cause the backreaction, change the inflaton dynamics, and make an impact on the primordial spectra, in particular, on the modes which are now observable in the CMB anisotropy spectrum. These factors can strongly constrain the final value of the generated field which would be 2--3 orders of magnitude less than in the case of increasing $\zeta$. 

\begin{figure}[ht!]
	\centering
	\includegraphics[height=4.5cm]{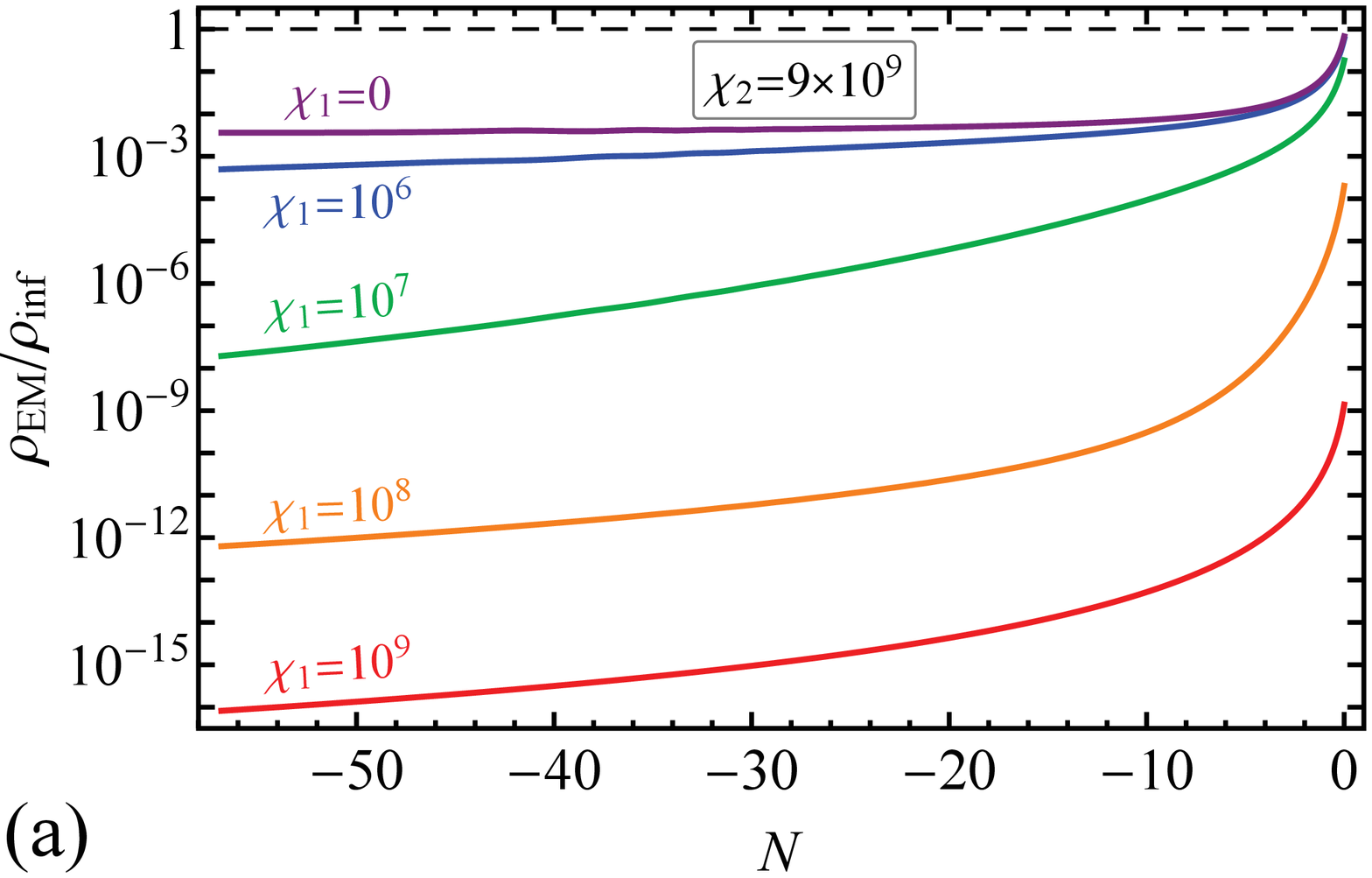}\hspace*{0.5cm}
	\includegraphics[height=4.5cm]{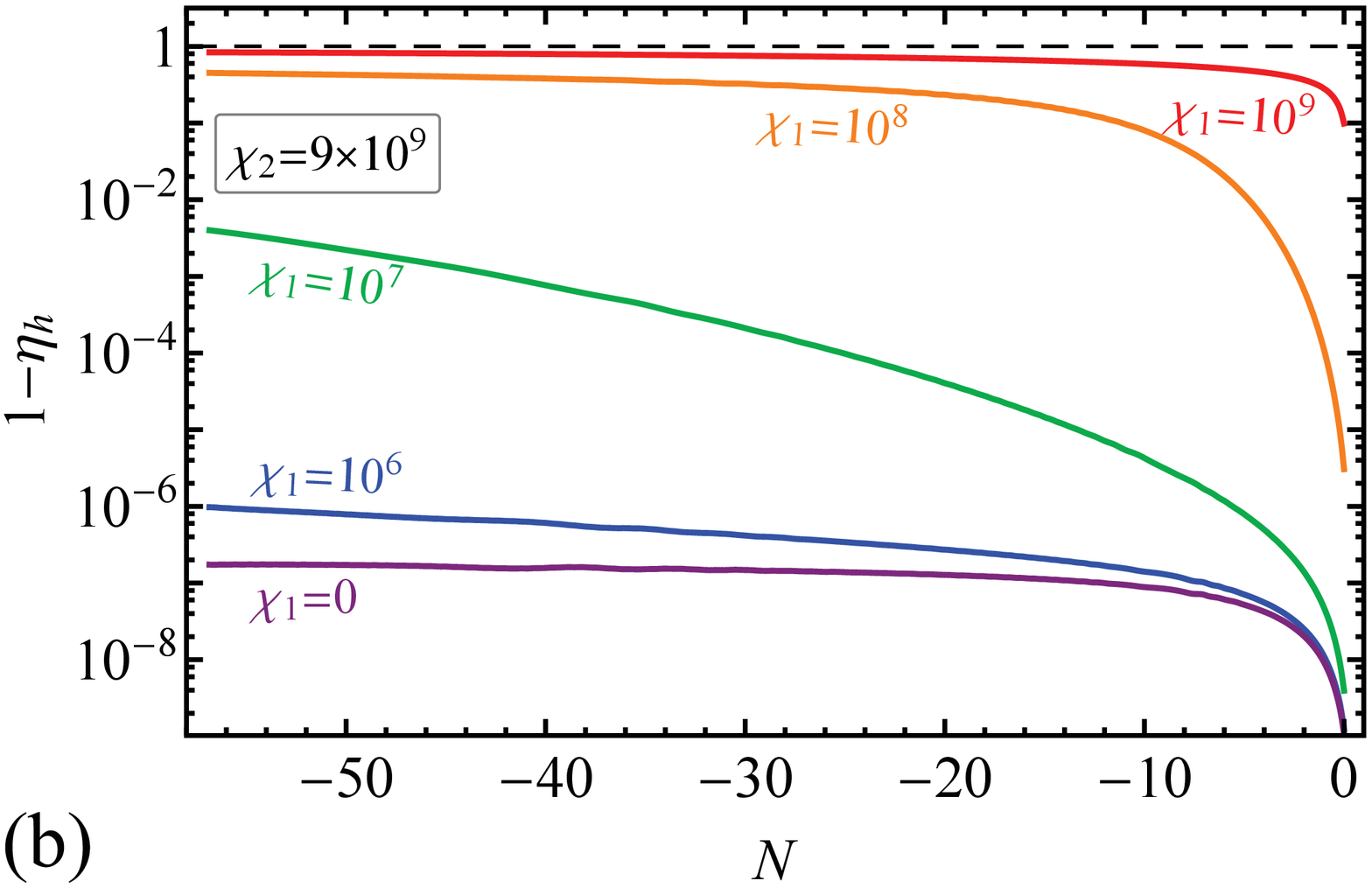}
	\caption{The same quantities as shown in Fig.~\ref{fig-m-k0}(a) and \ref{fig-m-k0}(b), respectively, in the case of metric Higgs inflation. The nonminimal coupling parameter $\chi_{2}=9\times 10^{9}$ and five different values of $\chi_{1}$ are used. Large values of the kinetic coupling parameter $\chi_{1}$ strongly suppress magnetogenesis, especially far from the end of inflation. For large $\chi_{1}$, both polarizations are enhanced comparably which leads to partially helical electromagnetic fields.}
	\label{fig-m-a9p9}
\end{figure}

\begin{figure}[ht!]
	\centering
	\includegraphics[height=4.5cm]{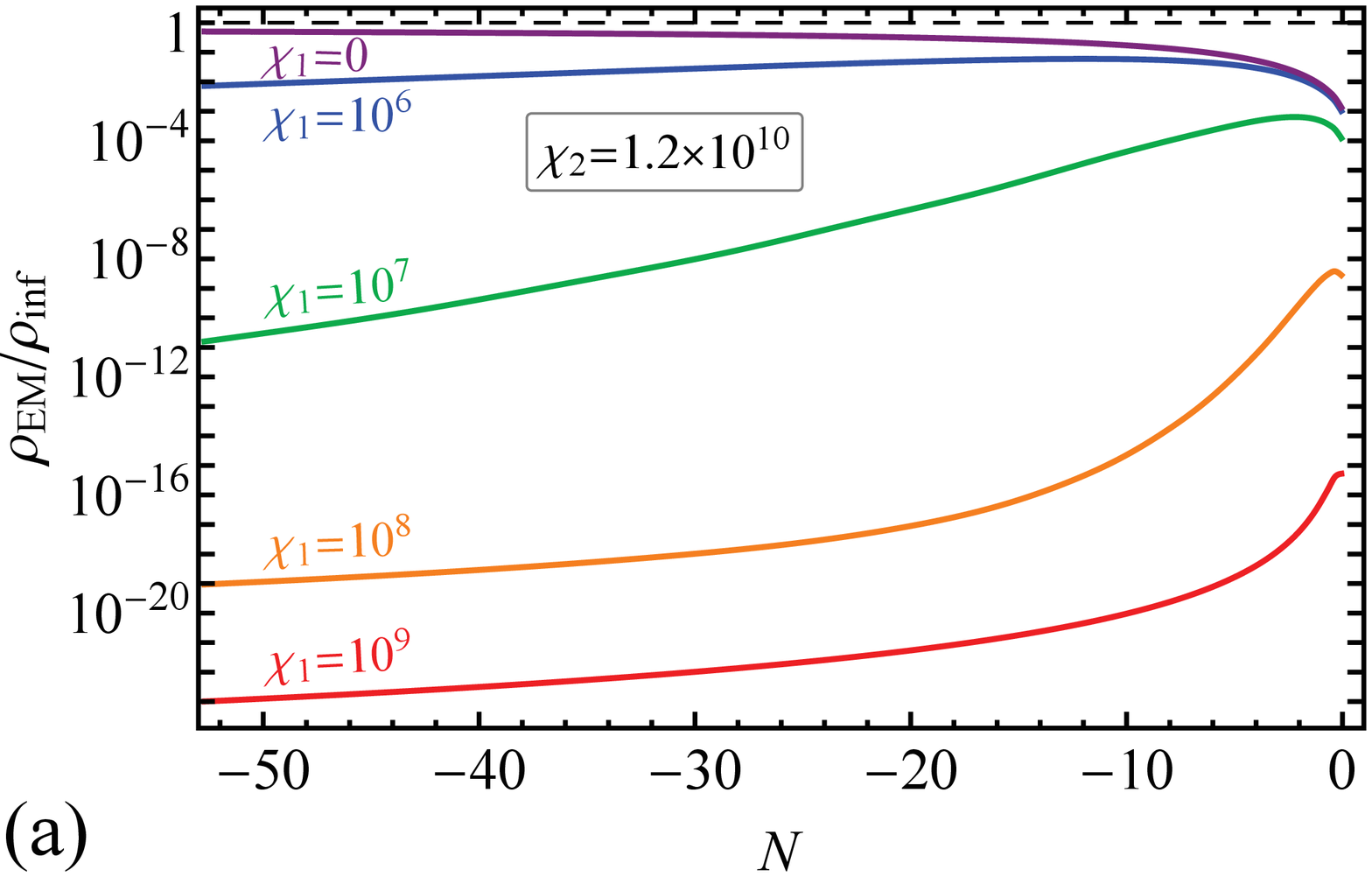}\hspace*{0.5cm}
	\includegraphics[height=4.5cm]{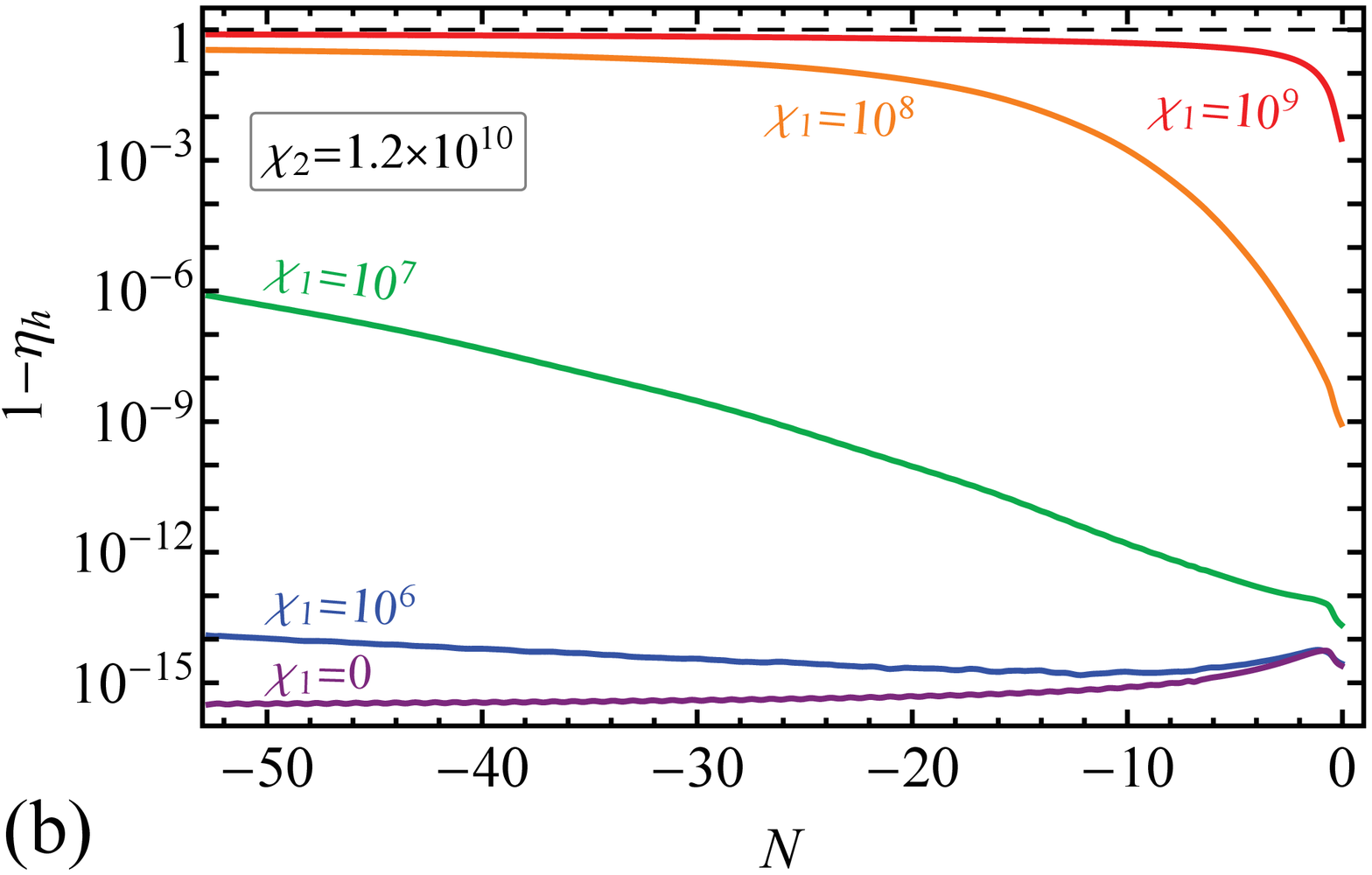}
	\caption{The same quantities as shown in Fig.~\ref{fig-m-k0}(a) and \ref{fig-m-k0}(b), respectively, in the case of Palatini Higgs inflation. The nonminimal coupling parameter $\chi_{2}=1.2\times 10^{10}$ and five different values of $\chi_{1}$ are used.}
	\label{fig-p-a12p9}
\end{figure}

Let us choose the maximal value of $\chi_{2}$ for which the backreaction does not occur and add the nonzero kinetic coupling $\chi_{1}$. The results for different values of $\chi_{1}$ are shown in Figs.~\ref{fig-m-a9p9} and \ref{fig-p-a12p9} for the metric and Palatini cases, respectively.

As we can see from these figures, adding nonzero kinetic coupling strongly suppresses the magnetogenesis, especially far from the end of inflation. This can also be understood from the mode equation (\ref{eq-mode-2}). The last term in brackets corresponding to the axial coupling contains the kinetic coupling function $I_{1}$ in the denominator. Far from the end of inflation, $I_{1}$ takes large values [see Eqs.~(\ref{I1-Higgs-2}) and (\ref{I1-Palatini})], which significantly reduce the value of $\zeta$ and slow down the magnetogenesis. Only when $\chi_{2}\gg \chi_{1}$ a satisfactory generation can be achieved. In such a case, the generated fields become almost maximally helical close to the end of inflation. 

\begin{figure}[ht!]
	\centering
	\includegraphics[height=4.5cm]{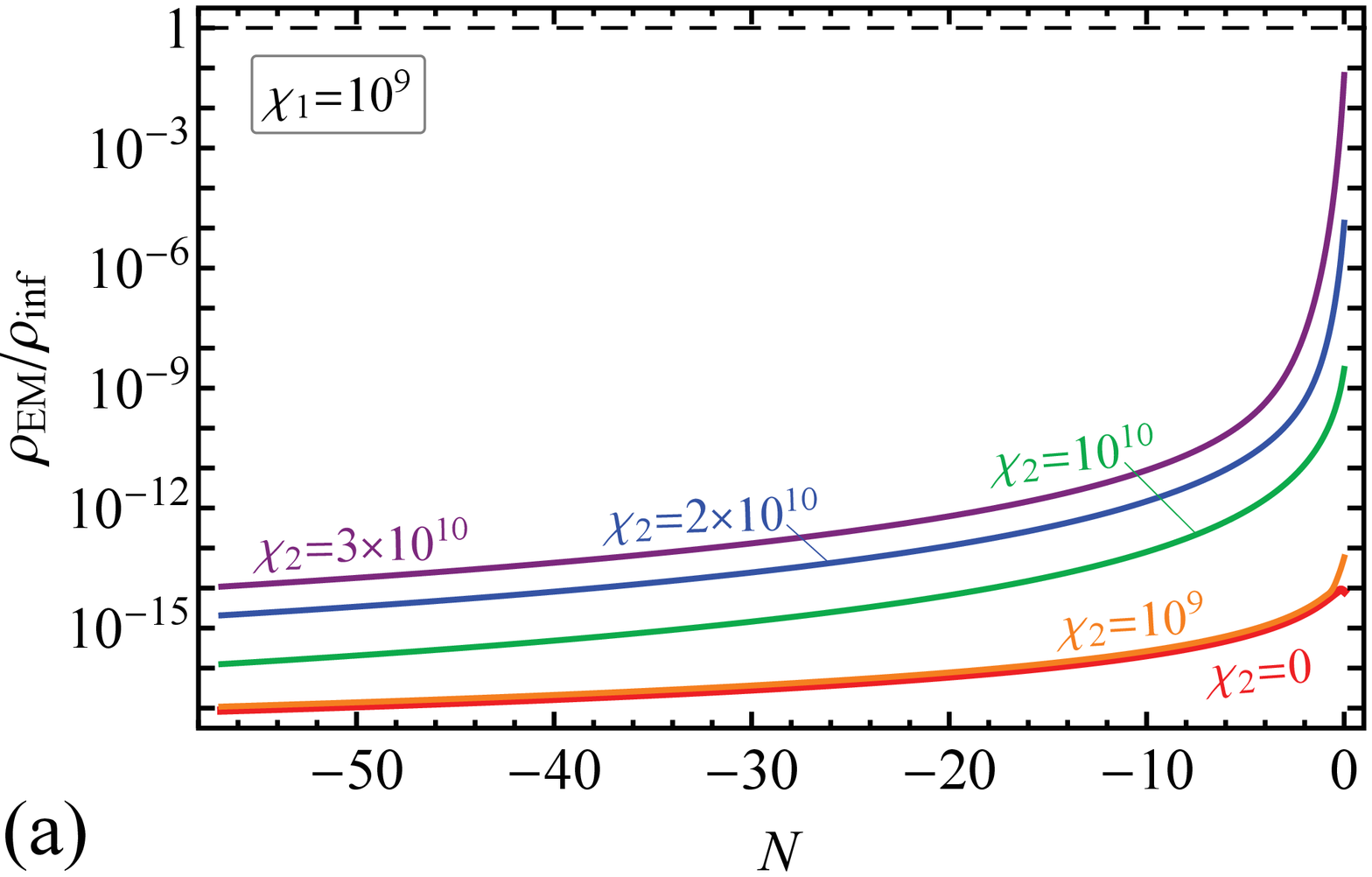}\hspace*{0.5cm}
	\includegraphics[height=4.5cm]{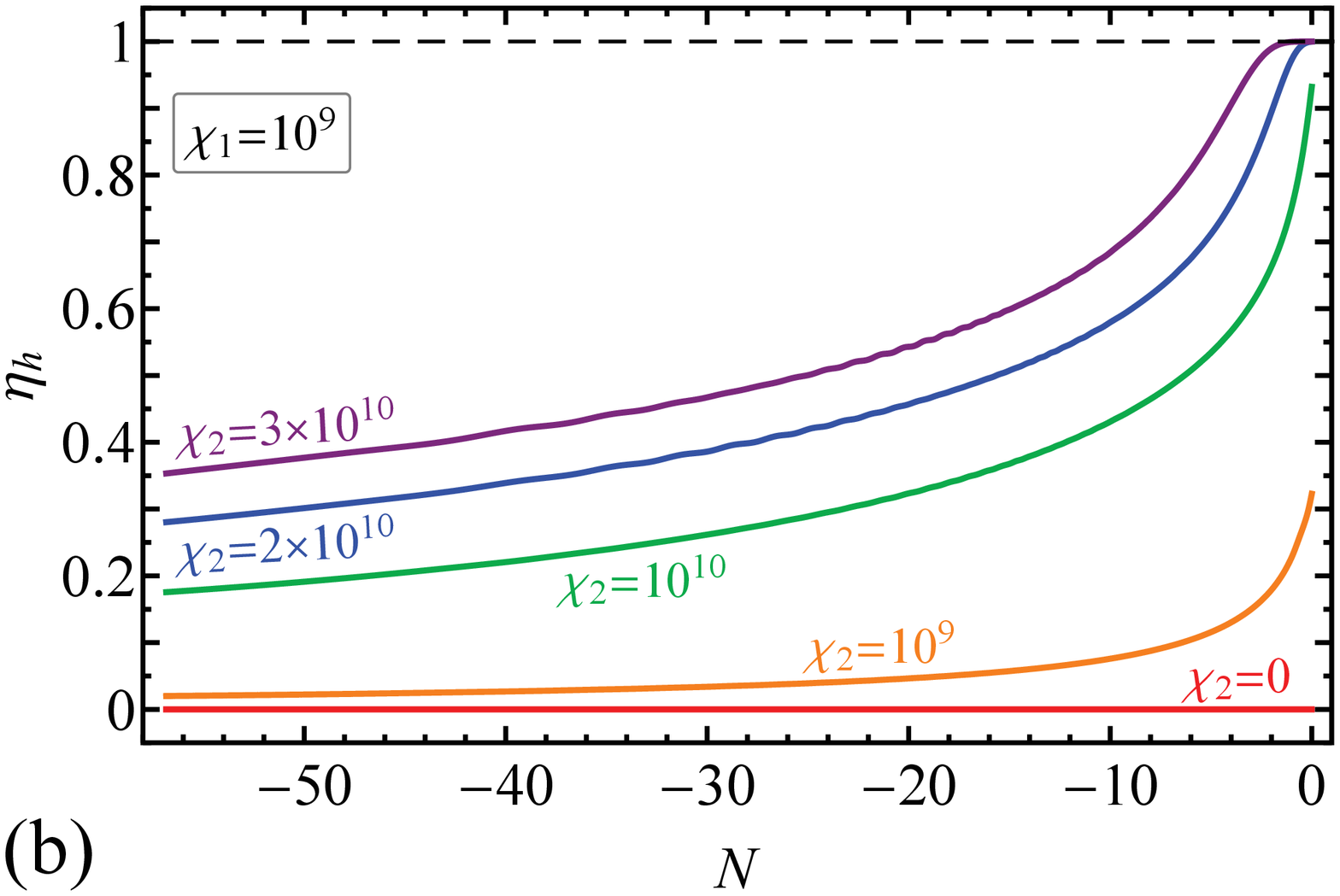}
	\caption{(a) The ratio of the electromagnetic energy density to that of the inflaton and (b) the helicality of the electromagnetic field $\eta_{h}$ generated during metric Higgs inflation as functions of the $e$-folding number counted from the end of inflation. The nonminimal coupling parameter $\chi_{1}=10^{9}$ and five different values of $\chi_{2}$ are used. During almost the whole inflation stage the magnetogenesis is suppressed by a large value of the coupling function $I_{1}$. The situation changes close to the end of inflation when a rapid growth occurs and the helicality approaches 1 (for sufficiently large value of $\chi_{2}$).}
	\label{fig-m-k1p9}
\end{figure}

\begin{figure}[ht!]
	\centering
	\includegraphics[height=4.5cm]{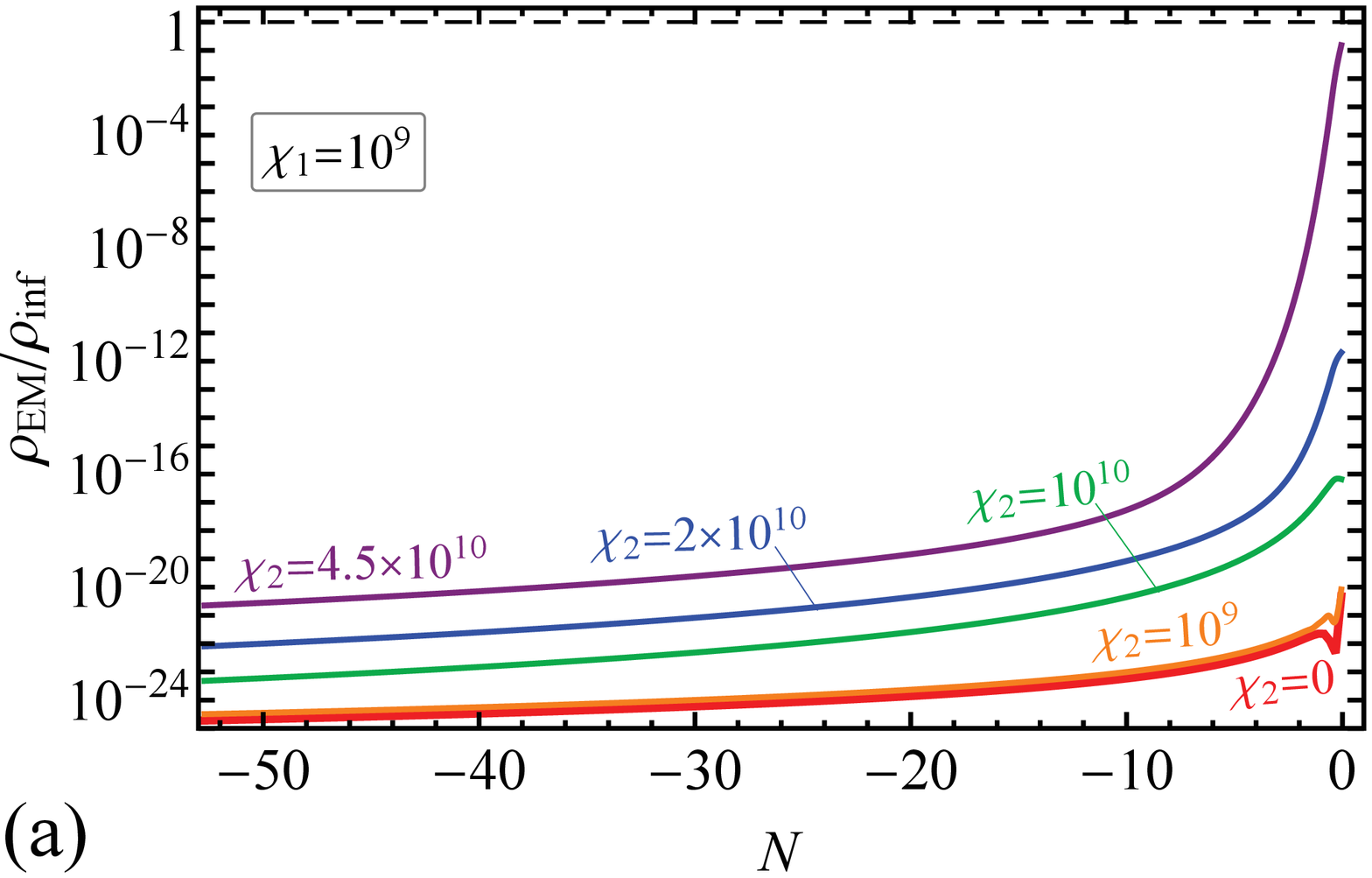}\hspace*{0.5cm}
	\includegraphics[height=4.5cm]{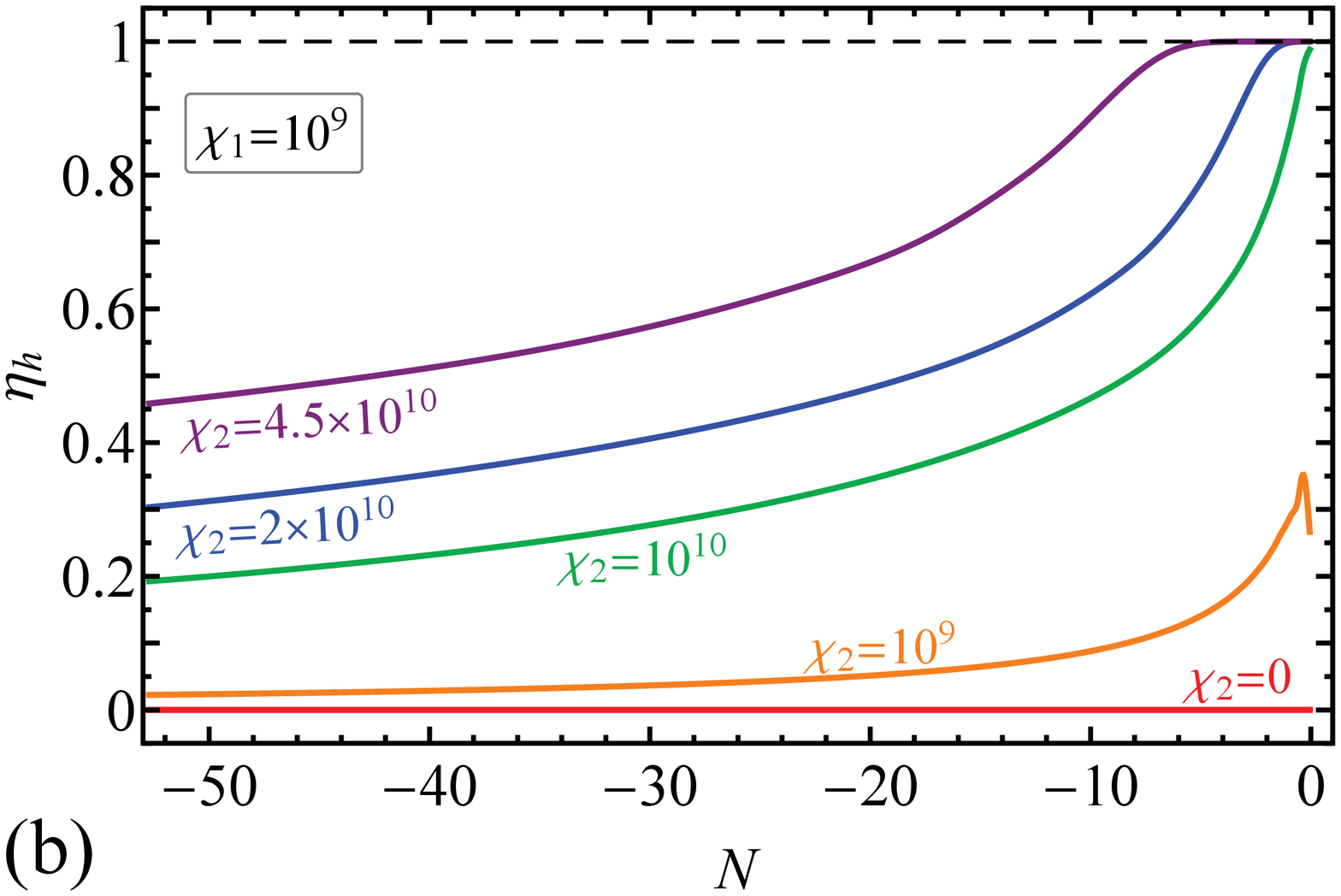}
	\caption{The same quantities as shown in Fig.~\ref{fig-m-k1p9}(a) and \ref{fig-m-k1p9}(b), respectively, in the case of Palatini Higgs inflation.}
	\label{fig-p-k1p9}
\end{figure}

There are also the second and third terms in brackets in Eq.~(\ref{eq-mode-2}), which correspond to the purely kinetic coupling of the electromagnetic field to the inflaton. The main difference in these terms from the last term is that they contain the same function both in the numerators and denominators. This means that it is not possible to increase these terms indefinitely by increasing the value of the coupling constant $\chi_{1}$. For large values of $\chi_{1}$ it just cancels out from these terms. For example, in the case of metric Higgs inflation, the coupling function $I_{1}\propto \exp(\sqrt{2/3}\phi/M_{p})$ for $\chi_{1}\gg \xi^{2}/\lambda$ and far from the end of inflation. It has the form of the Ratra coupling function $I_{1}=\exp(2\beta \phi/M_{p})$, which was considered in Refs.~\cite{Vilchinskii:2017,Sobol:2018}. It was shown that successful magnetogenesis occurs in such a model only for $\beta=\mathcal{O}(10)$. In our case, we have $\beta=1/\sqrt{6}<1$; therefore, the coupling function changes too slowly to cause a significant enhancement of electromagnetic modes. Indeed, it is straightforward to show that $I_{1}\propto \ln(a_{e}/a)$; i.e., it changes only logarithmically with the scale factor \cite{Savchenko:2018}, while for a successful generation one needs more rapid decrease; see, e.g., Refs.~\cite{Martin:2008,Demozzi:2009}. Thus, we conclude that even if the kinetic coupling parameter $\chi_{1}$ is so large that the unity in Eq.~(\ref{I1-Higgs-2}) can be neglected (the terms $\dot{I}_{1}/I_{1}$, $\ddot{I}_{1}/I_{1}$ are saturated), in the absence of the axial coupling it is impossible to generate significant electromagnetic fields. A similar conclusion can be made for the Palatini case. This is illustrated in Figs.~\ref{fig-m-k1p9} and \ref{fig-p-k1p9}, where the purely kinetic coupling case is shown by red lines.

Figures~\ref{fig-m-k1p9} and \ref{fig-p-k1p9} show that for a fixed value of the kinetic coupling $\chi_{1}$, the magnetogenesis is strongly suppressed unless the axial coupling is much greater, $\chi_{2}\gg \chi_{1}$. In the latter situation, the generated field remains very weak during almost all inflation stage, because $I_{1}$ is large and suppresses the generation. However, close to the end of inflation, $I_{1}$ decreases enough and the generation occurs very rapidly. This explains the shape of the spectrum in Fig.~\ref{fig-sp-mixed}(c): Only modes that cross the horizon close to the end of inflation undergo amplification due to the axial coupling.

\begin{figure}[ht!]
	\centering
	\includegraphics[height=4.5cm]{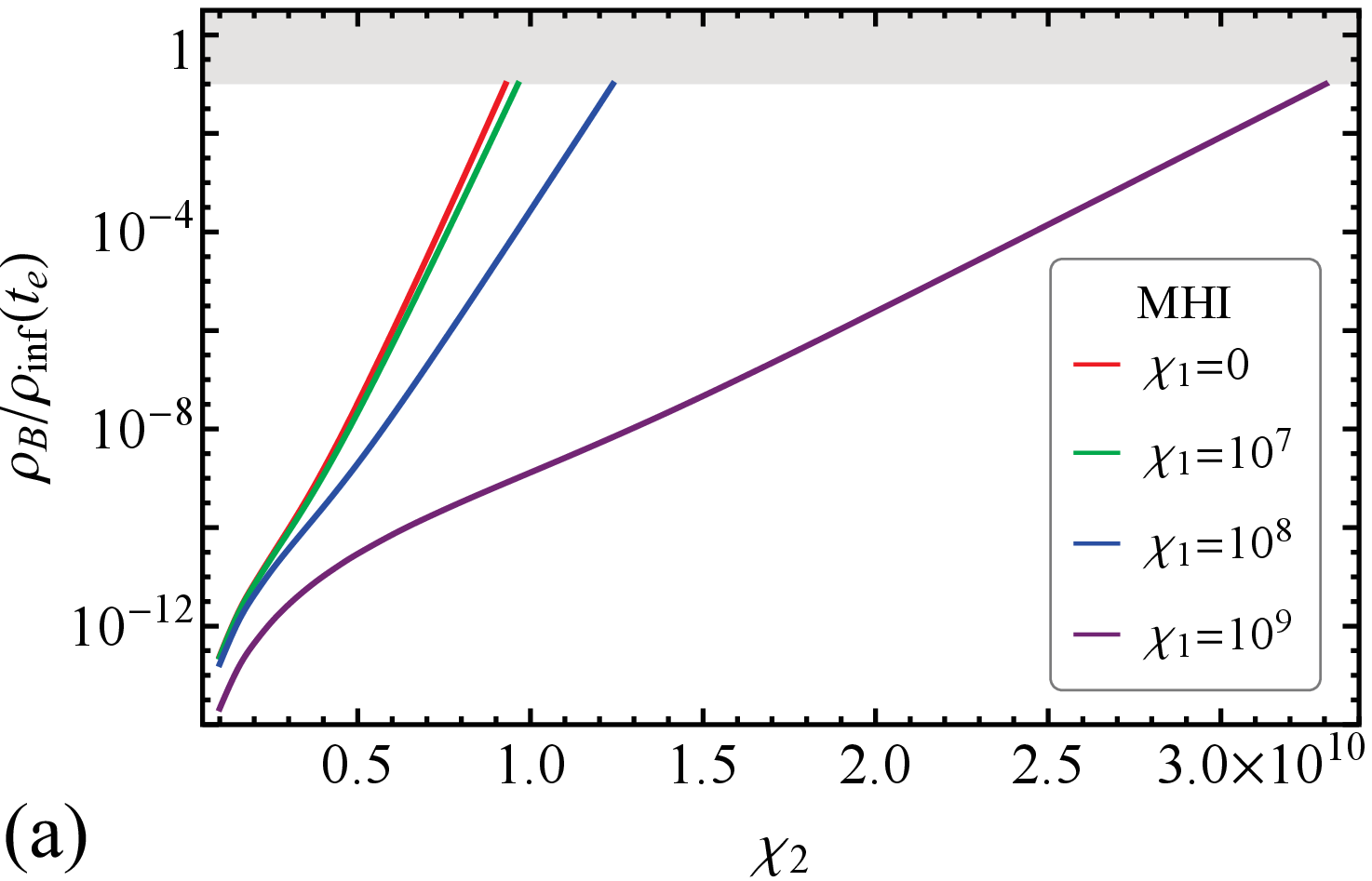}\hspace*{0.7cm}
	\includegraphics[height=4.5cm]{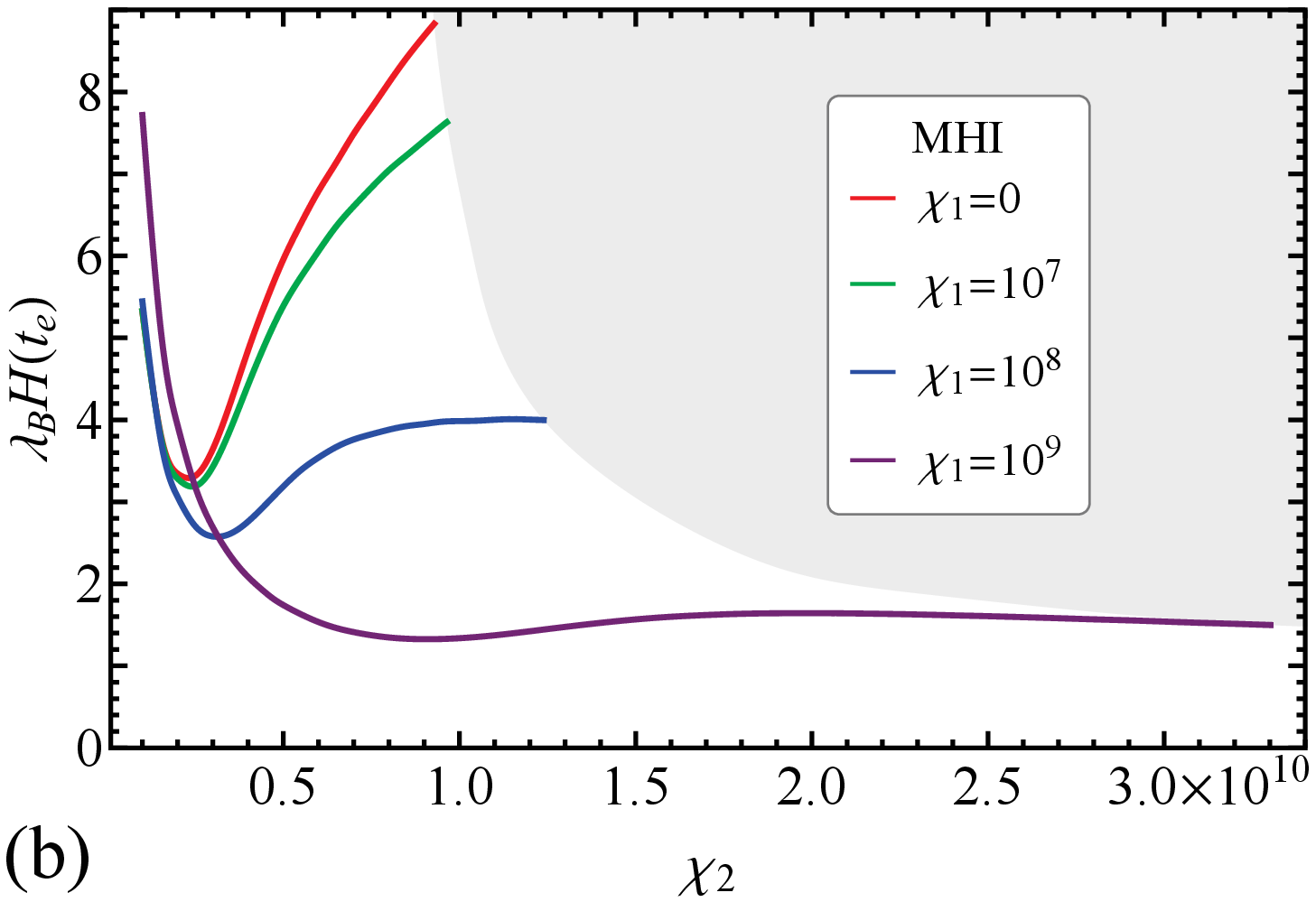}
	\caption{(a) The ratio of the magnetic energy density to that of the inflaton and (b) the magnetic correlation length compared to the Hubble radius at the end of metric Higgs inflation as functions of the axial coupling constant $\chi_{2}$ for four different values of the kinetic coupling constant $\chi_{1}$. The region which cannot be achieved without causing the backreaction during inflation is shaded.}
	\label{fig-end-inflation-m}
\end{figure}

\begin{figure}[ht!]
	\centering
	\includegraphics[height=4.5cm]{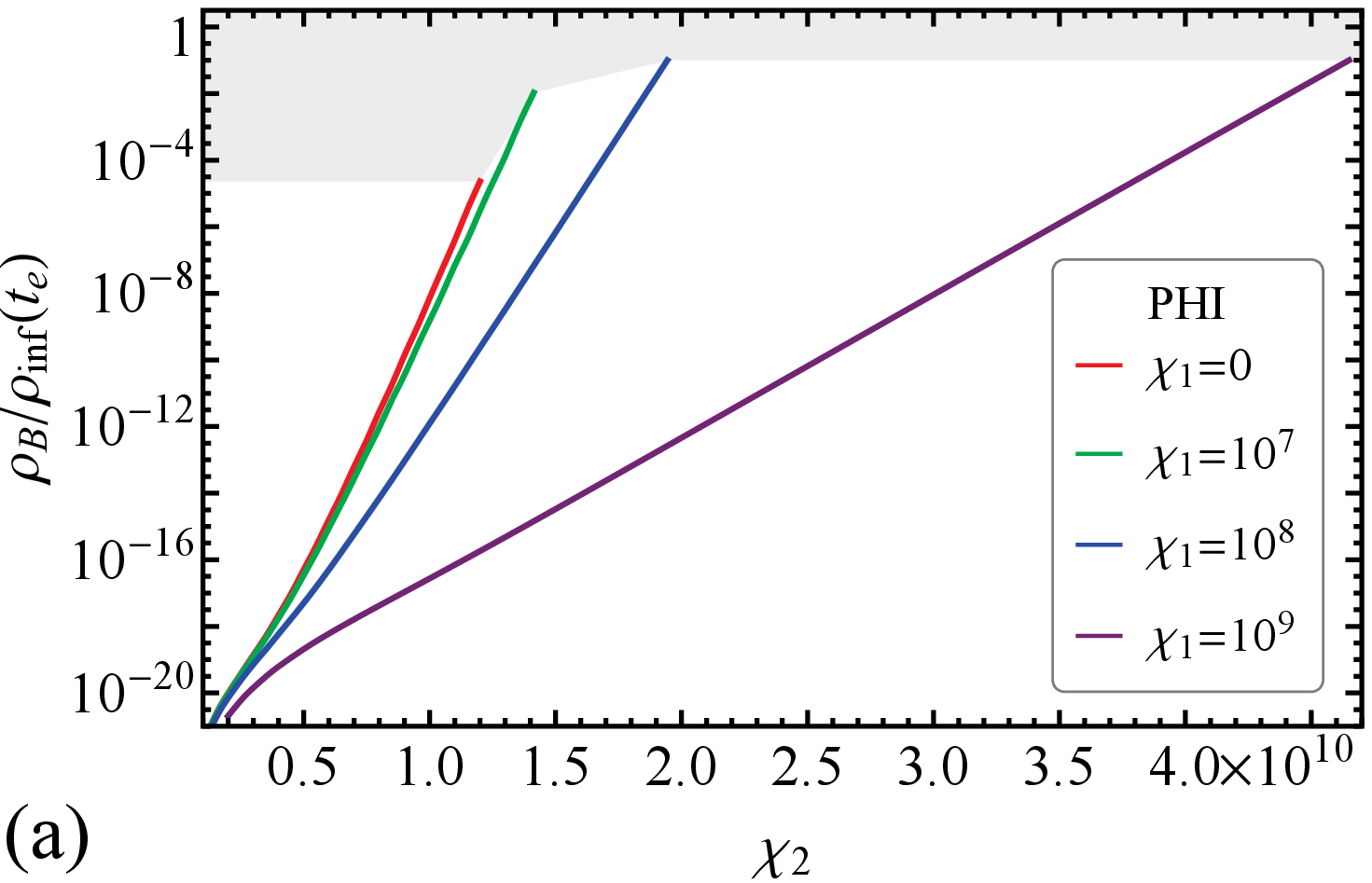}\hspace*{0.7cm}
	\includegraphics[height=4.5cm]{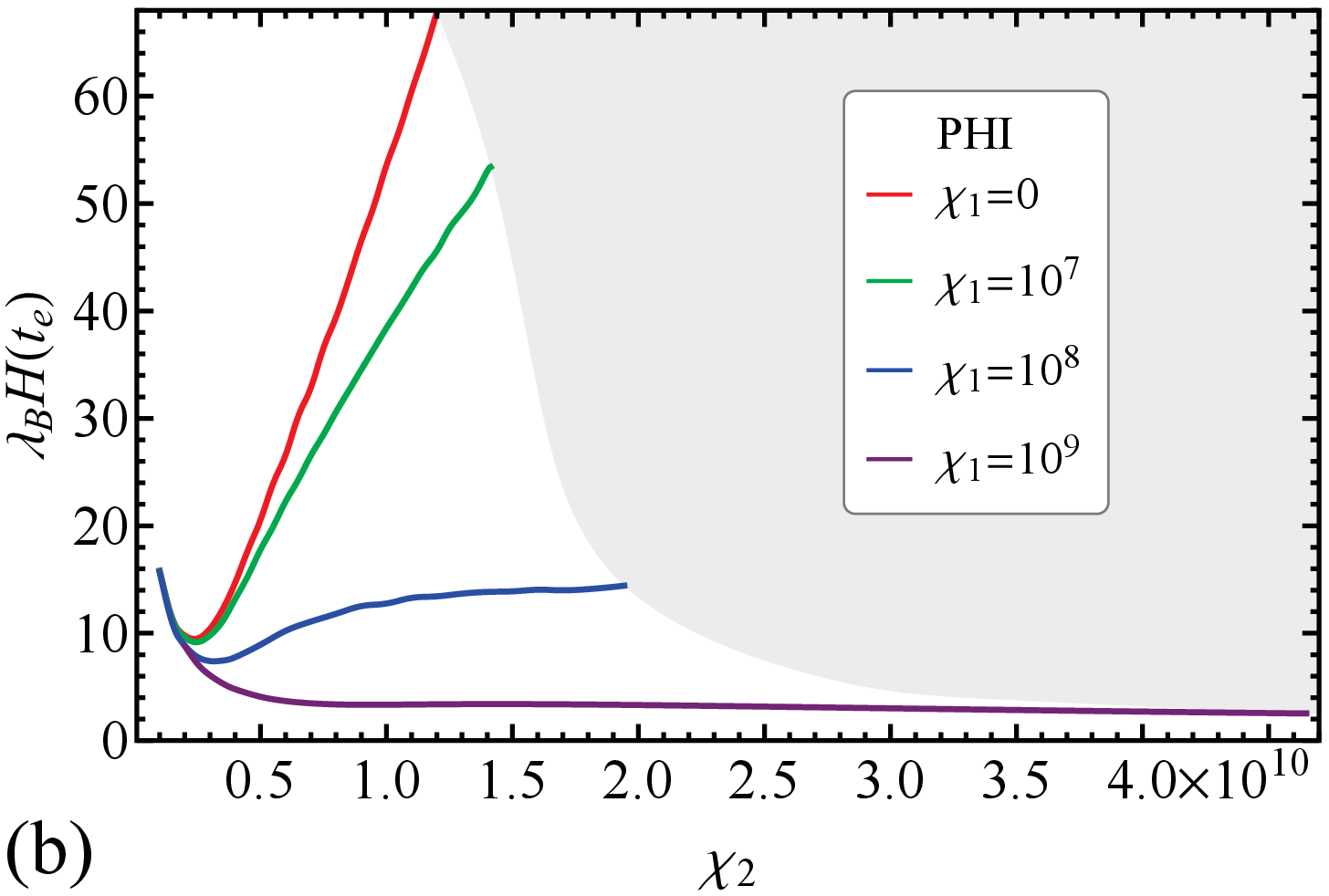}
	\caption{The same quantities as shown in Fig.~\ref{fig-end-inflation-m}(a) and \ref{fig-end-inflation-m}(b), respectively, in the case of Palatini Higgs inflation.}
	\label{fig-end-inflation-p}
\end{figure}

For further analysis, we need to find the magnetic component of the energy density and the magnetic correlation length at the end of inflation. We calculate them for different values of the coupling constants $\chi_{1,2}$ and show them in Figs.~\ref{fig-end-inflation-m} and \ref{fig-end-inflation-p}.


\subsection{Postinflationary evolution and the modern value of the magnetic field}

Finally, let us discuss the postinflationary evolution of the generated fields until the present epoch. As recent numerical simulations show \cite{Ema:2016,DeCross:2016,Rubio:2019}, the preheating stage following Higgs inflation appears to be very short and intensive, so that it is usually treated as instantaneous. During this stage, the Universe is filled with all particle species and the medium becomes highly conductive \cite{Ahonen:1996nq,Baym:1997gq,Ahonen:1998iz}. Therefore, the electric component of the generated field vanishes while the magnetic one is ``frozen'' in plasma \cite{Grasso:2001,Durrer:2013,Subramanian:2016}.

The joint evolution of plasma and the magnetic field strongly depends on the helicity of the latter. The common feature is that the shortest modes undergo Ohmic dissipation in view of a finite conductivity of the medium---the process of the so-called magnetic diffusion \cite{Grasso:2001}.  If the magnetic field has zero helicity, its long-wavelength modes evolve adiabatically $B_{k}\propto a^{-2}$, while the short-wavelength part of the spectrum gradually disappears transferring its energy into the kinetic motion of the plasma \cite{Banerjee:2004}. The shortest mode which survives is determined by the cosmic diffusion scale which is estimated as $L_{\rm diff}\sim 10^{-5}\,$pc at the present epoch \cite{Grasso:2001}.

For helical magnetic fields, however, the comoving helicity conservation enforces the transfer of energy from short- to long-wavelength modes thus increasing their chances to survive. This phenomenon is known as the inverse cascade of magnetic helicity, and it occurs in highly conductive and turbulent plasma \cite{Banerjee:2004} which existed in the Universe prior to recombination. Moreover, in the presence of chiral asymmetry in the fermionic sector, the inverse cascade occurs even without turbulence \cite{Joyce:1997uy,Boyarsky:2011uy,Tashiro:2012mf,Hirono:2015rla,Dvornikov:2016jth,Gorbar:2016klv,Brandenburg:2017rcb,Schober:2018wlo}. 

As we saw from the numerical results of the previous subsection, nonhelical magnetic fields generated for $\chi_{2}=0$ are extremely weak, because the kinetic coupling function $I_{1}$ evolves very slowly during inflation. The strongest fields can be generated only when the axial coupling $\chi_{2}$ is nonzero and much greater than the kinetic coupling $\chi_{1}$. In such a case, the generated fields are almost maximally helical. That is why in what follows we consider only this situation.

At the end of inflation, we know the magnitude $B(t_{e})=\sqrt{2\rho_{B}(t_{e})}$ and the correlation length $\lambda_{B}(t_{e})$ of the magnetic field; see Figs.~\ref{fig-end-inflation-m} and \ref{fig-end-inflation-p}. Let us introduce the corresponding comoving quantities which would be 
equal to the present-day values of $B$ and $\lambda_{B}$ if they were adiabatically rescaled until now, i.e.,
\begin{equation}
	\label{comoving-quantities}
	\tilde{B}(t_{e})=B(t_{e})\frac{a_{e}^{2}}{a_{0}^{2}}, \quad \tilde{\lambda}_{B}(t_{e})=\lambda_{B}(t_{e})\frac{a_{0}}{a_{e}},
\end{equation}
where $a_{e}$ and $a_{0}$ are the values of the scale factor at the end of inflation and today, respectively. Let us assume that the preheating is instantaneous so that $a_{\rm reh}\approx a_{e}$. Then, we can relate the scale factors using the comoving entropy conservation:
\begin{equation}
	\label{a0-ae}
	\frac{a_{0}}{a_{e}}=\frac{g_{\ast,{\rm reh}}^{1/3}T_{\rm reh}}{g_{\ast,0}^{1/3}T_{0}},
\end{equation}
where $T_{\rm reh}$ is the reheating temperature (which depends on the inflationary model; see Table~\ref{tab-parameters}), $T_{0}=2.725\,$K is the present value of the CMB temperature, $g_{\ast,{\rm reh}}=106.75$ is the total number of relativistic degrees of freedom in the Standard Model, and $g_{\ast,0}=43/11\approx 3.91$ is the corresponding effective number today. Note that if there were no magnetic diffusion and inverse cascade, the comoving quantities would give the present-day values of the magnetic field and its correlation length.

Because of the finite conductivity of plasma of the early Universe, the short-wavelength modes undergo dissipation or magnetic diffusion \cite{Grasso:2001}. However, the total comoving helicity must be conserved; therefore, the helicity stored in these short-wavelength modes is transferred (together with a part of the energy density) to the long-wavelength ones. The crucial characteristic of the inverse cascade is that such a transfer occurs only in the turbulent regime which establishes the equipartition of energy between the kinetic motion in plasma and the magnetic field \cite{Banerjee:2004}. In the following, we assume that this regime takes place in the early Universe starting from reheating and until recombination. It leads to the scaling of the comoving quantities with the conformal time $\tau$ \cite{Durrer:2013,Subramanian:2016,Banerjee:2004}:
\begin{equation}
	\tilde{B}\propto \tilde{\lambda}_{B}^{-1/2}\propto \tau^{-1/3}.
\end{equation}

As we show in Appendix~\ref{app-B}, the conformal times at reheating and recombination are related as
\begin{equation}
	\label{relation-conformal-times}
	\frac{\tau_{\rm r}}{\tau_{\rm reh}}=\kappa \frac{T_{\rm reh}}{T_{\rm r}}, \quad \kappa\approx 1.32.
\end{equation}
Here, $T_{\rm r}=4000\,$K is the temperature at the recombination epoch. Further, we assume that after recombination the evolution of the magnetic field is adiabatic so that its comoving characteristics do not change. Finally, we get the following present-day values of the magnetic field and its correlation length:
\begin{equation}
	B(t_{0})=\tilde{B}(t_{\rm r})=\sqrt{2\rho_{B}(t_{e})}\left(\frac{a_{0}}{a_{e}}\right)^{-2}\left(\frac{\tau_{\rm r}}{\tau_{\rm reh}}\right)^{-1/3}=\sqrt{\frac{\rho_{B}(t_{e})}{\rho_{\rm inf}(t_{e})}}\frac{\pi}{\sqrt{15}\kappa^{1/3}}\frac{g_{\ast,0}^{2/3}}{g_{\ast,{\rm reh}}^{1/6}}\frac{T_{0}^{2}T_{\rm r}^{1/3}}{T_{\rm reh}^{1/3}},
\end{equation}
\begin{equation}
	\lambda_{B}(t_{0})=\tilde{\lambda}_{B}(t_{\rm r})=\lambda_{B}(t_{e})\frac{a_{0}}{a_{e}}\left(\frac{\tau_{\rm r}}{\tau_{\rm reh}}\right)^{2/3}=\frac{\lambda_{B}(t_{e})}{H^{-1}(t_{e})}\frac{3\sqrt{10}\kappa^{2/3}}{\pi}\frac{1}{g_{\ast,0}^{1/3}g_{\ast,{\rm reh}}^{1/6}}\frac{M_p}{T_{0}T_{\rm r}^{2/3}T_{\rm reh}^{1/3}}.
\end{equation}
In the last equations, we normalized the magnetic energy density by that of the inflaton and the magnetic correlation length by the Hubble horizon size $l_{H}=1/H$ at the end of inflation. Now, taking into account the numerical values of reheating temperature and the number of degrees of freedom, we get the following estimates:
\begin{equation}
	B(t_{0})=5.3\times 10^{-15} \left(\frac{\rho_{B}(t_{e})}{0.1 \rho_{\rm inf}(t_{e})}\right)^{-1/2} \left(\frac{T_{\rm reh}}{10^{15}\,{\rm GeV}}\right)^{-1/3}\,{\rm G},
\end{equation}
\begin{equation}
	\lambda_{B}(t_{0})=1.4\times 10^{-5} \left(\frac{\lambda_{B}(t_{e})}{10 H^{-1}(t_{e})}\right) \left(\frac{T_{\rm reh}}{10^{15}\,{\rm GeV}}\right)^{-1/3}\,{\rm Mpc}.
\end{equation}

As we saw from the numerical results of the previous subsection, combining the parameters $\chi_{1}$ and $\chi_{2}$ it is possible to obtain the energy density of the generated field close to that of the inflaton. However, we are restricted by the ``no backreaction'' condition (\ref{weak-field}); therefore, we could take for definiteness $\rho_{B}=0.1\rho_{\rm inf}$ at the end of inflation. At the same time, the magnetic correlation length is of order $\lambda_{B}\sim 10H^{-1}$. These are the best possible values which can be achieved in both types of Higgs inflation; see Figs.~\ref{fig-end-inflation-m} and \ref{fig-end-inflation-p}. Since the typical correlation length is much less than Mpc today, for comparison with the results of blasars observations \cite{Tavecchio:2010,Ando:2010,Neronov:2010,Dolag:2010,Dermer:2011,Taylor:2011,Caprini:2015}, it is convenient to calculate the effective magnitude of the magnetic field $B_{\rm eff}=B\sqrt{\lambda_{B}/(1\,{\rm Mpc})}$. Using the reheating temperature from Table~\ref{tab-parameters}, we obtain the following estimates:
\begin{equation}
	B(t_{0})=4\times 10^{-15} \,{\rm G}, \qquad \lambda_{B}(t_{0})= 10\,{\rm pc}, \qquad B_{\rm eff}(t_{0})=1.3\times 10^{-17}\,{\rm G}, \qquad ({\rm MHI})
\end{equation}
\begin{equation}
	B(t_{0})=1.5\times 10^{-14} \,{\rm G}, \qquad \lambda_{B}(t_{0})= 42\,{\rm pc}, \qquad B_{\rm eff}(t_{0})= 10^{-16}\,{\rm G}. \qquad ({\rm PHI})
\end{equation}
The resulting value of Palatini Higgs inflation is slightly larger because of the lower reheating temperature.

\section{Conclusion}
\label{sec-concl}

In this work, we studied the generation of the electromagnetic field due to its nonminimal coupling to gravity during Higgs inflation. We started from the Jordan frame where, in addition to the nonminimal coupling of the Higgs field $\propto Rh^2$, we considered the parity-preserving $\propto RF_{\mu\nu}F^{\mu\nu}$ (or kinetic) and parity-violating $\propto R F_{\mu\mu}\tilde{F}^{\mu\nu}$ (or axial) terms in the action and performed the Weyl transformation to the Einstein frame treating the electromagnetic field as a perturbation (i.e., the Weyl transformation depended only on the Higgs field, like in the usual Higgs inflation). Then, neglecting the backreaction of the electromagnetic field on the dynamics of inflation, we identified the kinetic and axial coupling functions of the electromagnetic field to the inflaton. Thus, we reformulated this problem in terms of the well-known kinetic and axial coupling models; however, the coupling functions were not simply postulated (as it typically happens in the literature) but deduced from the structure of the nonminimal coupling.

We considered two formulations of gravity: the usual metric formulation and the Palatini one, which, being completely analogous in the Einstein-Hilbert theory, give rise to different inflationary models in the presence of the nonminimal coupling. However, despite the fact that the effective inflaton potentials and numerical values of parameters are very different in these two cases, both of them are in good accordance with the CMB observations \cite{Planck:2018-infl}. We showed that the magnetogenesis in these two formulations also has a few common features. First of all, the purely kinetic coupling ($\chi_{2}=0$) cannot lead to the generation of some sizeable electromagnetic fields, because the coupling function changes very slowly. Second, successful magnetogenesis may occur only in the case of purely axial ($\chi_{1}=0$) or axially dominated ($\chi_{2}\gg \chi_{1}$) coupling. There are, however, some qualitative differences in the two formulations. One of them is in the time behavior of the generated energy density in the purely axial case. In metric Higgs inflation, this quantity increases monotonically, while it decreases in the Palatini formulation. The latter case is unfavorable for magnetogenesis; however, adding nonzero $\chi_{1}$ of sufficient magnitude can improve the situation.

The strongest fields that can be generated in this model are always maximally helical that is a consequence of the axial coupling domination. Such fields have very good chances to survive until the present time because of the inverse cascade of magnetic helicity which occurs in the turbulent plasma of the early Universe. We estimated the maximal values of the present magnetic field and its coherence length which are equal to $10^{-14}-10^{-15}\,$G and $\sim 10-40\,$pc, respectively. The higher values correspond to the Palatini case, because it has lower energy scale of inflation and, therefore, is closer to us in time. Despite the fact that the generation of these fields avoids the strong coupling and the backreaction problems during inflation and does not contradict the upper bound $\sim 10^{-9}\,$G imposed by the CMB observations \cite{Planck-pmf}, they are potentially in conflict with recent constraints related to the baryon isocurvature modes of perturbations \cite{Kamada:2021}. However, this question requires more careful analysis and is beyond the scope of the present study.

\begin{acknowledgments}
	
	The work of E.~V.~G. and O.~M.~T. was supported by National Research Foundation of Ukraine Project No.~2020.02/0062.	
	The work of S.~I.~V. was supported by Swiss National Science Foundation Grant No.~SCOPE IZSEZ0-186551 and by German Academic Exchange Service Grant No.~57387479.
	The work of O.~O.~S. was supported by Swiss National Science Foundation Grant No.~200020B\_182864.
	
\end{acknowledgments}

\appendix

\section{Parameters of inflationary models}
\label{app-A}

In this appendix, we determine the parameters of the inflaton potential using the constraints imposed by the CMB observations \cite{Planck:2018-infl}.

First, we consider the general form for the inflaton potential
\begin{equation}
	V(\phi)=V_{0}\,v(x,\xi_{i}), \qquad x=\phi/M_{p},
\end{equation}
where $V_{0}$ is an amplitude of the potential, $\xi_{i}$ are dimensionless parameters of the potential, and $v$ is a given dimensionless function which determines the shape of the potential. The slow-roll inflation takes place until one of the following parameters reaches unity:
\begin{eqnarray}
	\epsilon&\equiv& \frac{M_{p}^{2}}{2}\left(\frac{V'_{\phi}}{V}\right)^{2}=\frac{1}{2}\left(\frac{v'_{x}}{v}\right)^{2},\label{epsilon}\\
	\eta&\equiv& M_{p}^{2}\frac{V''_{\phi\phi}}{V}=\frac{v''_{xx}}{v}.\label{eta}
\end{eqnarray}
Obviously, the corresponding value $\phi_{e}$ depends only on $\xi_{i}$ and not on $V_{0}$. The inflationary evolution is satisfactorily described by the Friedmann and Klein-Gordon equations in the slow-roll 
approximation:
\begin{equation}
	H^{2}=\left(\frac{\dot{a}}{a}\right)^{2}=\frac{1}{3M_{p}^{2}} V(\phi),\qquad
	3H\dot{\phi}+V'_{\phi}=0.\label{eqs-sr}
\end{equation}

Using these equations, one can relate the value of the inflaton field with the corresponding number of $e$-foldings to the end of inflation $N=\ln\frac{a_{e}}{a}$:
\begin{equation}
	\label{N-on-phi}
	\frac{d\phi}{d N}=M_{p}^{2}\frac{V'_{\phi}}{V}, \quad \Rightarrow \quad 
	N=-\int_{\phi/M_{p}}^{\phi_{e}/M_{p}}dx\frac{v(x,\xi_{i})}{v'_{x}(x,\xi_{i})}.
\end{equation}

To the first order in the slow-roll parameters, the spectral index of scalar perturbations and the tensor-to-scalar power ratio can be expressed as follows \cite{Martin:2013,Gorbunov-book-v2}:
\begin{equation}
	n_{s}=1-6\epsilon_{\star}+2\eta_{\star},\qquad
	r=16\epsilon_{\star}.\label{ns-r}
\end{equation}
They must be calculated at the moment of time $t_{\star}$ when the pivot scale $k_{\star}$ crosses the horizon. All quantities taken at this moment of time will be denoted by an asterisk.

The amplitude of the potential $V_{0}$ can be constrained by means of the amplitude of a primordial scalar power spectrum. Indeed, this amplitude can be predicted from the slow-roll inflationary dynamics and equals \cite{Gorbunov-book-v2}
\begin{equation}
	\label{amplitude}
	A_{s}=\left(\frac{H^{2}}{2\pi\dot{\phi}}\right)^{2}\Bigg|_{N_{\star}}=\frac{1}{12\pi^{2}}\frac{V_{0}}{M_{p}^{4}}\frac{v^{3}}{(v'_{x})^{2}}\Bigg|_{N_{\star}}.
\end{equation}

For the pivot scale $k_{\star}/a_{0}=0.05\,{\rm Mpc}^{-1}$, the Planck Collaboration reports the following numerical values of the parameters:
\begin{equation}
	\ln[10^{-10}A_{s}]=3.044\pm 0.014, \quad n_{s}=0.9649\pm 0.0042
\end{equation}
at 68\% C.L. The tensor-to-scalar power ratio is constrained only from above, $r<0.063$ at 95\% C.L.

The pivot scale exits the Hubble horizon during inflation when $k_{\star}=a_{\star}H_{\star}$. Thus, we have the following relation:
\begin{equation}
	H_{\star}=\frac{k_{\star}}{a_{\star}}=\frac{k_{\star}}{a_{0}}\times\frac{a_{0}}{a_{\rm reh}}\times\frac{a_{\rm reh}}{a_{e}}\times\frac{a_{e}}{a_{\star}},
\end{equation}
where indices $e$, ${\rm reh}$, and $0$ correspond to quantities calculated at the end of inflation, at the end of reheating, and at the present time, respectively. The last multiplier is equal to $\exp(N_{\star})$. For simplicity, we assume that the reheating is instantaneous. Therefore, ${a_{\rm reh}}/{a_{e}}=1$ and the reheating temperature is given by
\begin{equation}
	T_{\rm reh}=\left(\frac{30 V(\phi_{e})}{\pi^{2}g_{\ast}}\right)^{1/4},
\end{equation}
where $g_{\ast}=106.75$ is the total effective number of the relativistic degrees of freedom in the Standard Model. The relation between $a_{\rm reh}$ and $a_{0}$ is given by Eq.~(\ref{a0-ae}). Then, using the fact that $H_{\star}^{2}=V(\phi_{\star})/(3M_{p}^{2})$, we get the following equation for $N_{\star}$:
\begin{equation}
	\label{N-star}
	N_{\star}-\frac{1}{4}\ln r(N_{\star})=\ln\left(\frac{g_{0}^{1/3}T_{0}}{k_{\star}/a_{0}}\right)+\frac{1}{4}\ln\left(\frac{\pi^{4}}{180}\frac{A_{s}}{g_{\ast}^{1/3}}\right)+\frac{1}{4}\ln\frac{V_{\star}}{V_{e}}\approx 56.8,
\end{equation}
where the last term on the right-hand side can be neglected for the plateau models. Using the explicit dependence $r(N_{\star})$ from Eq.~(\ref{ns-r}), one can solve Eq.~(\ref{N-star}) for $N_{\star}$. Finally, the amplitude of the potential can be determined from Eq.~(\ref{amplitude}).

\subsection{Metric Higgs inflation}

For metric Higgs inflation the function $v$ has no internal parameters
\begin{equation}
	v(x)=\Big[1-\exp\Big(-\sqrt{\frac{2}{3}}x\Big)\Big]^{2}.
\end{equation}
The slow-roll parameters are the following:
\begin{equation}
	\epsilon= \frac{4}{3}\frac{1}{\left[\exp\left(\sqrt{\frac{2}{3}}x\right)-1\right]^{2}},\quad
	\eta=-\frac{4}{3}\frac{\exp\left(\sqrt{\frac{2}{3}}x\right)-2}{\left[\exp\left(\sqrt{\frac{2}{3}}x\right)-1\right]^{2}}.
\end{equation}
Inflation ends when $\epsilon=1$, and the corresponding value of the inflaton reads as
\begin{equation}
	\frac{\phi_{e}}{M_{p}}=\sqrt{\frac{3}{2}}\ln\left(1+\frac{2}{\sqrt{3}}\right)\approx 0.94.
\end{equation}
Using Eq.~(\ref{N-on-phi}), we get the following equation:
\begin{equation}
	\label{xast-alpha-transcend}
	\exp\left(\sqrt{\frac{2}{3}}x_{\star}\right)-\sqrt{\frac{2}{3}}x_{\star}=\frac{4}{3}N_{\star}+1+\frac{2}{\sqrt{3}}-\ln\left(1+\frac{2}{\sqrt{3}}\right),
\end{equation}
whose approximate solution for $N_{\star}\gg 1$ has the form
\begin{equation}
	x_{\star}\approx \sqrt{\frac{3}{2}}\ln\left[\frac{4}{3}N_{\star}\right].
\end{equation}
Using this solution, we calculate the spectral characteristics of primordial perturbations
\begin{eqnarray}
	n_{s}&=&1-\frac{8}{3}\frac{\exp\left(\sqrt{\frac{2}{3}}x_{\star}\right)+1}{\left[\exp\left(\sqrt{\frac{2}{3}}x_{\star}\right)-1\right]^{2}}\approx 1-\frac{2}{N_{\star}},\\
	r&=&\frac{64}{3}\frac{1}{\left[\exp\left(\sqrt{\frac{2}{3}}x_{\star}\right)-1\right]^{2}}\approx \frac{12}{N_{\star}^{2}}.
\end{eqnarray}

Solving Eq.~(\ref{N-star}), we get $N_{\star}\approx 55.4$. The corresponding values of the inflationary parameters are listed in Table~\ref{tab-parameters}.

\subsection{Palatini Higgs inflation}

For Palatini Higgs inflation, the shape of the potential depends only on one parameter $\xi$:
\begin{equation}
	v(x,\xi)={\rm tanh}^{4}(\sqrt{\xi}x).
\end{equation}
The slow-roll parameters read as
\begin{equation}
	\epsilon=\frac{32\xi}{{\rm sinh}^{2}(2\sqrt{\xi}x)}, \quad \eta=\frac{-16\xi[{\rm cosh}(2\sqrt{\xi}x)-4]}{{\rm sinh}^{2}(2\sqrt{\xi}x)}
\end{equation}

Despite the fact that $|\eta|>\epsilon$ for $\xi\gg 1$, inflation still lasts until $\epsilon=1$ is reached, implying
\begin{equation}
\frac{\phi_{e}}{M_{p}}=\frac{1}{2\sqrt{\xi}}{\rm arccosh}(\sqrt{32\xi+1}).
\end{equation}
Integrating Eq.~(\ref{N-on-phi}), it is straightforward to get
\begin{equation}
	x_{\star}=\frac{1}{2\sqrt{\xi}}{\rm arccosh}\left(16\xi N_{\star}+\sqrt{1+32\xi}\right),
\end{equation}
i.e., during the whole inflation stage the scalar field remains sub-Planckian $\phi\sim M_{p}/\sqrt{\xi}\ll M_{p}$. The spectral index of the scalar perturbations and the tensor-to-scalar power ratio are equal to
\begin{eqnarray}
	n_{s}&=&1-\frac{16N_{\star}\xi+2+\sqrt{1+32\xi}}{8N_{\star}^{2}\xi+N_{\star}\sqrt{1+32\xi}+1}\approx 1-\frac{2}{N_{\star}},\\
	r&=&\frac{16}{8N_{\star}^{2}\xi+N_{\star}\sqrt{1+32\xi}+1}\approx \frac{2}{\xi N^{2}_{\star}}.
\end{eqnarray}
From Eq.~(\ref{N-star}), we obtain that $N_{\star}\approx 54.9-(1/4)\ln\xi\approx 50.9$ for $\xi=10^{7}$. The numerical values for the amplitude of the potential and other parameters are listed in Table~\ref{tab-parameters}.

\section{Relation between conformal times}
\label{app-B}

In this appendix, we derive Eq.~(\ref{relation-conformal-times}) which relates the conformal times at the end of reheating $\tau_{\rm reh}$ and recombination $\tau_{\rm r}$. First, let us determine the Hubble parameter deep 
in the radiation domination epoch
\begin{equation}
	H^{2}=\frac{1}{3M_{p}^{2}}\rho_{\rm tot}=\frac{1}{3M_{p}^{2}}\frac{\pi^{2}}{30}g_{\ast}T^{4}=\frac{1}{3M_{p}^{2}}\frac{\pi^{2}}{30}\frac{g_{\ast,0}^{4/3}}{g_{\ast}^{1/3}}\frac{T_{0}^{4}a_{0}^{4}}{a^{4}},
\end{equation}
where $g_{\ast}$ is the effective number of relativistic degrees of freedom at temperature $T$, and we used the comoving entropy conservation in the last step. For temperatures $T>200\,$GeV, $g_{\ast}={\rm const}$, and the Hubble parameter changes in time as $H=H_{1}a_{0}^{2}/a^{2}$, where $H_{1}=(\pi/3\sqrt{10})(g_{\ast,0}^{2/3}/g_{\ast}^{1/6})(T_{0}^{2}/M_{p})={\rm const}$. Then, the conformal time equals
\begin{equation}
	\label{tau-RD}
	\tau=\int_{0}^{t}\frac{dt'}{a(t')}\approx \int_{0}^{a}\frac{d\tilde{a}}{\tilde{a}^{2}H(\tilde{a})}= \frac{1}{H_{1}a_{0}^{2}}a=\frac{3\sqrt{10}}{\pi}\frac{M_{p}}{a_{0}T_{0}g_{\ast,0}^{1/3}}\frac{1}{g_{\ast}^{1/6}T}.
\end{equation}
Inserting here $T=T_{\rm reh}$ and $g_{\ast,{\rm reh}}$, we would get $\tau_{\rm reh}$.

Further, let us determine the conformal time at recombination. This epoch is very close to the matter-radiation equality, that is why we should keep both matter and radiation contributions to the Hubble parameter
\begin{equation}
	H=H_{0}\sqrt{\Omega_{M}\left(\frac{a_{0}}{a}\right)^{3}+\Omega_{\rm rad}\left(\frac{a_{0}}{a}\right)^{4}}.
\end{equation}
Here, $\Omega_{M}=\rho_{M,0}/\rho_{c}$ is the matter density parameter, $\rho_{M,0}$ is the total nonrelativistic matter energy density today, and $\rho_{c}=3M_{p}^{2}H_{0}^{2}$ is the critical density. The radiation density parameter is equal to \cite{Gorbunov-book-v1}
\begin{equation}
	\label{Omega-rad}
	\Omega_{\rm rad}=\frac{\rho_{{\rm rad},0}}{\rho_{c}}=\frac{\pi^{2}\hat{g}_{\ast,0}T_{0}^{4}}{90M_{p}^{2}H_{0}^{2}},
\end{equation}
where $\hat{g}_{\ast,0}=2+(7/8)\times 6\times (4/11)^{4/3}\approx 3.36$ is the present-day effective number of relativistic degrees of freedom for the energy density. Then, the conformal time is given by
\begin{equation}
	\tau=\int_{0}^{a}\frac{d\tilde{a}}{\tilde{a}^{2}H(\tilde{a})}=\frac{1}{a_{0}H_{0}\sqrt{\Omega_{M}}}\int_{0}^{a/a_{0}}\frac{dx}{\sqrt{x+\Omega_{\rm rad}/\Omega_{M}}}=\frac{2}{a_{0}H_{0}\sqrt{\Omega_{\rm rad}}}\frac{T_{0}}{T_{\rm eq}}\left(\sqrt{\frac{T_{\rm eq}}{T}+1}-1\right),
\end{equation}
where we used the fact that for temperatures below 0.5~MeV the simple relation $aT=a_{0}T_{0}$ holds, and introduced the temperature at matter-radiation equality $T_{\rm eq}=T_{0} \Omega_{M}/\Omega_{\rm rad}$. Then, using Eq.~(\ref{Omega-rad}), we obtain
\begin{equation}
	\label{tau-r}
	\tau_{\rm r}=\frac{6\sqrt{10}}{\pi}\frac{M_{p}}{a_{0}T_{0}\hat{g}_{\ast,0}^{1/2}}\frac{1}{T_{\rm eq}}\left(\sqrt{\frac{T_{\rm eq}}{T_{\rm r}}+1}-1\right).
\end{equation}
Finally, taking the ratio of Eqs.(\ref{tau-r}) and (\ref{tau-RD}) we get Eq.~(\ref{relation-conformal-times}), where the dimensionless factor $\kappa$ equals
\begin{equation}
	\kappa=2\frac{g_{\ast,0}^{1/3}g_{\ast,{\rm reh}}^{1/6}}{\hat{g}_{\ast,0}^{1/2}}\frac{T_{\rm r}}{T_{\rm eq}}\left(\sqrt{\frac{T_{\rm eq}}{T_{\rm r}}+1}-1\right)\approx 1.32
\end{equation}
for $T_{\rm r}=4000\,$K and $T_{\rm eq}=6.57\times10^{4}\,\Omega_{M}h^2\,{\rm K}=9400\,$K [the latter value is based on the best fit results for $\Omega_{M}=0.3153$ and $H_{0}=67.36\,{\rm km/(sec\, Mpc)}$ measured by the Planck Collaboration \cite{Planck:2018-infl}].


\begin{thebibliography}{999}
	
	
	\bibitem{Tavecchio:2010} F.~Tavecchio, G.~Ghisellini, L.~Foschini, G.~Bonnoli, G.~Ghirlanda, and P.~Coppi,  {The intergalactic magnetic field constrained by Fermi/Large Area Telescope observations of the TeV blazar 1ES 0229+200}, \href{https://doi.org/10.1111/j.1745-3933.2010.00884.x}{Mon. Not. R. Astron. Soc. \textbf{406}, L70 (2010)} [\href{https://arxiv.org/abs/1004.1329}{arXiv: 1004.1329 [astro-ph.CO]}]. 
	
	\bibitem{Ando:2010}	S.~Ando and A.~Kusenko, Evidence for gamma-ray halos around active galactic nuclei and the first measurement of intergalactic magnetic fields,
	\href{https://doi.org/10.1088/2041-8205/722/1/L39}{Astrophys. J. Lett. \textbf{722}, L39 (2010)}	[\href{https://arxiv.org/abs/1005.1924}{arXiv: 1005.1924 [astro-ph.HE]}]. 
	
	\bibitem{Neronov:2010} A.~Neronov and I.~Vovk,  {Evidence for strong extragalactic magnetic fields from Fermi observations of TeV blazars}, \href{https://doi.org/10.1126/science.1184192}{Science \textbf{328}, 73 (2010) [\href{https://arxiv.org/abs/1006.3504}{arXiv: 1006.3504 [astro-ph.HE]}].} 
	
	\bibitem{Dolag:2010} K.~Dolag, M.~Kachelriess, S.~Ostapchenko, and R.~Tomas,
	Lower limit on the strength and filling factor of extragalactic magnetic fields,
	\href{https://doi.org/10.1088/2041-8205/727/1/L4}{Astrophys. J. Lett. \textbf{727}, L4 (2011)} [\href{https://arxiv.org/abs/1009.1782}{arXiv: 1009.1782 [astro-ph.HE]}]. 
	
	\bibitem{Dermer:2011} C.D.~Dermer, M.~Cavadini, S.~Razzaque, J.D.~Finke, J.~Chiang, and B.~Lott,  {Time delay of cascade radiation for TeV blazars and the measurement of the intergalactic magnetic field}, \href{https://doi.org/10.1088/2041-8205/733/2/L21}{Astrophys. J. Lett. \textbf{733}, L21 (2011)} [\href{https://arxiv.org/abs/1011.6660}{arXiv: 1011.6660 [astro-ph.HE]}]. 
	
	\bibitem{Taylor:2011} A.M.~Taylor, I.~Vovk, and A.~Neronov,  {Extragalactic magnetic fields constraints from simultaneous GeV-TeV observations of blazars}, \href{https://doi.org/10.1051/0004-6361/201116441}{Astron. Astrophys. \textbf{529}, A144 (2011)} [\href{https://arxiv.org/abs/1101.0932}{arXiv: 1101.0932 [astro-ph.HE]}]. 
	
	\bibitem{Caprini:2015} C.~Caprini and S.~Gabici,  {Gamma-ray observations of blazars and the intergalactic magnetic field spectrum}, \href{https://doi.org/10.1103/PhysRevD.91.123514}{Phys. Rev. D \textbf{91}, 123514 (2015)} [\href{https://arxiv.org/abs/1504.00383}{arXiv: 1504.00383 [astro-ph.CO]}]. 
	
	
	
	
	\bibitem{Kronberg:1994} P.P.~Kronberg, {Extragalactic magnetic fields}, \href{https://doi.org/10.1088/0034-4885/57/4/001}{Rep. Prog. Phys. \textbf{57}, 325 (1994).} 
	
	\bibitem{Grasso:2001} D.~Grasso and H.R.~Rubinstein, {Magnetic fields in the early Universe}, \href{https://doi.org/10.1016/S0370-1573(00)00110-1}{Phys. Rep. \textbf{348}, 163 (2001)} [\href{https://arxiv.org/abs/astro-ph/0009061}{arXiv: astro-ph/0009061}]. 
	
	\bibitem{Widrow:2002} L.M.~Widrow, {Origin of galactic and extragalactic magnetic fields}, \href{https://doi.org/10.1103/RevModPhys.74.775}{Rev. Mod. Phys. \textbf{74}, 775 (2002)} [\href{https://arxiv.org/abs/astro-ph/0207240}{arXiv: astro-ph/0207240}]. 
	
	\bibitem{Giovannini:2004} M.~Giovannini, {The magnetized Universe}, \href{https://doi.org/10.1142/S0218271804004530}{Int. J. Mod. Phys. D \textbf{13}, 391 (2004)} [\href{https://arxiv.org/abs/astro-ph/0312614}{arXiv: astro-ph/0312614}]. 
	
	\bibitem{Kandus:2011} A.~Kandus, K.E.~Kunze, and C.~G.~Tsagas, {Primordial magnetogenesis}, \href{https://doi.org/10.1016/j.physrep.2011.03.001}{Phys. Rep. \textbf{505}, 1 (2011)} [\href{https://arxiv.org/abs/1007.3891}{arXiv: 1007.3891 [astro-ph.CO]}]. 
	
	\bibitem{Durrer:2013} R.~Durrer and A.~Neronov, {Cosmological magnetic fields: Their generation, evolution and observation}, \href{https://doi.org/10.1007/s00159-013-0062-7}{Astron. Astrophys. Rev. \textbf{21}, 62 (2013)} [\href{https://arxiv.org/abs/1303.7121}{arXiv: 1303.7121 [astro-ph.CO]}]. 
	
	\bibitem{Subramanian:2016} K.~Subramanian, {The origin, evolution and signatures of primordial magnetic fields}, \href{https://doi.org/10.1088/0034-4885/79/7/076901}{Rep. Prog. Phys. \textbf{79}, 076901 (2016)} [\href{https://arxiv.org/abs/1504.02311}{arXiv: 1504.02311 [astro-ph.CO]}]. 
	
	
	
	
	\bibitem{Zeldovich:1980book} Ya.B.~Zeldovich, A.A.~Ruzmaikin, and D.D.~Sokoloff, \textit{Magnetic Fields in Astrophysics} (Gordon and Breach, New York, 1990). 
	
	\bibitem{Lesch:1995} H.~Lesch and M.~Chiba, Protogalactic evolution and magnetic fields, Astron. Astrophys. \textbf{297}, 305 (1995) [\href{https://arxiv.org/abs/astro-ph/9411072}{arXiv: astro-ph/9411072}].
	
	\bibitem{Kulsrud:1997} R.~Kulsrud, S.C.~Cowley, A.V.~Gruzinov, and R.~N.~Sudan, Dynamos and cosmic magnetic fields, \href{https://doi.org/10.1016/S0370-1573(96)00061-0}{Phys. Rep. \textbf{283}, 213 (1997)}. 
		
	\bibitem{Colgate:2001} S.A.~Colgate and H.~Li, The origin of the magnetic fields of the universe: The plasma astrophysics of the free energy of the Universe, \href{https://doi.org/10.1063/1.1351827}{Phys. Plasmas \textbf{8}, 2425 (2001)} [\href{https://arxiv.org/abs/astro-ph/0012484}{arXiv: astro-ph/0012484}]. 
	
	
	\bibitem{Harrison:1970} E.R.~Harrison, Fluctuations at the threshold of classical cosmology,
	\href{https://doi.org/10.1103/PhysRevD.1.2726}{Phys. Rev. D \textbf{1}, 2726 (1970)}.
	
	\bibitem{Zeldovich:1972} Ya.B.~Zeldovich, A hypothesis, unifying the structure and the entropy of the Universe, \href{https://doi.org/10.1093/mnras/160.1.1P}{Mon. Not. R. Astron. Soc. \textbf{160}, 1P (1972)}. 
	
	\bibitem{Chibisov:1982} G.V.~Chibisov and V.F.~Mukhanov, Galaxy formation and phonons,
	\href{https://doi.org/10.1093/mnras/200.3.535}{Mon. Not. R. Astron. Soc. \textbf{200}, 535 (1982)}. 
	
	\bibitem{Mukhanov:1992} V.F.~Mukhanov, H.A.~Feldman, and R.H.~Brandenberger,
	Theory of cosmological perturbations. Part 1. Classical perturbations. Part 2. Quantum theory of perturbations. Part 3. Extensions,
	\href{https://doi.org/10.1016/0370-1573(92)90044-Z}{Phys. Rep. \textbf{215}, 203 (1992)}.
	
	\bibitem{Durrer:book} R.~Durrer, \textit{The Cosmic Microwave Background} (Cambridge University Press, New York, 2008).

	
	
	
	\bibitem{Turner:1988} M.S.~Turner and L.M.~Widrow,  {Inflation-produced, large-scale magnetic fields}, \href{https://doi.org/10.1103/PhysRevD.37.2743}{Phys. Rev. D \textbf{37}, 2743 (1988).} 
	
	\bibitem{Ratra:1992} B.~Ratra, Cosmological ``seed'' magnetic field from inflation, Astrophys. J. \textbf{391}, L1 (1992).
	
	\bibitem{Garretson:1992} W.~D.~Garretson, G.~B.~Field, and S.~M.~Carroll,  {Primordial magnetic fields from pseudo-Goldstone bosons}, \href{https://doi.org/10.1103/PhysRevD.46.5346}{Phys. Rev. D {\bf 46}, 5346 (1992)} [\href{https://arxiv.org/abs/hep-ph/9209238}{arXiv: hep-ph/9209238}]. 
	
	\bibitem{Dolgov:1993} A.D.~Dolgov,  {Breaking of conformal invariance and electromagnetic field generation in the Universe}, \href{https://doi.org/10.1103/PhysRevD.48.2499}{Phys. Rev. D \textbf{48}, 2499 (1993)} [\href{https://arxiv.org/abs/hep-ph/9301280}{arXiv: hep-ph/9301280}]. 
	


	\bibitem{Giovannini:2001} M.~Giovannini,  {Variation of the gauge couplings during inflation}, \href{https://doi.org/10.1103/PhysRevD.64.061301}{Phys. Rev. D \textbf{64}, 061301(R) (2001)} [\href{https://arxiv.org/abs/astro-ph/0104290}{arXiv: astro-ph/0104290}]. 
	
	\bibitem{Bamba:2004} K.~Bamba and J.~Yokoyama,  {Large-scale magnetic fields from inflation in dilaton electromagnetism}, \href{https://doi.org/10.1103/PhysRevD.69.043507}{Phys. Rev. D \textbf{69}, 043507 (2004)} [\href{https://arxiv.org/abs/astro-ph/0310824}{arXiv: astro-ph/0310824}]. 
	
	\bibitem{Martin:2008} J.~Martin and J.~Yokoyama,  {Generation of large scale magnetic fields in single-field inflation}, \href{https://doi.org/10.1088/1475-7516/2008/01/025}{J. Cosmol. Astropart. Phys. 01 (2008) 025} [\href{https://arxiv.org/abs/0711.4307}{arXiv: 0711.4307 [astro-ph]}]. 
	
	\bibitem{Kanno:2009} S.~Kanno, J.~Soda, and M.~Watanabe,  {Cosmological magnetic fields from inflation and backreaction}, \href{https://doi.org/10.1088/1475-7516/2009/12/009}{J. Cosmol. Astropart. Phys. 12 (2009) 009} [\href{https://arxiv.org/abs/0908.3509}{arXiv: 0908.3509 [astro-ph.CO]}]. 
	
	\bibitem{Demozzi:2009} V.~Demozzi, V.M.~Mukhanov, and H.~Rubinstein,  {Magnetic fields from inflation?}, \href{https://doi.org/10.1088/1475-7516/2009/08/025}{J. Cosmol. Astropart. Phys. 08 (2009) 025} [\href{https://arxiv.org/abs/0907.1030}{arXiv: 0907.1030 [astro-ph.CO]}]. 
	
	\bibitem{Ferreira:2013} R.J.Z.~Ferreira, R.K.~Jain, and M.S.~Sloth,  {Inflationary magnetogenesis without the strong coupling problem}, \href{https://doi.org/10.1088/1475-7516/2013/10/004}{J. Cosmol. Astropart. Phys. 10 (2013) 004} [\href{https://arxiv.org/abs/1305.7151}{arXiv: 1305.7151 [astro-ph.CO]}]. 
	
	\bibitem{Ferreira:2014} R.J.Z.~Ferreira, R.K.~Jain, and M.S.~Sloth,  {Inflationary magnetogenesis without the strong coupling problem. II. Constraints from CMB anisotropies and $B$-modes}, \href{https://doi.org/10.1088/1475-7516/2014/06/053}{J. Cosmol. Astropart. Phys. 06 (2014) 053} [\href{https://arxiv.org/abs/1403.5516}{arXiv: 1403.5516 [astro-ph.CO]}]. 
	
	\bibitem{Vilchinskii:2017} S.~Vilchinskii, O.~Sobol, E.V.~Gorbar, and I.~Rudenok,  {Magnetogenesis during inflation and preheating in the Starobinsky model}, \href{https://doi.org/10.1103/PhysRevD.95.083509}{Phys. Rev. D \textbf{95}, 083509 (2017)} [\href{https://arxiv.org/abs/1702.02774}{arXiv: 1702.02774 [astro-ph.CO]}]. 
	
	\bibitem{Sharma:2017b} R.~Sharma, S.~Jagannathan, T.R.~Seshadri, and K.~Subramanian,  {Challenges in inflationary magnetogenesis: Constraints from strong coupling, backreaction, and the Schwinger effect}, \href{https://doi.org/10.1103/PhysRevD.96.083511}{Phys. Rev. D \textbf{96}, 083511 (2017)} [\href{https://arxiv.org/abs/1708.08119}{arXiv: 1708.08119 [astro-ph.CO]}]. 
	
	\bibitem{Savchenko:2018} O.~Savchenko and Yu.~Shtanov, Magnetogenesis by non-minimal coupling to gravity in the Starobinsky inflationary model, \href{https://doi.org/10.1088/1475-7516/2018/10/040}{J. Cosmol. Astropart. Phys. 10 (2018) 040} [\href{https://arxiv.org/abs/1808.06193}{arXiv: 1808.06193 [astro-ph.CO]}]. 
		
	\bibitem{Sobol:2018} O.O.~Sobol, E.V.~Gorbar, M.~Kamarpour, and S.I.~Vilchinskii,  Influence of backreaction of electric fields and Schwinger effect on inflationary magnetogenesis, \href{https://doi.org/10.1103/PhysRevD.98.063534}{Phys. Rev. D \textbf{98}, 063534 (2018)} [\href{https://arxiv.org/abs/1807.09851}{arXiv: 1807.09851 [hep-ph]}]. 
	
	\bibitem{Shtanov:2020} Y.~Shtanov and M.~Pavliuk, Model-independent constraints in inflationary magnetogenesis,
	\href{https://doi.org/10.1088/1475-7516/2020/08/042}{J. Cosmol. Astropart. Phys. 08 (2020) 042} [\href{https://arxiv.org/abs/2004.00947}{arXiv: 2004.00947 [astro-ph.CO]}]. 
	
	\bibitem{Talebian:2020}	A.~Talebian, A.~Nassiri-Rad, and H.~Firouzjahi, Revisiting magnetogenesis during inflation, \href{https://doi.org/10.1103/PhysRevD.102.103508}{Phys. Rev. D \textbf{102}, 103508 (2020)} [\href{https://arxiv.org/abs/2007.11066}{arXiv: 2007.11066 [gr-qc]}]. 
	
	\bibitem{Durrer:2011} R.~Durrer, L.~Hollenstein, and R.K.~Jain, Can slow roll inflation induce relevant helical magnetic fields?, \href{https://doi.org/10.1088/1475-7516/2011/03/037}{J. Cosmol. Astropart. Phys. 03 (2011) 037} [\href{http://arxiv.org/abs/1005.5322}{arXiv: 1005.5322 [astro-ph.CO]}]. 
	
	\bibitem{Anber:2006} M.M.~Anber and L.~Sorbo, $N$-flationary magnetic fields, \href{https://doi.org/10.1088/1475-7516/2006/10/018}{J. Cosmol. Astropart. Phys. 10 (2006) 018} [\href{https://arxiv.org/abs/astro-ph/0606534}{arXiv: astro-ph/0606534}]. 
	
	\bibitem{Anber:2010} M.M.~Anber and L.~Sorbo, Naturally inflating on steep potentials through electromagnetic dissipation, \href{https://doi.org/10.1103/PhysRevD.81.043534}{Phys. Rev. D \textbf{81}, 043534 (2010)} [\href{https://arxiv.org/abs/0908.4089}{arXiv: 0908.4089 [hep-th]}]. 
	
	\bibitem{Barnaby:2012} N.~Barnaby, E.~Pajer, and M.~Peloso, Gauge field production in axion inflation: Consequences for monodromy, non-Gaussianity in the CMB, and gravitational waves at interferometers, \href{https://doi.org/10.1103/PhysRevD.85.023525}{Phys. Rev. D \textbf{85}, 023525 (2012)} [\href{https://arxiv.org/abs/1110.3327}{arXiv: 1110.3327 [astro-ph.CO]}]. 
	
	\bibitem{Caprini:2014} C.~Caprini and L.~Sorbo, Adding helicity to inflationary	magnetogenesis, \href{https://doi.org/10.1088/1475-7516/2014/10/056}{J. Cosmol. Astropart. Phys. 10 (2014) 056} [\href{https://arxiv.org/abs/1407.2809}{arXiv: 1407.2809 [astro-ph.CO]}]. 
	
	\bibitem{Anber:2015} M.M.~Anber and E.~Sabancilar, Hypermagnetic fields and baryon asymmetry from pseudoscalar inflation, \href{https://doi.org/10.1103/PhysRevD.92.101501}{Phys. Rev. D \textbf{92}, 101501(R) (2015)} [\href{https://arxiv.org/abs/1507.00744}{arXiv: 1507.00744 [hep-th]}]. 
	
	\bibitem{Ng:2015} K.-W.~Ng, S.-L.~Cheng, and W.~Lee, Inflationary dilaton-axion magnetogenesis, \href{https://doi.org/10.6122/CJP.20150909}{Chin. J. Phys. (Taipei) \textbf{53}, 110105 (2015)} [\href{https://arxiv.org/abs/1409.2656}{arXiv:1409.2656 [astro-ph.CO]}]. 
	
	\bibitem{Fujita:2015} T.~Fujita, R.~Namba, Y.~Tada, N.~Takeda, and H.~Tashiro, Consistent generation of magnetic fields in axion inflation models, \href{https://doi.org/10.1088/1475-7516/2015/05/054}{J. Cosmol. Astropart. Phys. 05 (2015) 054} [\href{http://arxiv.org/abs/1503.05802}{arXiv: 1503.05802 [astro-ph.CO]}]. 
	
	\bibitem{Adshead:2015} P.~Adshead, J.T.~Giblin, Jr., T.R.~Scully, and E.I.~Sfakianakis, Gauge-preheating and the end of axion inflation, \href{https://doi.org/10.1088/1475-7516/2015/12/034}{J. Cosmol. Astropart. Phys. 12 (2015) 034} [\href{https://arxiv.org/abs/1502.06506}{arXiv: 1502.06506 [astro-ph.CO]}]. 
	
	\bibitem{Adshead:2016} P.~Adshead, J.T.~Giblin, Jr., T.R.~Scully, and E.I.~Sfakianakis, Magnetogenesis from axion inflation, \href{https://doi.org/10.1088/1475-7516/2016/10/039}{J. Cosmol. Astropart. Phys. 10 (2016) 039} [\href{https://arxiv.org/abs/1606.08474}{arXiv: 1606.08474 [astro-ph.CO]}]. 
	
	\bibitem{Notari:2016} A.~Notari and K.~Tywoniuk, Dissipative axial inflation, \href{https://doi.org/10.1088/1475-7516/2016/12/038}{J. Cosmol. Astropart. Phys. 12 (2016) 038} [\href{http://arxiv.org/abs/1608.06223}{arXiv: 1608.06223 [hep-th]}]. 
	
	\bibitem{Domcke:2018eki} V.~Domcke and K.~Mukaida, Gauge field and fermion production during axion inflation, \href{https://doi.org/10.1088/1475-7516/2018/11/020}{J. Cosmol. Astropart. Phys. 11 (2018) 020} [\href{https://arxiv.org/abs/1806.08769}{arXiv: 1806.08769 [hep-ph]}].	
		
	\bibitem{Cuissa:2018} J.R.C.~Cuissa and D.G.~Figueroa, Lattice formulation of axion inflation. Application to preheating,
	\href{https://doi.org/10.1088/1475-7516/2019/06/002}{J. Cosmol. Astropart. Phys. 06 (2019) 002} [\href{https://arxiv.org/abs/1812.03132}{arXiv: 1812.03132 [astro-ph.CO]}].
	
	\bibitem{Shtanov:2019} Yu.~Shtanov, Viable inflationary magnetogenesis with helical coupling, \href{https://doi.org/10.1088/1475-7516/2019/10/008}{J. Cosmol. Astropart. Phys. 10 (2019) 008} [\href{https://arxiv.org/abs/1902.05894}{arXiv: 1902.05894 [astro-ph.CO]}].	
	
	\bibitem{Shtanov:2019b}	Y.V.~Shtanov and M.V.~Pavliuk, Inflationary magnetogenesis with helical coupling, \href{https://doi.org/10.15407/ujpe64.11.1009}{Ukr. J. Phys. \textbf{64}, 1009 (2019)} [\href{https://arxiv.org/abs/1911.10424}{arXiv: 1911.10424 [astro-ph.CO]}]. 
	
	\bibitem{Sobol:2019} O.O.~Sobol, E.V.~Gorbar, and S.I.~Vilchinskii, Backreaction of electromagnetic fields and the Schwinger effect in pseudoscalar inflation magnetogenesis, \href{https://doi.org/10.1103/PhysRevD.100.063523}{Phys.\ Rev.\ D {\bf 100}, 063523 (2019)} [\href{http://arxiv.org/abs/1907.10443}{arXiv: 1907.10443 [astro-ph.CO]}]. 
	
	\bibitem{Kamarpour:2019} M.~Kamarpour, Magnetogenesis in Higgs inflation, \href{https://doi.org/10.1007/s10714-021-02824-0}{Gen. Relativ. Gravit. \textbf{53}, 53 (2021)} [\href{http://arxiv.org/abs/1912.12540}{arXiv: 1912.12540 [gr-qc]}]. 
	
	\bibitem{Domcke:2020zez} V.~Domcke, V.~Guidetti, Y.~Welling, and A.~Westphal, Resonant backreaction in axion inflation, \href{https://doi.org/10.1088/1475-7516/2020/09/009}{J. Cosmol. Astropart. Phys. 09 (2020) 009} [\href{https://arxiv.org/abs/2002.02952}{arXiv: 2002.02952 [astro-ph.CO]}].	
	
	\bibitem{Bamba:2008} K.~Bamba and S.D.~Odintsov, Inflation and late-time cosmic acceleration in non-minimal Maxwell-$F(R)$ gravity and the generation of large-scale magnetic fields,
	\href{https://doi.org/10.1088/1475-7516/2008/04/024}{J. Cosmol. Astropart. Phys. 04 (2008) 024} [\href{https://arxiv.org/abs/0801.0954}{arXiv: 0801.0954 [astro-ph]}].	
	
	\bibitem{Bamba:2020} K.~Bamba, E.~Elizalde, S.D.~Odintsov, and T.~Paul, Inflationary magnetogenesis with reheating phase from higher curvature coupling,
	\href{https://doi.org/10.1088/1475-7516/2021/04/009}{J. Cosmol. Astropart. Phys. 04 (2021) 009} [\href{https://arxiv.org/abs/2012.12742}{arXiv: 2012.12742 [gr-qc]}].	
	
	\bibitem{Maity:2021} D.~Maity, S.~Pal, and T.~Paul, Effective theory of inflationary magnetogenesis and constraints on reheating, \href{https://doi.org/10.1088/1475-7516/2021/05/045}{J. Cosmol. Astropart. Phys. 05 (2021) 045} [\href{https://arxiv.org/abs/2103.02411}{arXiv: 2103.02411 [hep-th]}]. 
	
	
	

	

	\bibitem{Martin:2013} J.~Martin, C.~Ringeval, and V.~Vennin, Encyclop\ae{}dia inflationaris, \href{https://doi.org/10.1016/j.dark.2014.01.003}{Phys. Dark Univ. \textbf{5-6}, 75 (2014)} [\href{https://arxiv.org/abs/1303.3787}{arXiv: 1303.3787 [astro-ph.CO]}].	
	
	\bibitem{Planck:2018-infl} Y.~Akrami  \textit{et al.} (Planck Collaboration),  {Planck 2018 results. X. Constraints on inflation}, \href{https://doi.org/10.1051/0004-6361/201833887}{Astron. Astrophys. \textbf{641}, A10 (2020)} [\href{https://arxiv.org/abs/1807.06211}{arXiv: 1807.06211 [astro-ph.CO]}]. 
	
	\bibitem{Starobinsky:1980} A.A.~Starobinsky,  {A new type of isotropic cosmological models without singularity}, \href{https://doi.org/10.1016/0370-2693(80)90670-X}{Phys. Lett. \textbf{91B}, 99 (1980)}. 
	
	\bibitem{Ferrara:2013} S.~Ferrara, R.~Kallosh, A.~Linde, and M.~Porrati, Minimal supergravity models of inflation, \href{https://doi.org/10.1103/PhysRevD.88.085038}{Phys. Rev. D \textbf{88}, 085038 (2013)} [\href{https://arxiv.org/abs/1307.7696}{arXiv: 1307.7696 [hep-th]}]. 
	
	\bibitem{Kallosh:2013} R.~Kallosh, A.~Linde, and D.~Roest, Superconformal inflationary $\alpha$-attractors, \href{https://doi.org/10.1007/JHEP11(2013)198}{J. High Energy Phys. 11 (2013) 198} [\href{https://arxiv.org/abs/1311.0472}{arXiv: 1311.0472 [hep-th]}]. 
	
	\bibitem{Bezrukov:2007} F.L.~Bezrukov and M.~Shaposhnikov, The Standard Model Higgs boson as the inflaton, \href{https://doi.org/10.1016/j.physletb.2007.11.072}{Phys. Lett. B \textbf{659}, 703 (2008)} [\href{https://arxiv.org/abs/0710.3755}{arXiv: 0710.3755 [hep-th]}]. 
	
	\bibitem{Bauer:2008} F.~Bauer and D.A.~Demir, Inflation with non-minimal coupling: Metric versus Palatini formulations, \href{https://doi.org/10.1016/j.physletb.2008.06.014}{Phys. Lett. B \textbf{665}, 222 (2008)} [\href{https://arxiv.org/abs/0803.2664}{arXiv: 0803.2664 [hep-ph]}]. 
	
	\bibitem{Linde:1982} A.~D.~Linde, A new inflationary universe scenario: A possible solution of the horizon, flatness, homogeneity, isotropy and primordial monopole problems, \href{https://doi.org/10.1016/0370-2693(82)91219-9}{Phys. Lett. \textbf{108B}, 389 (1982)}. 
	
	\bibitem{Kallosh:2019} R.~Kallosh and A.~Linde, On hilltop and brane inflation after Planck, \href{https://doi.org/10.1088/1475-7516/2019/09/030}{J. Cosmol. Astropart. Phys. 09 (2019) 030} [\href{https://arxiv.org/abs/1906.02156}{arXiv: 1906.02156 [hep-th]}].	
	
	
	
	\bibitem{Palatini:1919} A.~Palatini, Deduzione invariantiva delle equazioni gravitazionali dal principio di Hamilton (Invariant derivation of gravitational equations from Hamilton's principle), \href{https://doi.org/10.1007/BF03014670}{Rend. Circ. Matem. Palermo \textbf{43}, 203 (1919)}. 
	
	\bibitem{Einstein:1925} A.~Einstein, Einheitliche Feldtheorie von Gravitation und Elektrizit{\"{a}}t (Unified field theory of gravitation and electricity), \href{https://doi.org/10.1002/3527608958.ch30}{Sitzungsber. Preuss. Akad. Wiss. Phys. Math. Kl. \textbf{414} (1925)}. 
	
	\bibitem{Rubio:2018} J.~Rubio, Higgs inflation, \href{https://doi.org/10.3389/fspas.2018.00050}{Front. Astron. Space Sci. \textbf{5}, 50 (2019)} [\href{https://arxiv.org/abs/1807.02376}{arXiv: 1807.02376 [hep-ph]}].	
	 
	\bibitem{Tenkanen:2020}	T.~Tenkanen, Tracing the high energy theory of gravity: An introduction to Palatini inflation, \href{https://doi.org/10.1007/s10714-020-02682-2}{Gen. Relativ. Grav. \textbf{52}, 33 (2020)} [\href{https://arxiv.org/abs/2001.10135}{arXiv: 2001.10135 [astro-ph.CO]}].	
	
	\bibitem{Shaposhnikov:2020}	M.~Shaposhnikov, A.~Shkerin, and S.~Zell, Quantum effects in Palatini Higgs inflation,
	\href{https://doi.org/10.1088/1475-7516/2020/07/064}{J. Cosmol. Astropart. Phys. 07 (2020) 064} [\href{https://arxiv.org/abs/2002.07105}{arXiv: 2002.07105 [hep-ph]}].	
	
	
	
	\bibitem{Frion:2020} E.~Frion, N.~Pinto-Neto, S.D.P.~Vitenti, and S.E.~Perez Bergliaffa, Primordial magnetogenesis in a bouncing universe,
	\href{https://doi.org/10.1103/PhysRevD.101.103503}{Phys. Rev. D \textbf{101}, 103503 (2020)} [\href{https://arxiv.org/abs/2004.07269}{arXiv: 2004.07269 [gr-qc]}].	
	
	
	
	
	\bibitem{Calmet:2016} X.~Calmet and I.~Kuntz, Higgs Starobinsky inflation,
	\href{https://doi.org/10.1140/epjc/s10052-016-4136-3}{Eur. Phys. J. C \textbf{76}, 289 (2016)} [\href{https://arxiv.org/abs/1605.02236}{arXiv: 1605.02236 [hep-th]}].	
	
	\bibitem{Antoniadis:2018} I.~Antoniadis, A.~Karam, A.~Lykkas, and K.~Tamvakis, Palatini inflation in models with an $R^2$ term,
	\href{https://doi.org/10.1088/1475-7516/2018/11/028}{J. Cosmol. Astropart. Phys. 11 (2018) 028} [\href{https://arxiv.org/abs/1810.10418}{arXiv: 1810.10418 [gr-qc]}].	
	
	\bibitem{Gialamas:2020}	I.D.~Gialamas, A.~Karam, A.~Lykkas, and T.~D.~Pappas, Palatini-Higgs inflation with nonminimal derivative coupling,
	\href{https://doi.org/10.1103/PhysRevD.102.063522}{Phys. Rev. D \textbf{102}, 063522 (2020)} [\href{https://arxiv.org/abs/2008.06371}{arXiv: 2008.06371 [gr-qc]}].	
	
		
	


		
	\bibitem{Momot:2019} 
	E.V.~Gorbar, A.I.~Momot, O.O.~Sobol, and S.I.~Vilchinskii,
	Kinetic approach to the Schwinger effect during inflation,
	\href{https://doi.org/10.1103/PhysRevD.100.123502}{Phys.\ Rev.\ D {\bf 100}, 123502 (2019)} [\href{http://arxiv.org/abs/1909.10332}{arXiv: 1909.10332 [gr-qc]}].
	
	
	
	\bibitem{Joyce:1997uy} M.~Joyce and M.~E.~Shaposhnikov, {Primordial Magnetic Fields, Right-Handed Electrons, and the Abelian Anomaly},
	\href{https://doi.org/10.1103/PhysRevLett.79.1193}{Phys. Rev. Lett. \textbf{79}, 1193 (1997)} [\href{https://arxiv.org/abs/astro-ph/9703005}{arXiv: astro-ph/9703005}]. 
	
	\bibitem{Boyarsky:2011uy} A.~Boyarsky, J.~Fr\"{o}hlich, and O.~Ruchayskiy, {Self-Consistent Evolution of Magnetic Fields and Chiral Asymmetry in the Early Universe},
	\href{https://doi.org/10.1103/PhysRevLett.108.031301}{Phys. Rev. Lett. \textbf{108}, 031301 (2012)} [\href{https://arxiv.org/abs/1109.3350}{arXiv:1109.3350 [astro-ph.CO]}]. 
	
	
	\bibitem{Bunch:1978} 
	T.S.~Bunch and P.C.W.~Davies, Quantum field theory in de Sitter space: Renormalization by point splitting,
	\href{https://doi.org/10.1098/rspa.1978.0060}{Proc.\ R.\ Soc.\ Lond.\ A {\bf 360}, 117 (1978)}. 
	



	\bibitem{Ema:2016} Y.~Ema, R.~Jinno, K.~Mukaida, and K.~Nakayama, Violent preheating in inflation with nonminimal coupling,
	\href{https://doi.org/10.1088/1475-7516/2017/02/045}{J. Cosmol. Astropart. Phys. 02 (2017) 045} [\href{https://arxiv.org/abs/1609.05209}{arXiv: 1609.05209 [hep-ph]}].	
	
	\bibitem{DeCross:2016}
	M.P.~DeCross, D.I.~Kaiser, A.~Prabhu, C.~Prescod-Weinstein, and E.I.~Sfakianakis, Preheating after multifield inflation with nonminimal couplings, III: Dynamical spacetime results,
	\href{https://doi.org/10.1103/PhysRevD.97.023528}{Phys. Rev. D \textbf{97}, 023528 (2018)} [\href{https://arxiv.org/abs/1610.08916}{arXiv: 1610.08916 [astro-ph.CO]}].	
	
	\bibitem{Rubio:2019} J.~Rubio and E.S.~Tomberg, Preheating in Palatini Higgs inflation,
	\href{https://doi.org/10.1088/1475-7516/2019/04/021}{J. Cosmol. Astropart. Phys. 04 (2019) 021} [\href{https://arxiv.org/abs/1902.10148}{arXiv: 1902.10148 [hep-ph]}].	
	
	
	\bibitem{Ahonen:1996nq} J.~Ahonen and K.~Enqvist, Electrical conductivity in the early Universe, \href{https://doi.org/10.1016/0370-2693(96)00633-8}{Phys. Lett. B \textbf{382}, 40 (1996)} [\href{https://arxiv.org/abs/hep-ph/9602357}{arXiv: hep-ph/9602357}]. 
	
	\bibitem{Baym:1997gq} G.~Baym and H.~Heiselberg, Electrical conductivity in the early Universe, \href{https://doi.org/10.1103/PhysRevD.56.5254}{Phys. Rev. D \textbf{56}, 5254 (1997)} [\href{https://arxiv.org/abs/astro-ph/9704214}{arXiv: astro-ph/9704214}]. 
	
	\bibitem{Ahonen:1998iz} J.~Ahonen, Transport coefficients in the early Universe, \href{https://doi.org/10.1103/PhysRevD.59.023004}{Phys. Rev. D \textbf{59}, 023004 (1998)} [\href{https://arxiv.org/abs/hep-ph/9801434}{arXiv: hep-ph/9801434}]. 
	
	
	\bibitem{Banerjee:2004} R.~Banerjee and K.~Jedamzik, Evolution of cosmic magnetic fields: From the very early Universe, to recombination, to the present, \href{https://doi.org/10.1103/PhysRevD.70.123003}{Phys. Rev. D \textbf{70}, 123003 (2004)} [\href{https://arxiv.org/abs/astro-ph/0410032}{arXiv: astro-ph/0410032}]. 
	
	
	
	\bibitem{Tashiro:2012mf} H.~Tashiro, T.~Vachaspati, and A.~Vilenkin, {Chiral effects and cosmic magnetic fields},
	\href{https://doi.org/10.1103/PhysRevD.86.105033}{Phys. Rev. D \textbf{86}, 105033 (2012)} [\href{https://arxiv.org/abs/1206.5549}{arXiv:1206.5549 [astro-ph.CO]}]. 
	
	\bibitem{Hirono:2015rla} Y.~Hirono, D.~Kharzeev, and Y.~Yin, Self-similar inverse cascade of magnetic helicity driven by the chiral anomaly,
	\href{https://doi.org/10.1103/PhysRevD.92.125031}{Phys. Rev. D \textbf{92}, 125031 (2015)} [\href{https://arxiv.org/abs/1509.07790}{arXiv:1509.07790 [hep-th]}]. 
	
	\bibitem{Dvornikov:2016jth} M.~Dvornikov and V.~B.~Semikoz, Influence of the turbulent motion on the chiral magnetic effect in the early Universe,
	\href{https://doi.org/10.1103/PhysRevD.95.043538}{Phys. Rev. D \textbf{95}, 043538 (2017)} [\href{https://arxiv.org/abs/1612.05897}{arXiv:1612.05897 [astro-ph.CO]}]. 
	
	\bibitem{Gorbar:2016klv} E.~V.~Gorbar, I.~Rudenok, I.~A.~Shovkovy, and S.~Vilchinskii, Anomaly-driven inverse cascade and inhomogeneities in a magnetized chiral plasma in the early Universe,
	\href{https://doi.org/10.1103/PhysRevD.94.103528}{Phys. Rev. D \textbf{94}, 103528 (2016)} [\href{https://arxiv.org/abs/1610.01214}{arXiv:1610.01214 [hep-ph]}]. 
	
	\bibitem{Brandenburg:2017rcb} A.~Brandenburg, J.~Schober, I.~Rogachevskii, T.~Kahniashvili, A.~Boyarsky, J.~Fr\"{o}hlich, O.~Ruchayskiy, and N.~Kleeorin, The turbulent chiral-magnetic cascade in the early Universe,
	\href{https://doi.org/10.3847/2041-8213/aa855d}{Astrophys. J. Lett. \textbf{845}, L21 (2017)} [\href{https://arxiv.org/abs/1707.03385}{arXiv:1707.03385 [astro-ph.CO]}]. 
	
	\bibitem{Schober:2018wlo} J.~Schober, A.~Brandenburg, and I.~Rogachevskii, Chiral fermion asymmetry in high-energy plasma simulations,
	\href{https://doi.org/10.1080/03091929.2019.1591393}{Geophys. Astrophys. Fluid Dyn. \textbf{114}, 106 (2020)} [\href{https://arxiv.org/abs/1808.06624}{arXiv:1808.06624 [physics.plasm-ph]}]. 
	
	\bibitem{Planck-pmf} P.A.R.~Ade \textit{et al}. (Planck Collaboration), Planck 2015 results. XIX. Constraints on primordial magnetic fields,
	\href{https://doi.org/10.1051/0004-6361/201525821}{Astron. Astrophys. {\bf 594}, A19 (2016)} [\href{https://arxiv.org/abs/1502.01594}{arXiv: 1502.01594 [astro-ph.CO]}]. 
	
	\bibitem{Kamada:2021} K.~Kamada, F.~Uchida, and J.~Yokoyama, Baryon isocurvature constraints on the primordial hypermagnetic fields,
	\href{https://doi.org/10.1088/1475-7516/2021/04/034}{J. Cosmol. Astropart. Phys. 04 (2021) 034} [\href{https://arxiv.org/abs/2012.14435}{arXiv: 2012.14435 [astro-ph.CO]}].	
	
	\bibitem{Gorbunov-book-v2} D.S.~Gorbunov and V.A.~Rubakov, \textit{Introduction to the Theory of the Early Universe: Cosmological Perturbations and Inflation Theory} (World Scientific, Singapore, 2011), Vol. 2.
	
	\bibitem{Gorbunov-book-v1} D.S.~Gorbunov and V.A.~Rubakov, \textit{Introduction to the Theory of the Early Universe: Hot Big Bang Theory} (World Scientific, Singapore, 2011), Vol. 1.

\end{thebibliography}
\end{document}